\pdfoutput=1 
\documentclass[letterpaper,twocolumn,10pt]{article}
\usepackage{usenix2019, epsfig, endnotes}
\usepackage{microtype}
\usepackage{color}
\usepackage{multirow}
\usepackage{url}
\usepackage{algorithm}
\usepackage[noend]{algpseudocode}
\usepackage{subfigure}
\usepackage{enumitem} 
\usepackage{times}
\usepackage{xspace}
\usepackage[table]{xcolor}
\usepackage{latexsym}
\usepackage{amsmath}
\usepackage{amsthm}
\usepackage{tabularx}
\usepackage{balance}
\usepackage[hidelinks]{hyperref}
\usepackage[small,compact]{titlesec}
\usepackage[square,sort,comma,numbers]{natbib}

\setlist{nolistsep}
\newcommand{\para}[1]{\medskip\noindent\textbf{#1}}

\newcommand{\paraf}[1]{\noindent\textbf{#1}}
\newcommand{\cut}[1]{}

\newcommand{\sysname}{RackSched\xspace}

\newtheorem{theorem}{Theorem}

\begin{document}
\date{}

\title{\Large \bf \sysname: A Microsecond-Scale Scheduler for Rack-Scale Computers\\(Technical Report)}

\author{
{\rm Hang Zhu}\\
Johns Hopkins University
\and
{\rm Kostis Kaffes}\\
Stanford University
\and
{\rm Zixu Chen}\\
Johns Hopkins University
\and
{\rm Zhenming Liu}\\
College of William and Mary
\and
{\rm Christos Kozyrakis}\\
Stanford University
\and
{\rm Ion Stoica}\\
UC Berkeley
\and
{\rm Xin Jin}\\
Johns Hopkins University
}

\maketitle

\begin{abstract}
Low-latency online services have strict Service Level Objectives (SLOs) that
require datacenter systems to support high throughput at microsecond-scale tail
latency. Dataplane operating systems have been designed to \emph{scale up}
multi-core servers with minimal overhead for such SLOs. However, as application
demands continue to increase, scaling up is not enough, and serving larger
demands requires these systems to scale out to multiple servers in a rack.

We present \sysname, the first rack-level microsecond-scale scheduler that
provides the abstraction of a rack-scale computer (i.e., a huge server with
hundreds to thousands of cores) to an external service with network-system
co-design. The core of \sysname is a two-layer scheduling framework that
integrates inter-server scheduling in the top-of-rack (ToR) switch with
intra-server scheduling in each server. We use a combination of analytical
results and simulations to show that it provides near-optimal performance as
centralized scheduling policies, and is robust for both low-dispersion and
high-dispersion workloads. We design a custom switch data plane for the
inter-server scheduler, which realizes power-of-k-choices, ensures request
affinity, and tracks server loads accurately and efficiently. We implement a
\sysname prototype on a cluster of commodity servers connected by a Barefoot
Tofino switch. End-to-end experiments on a twelve-server testbed show that
\sysname improves the throughput by up to 1.44$\times$, and scales out the
throughput near linearly, while maintaining the same tail latency as one server
until the system is saturated.
\end{abstract}

\section{Introduction}
\label{sec:introduction}

Online services such as search, social networking and
e-commerce have strict end-to-end user-facing Service Level Objectives
(SLOs)~\cite{Dean:tail, microseconds}. To meet such SLOs, datacenter systems behind these services
are expected to provide high throughput with low \emph{tail latency} in the range of tens
to hundreds of \emph{microseconds}~\cite{microseconds}. Example systems include
key-value stores~\cite{rocksdb, redis, memcached}, transactional databases~\cite{silo, voldtb}, search ranking and
sorting~\cite{barroso2003web}, microservices and function-as-a-service frameworks~\cite{boucher2018putting},
and graph stores~\cite{facebook-tao-sigmod12,splinter}.

With the end of Moore's law and Dennard's scaling,
applications can no longer rely on single-threaded code to execute faster on newer
processors with increased clock rates and instruction-level
parallelism~\cite{hennessy2019new}. This leads to the rise of multi-core
machines to scale up computation. Meeting microsecond-scale tail latency is
challenging given that the request processing times with single-threaded code on
bare metal hardware are already in the same time scale. This means that the
operating system (OS) should impose minimal overhead when it manages resources
and scales up these applications on multi-core machines.

This calls for new software architectures to efficiently utilize the resources
of multi-core machines. One critical component of such architectures is
scheduling. Dataplane operating systems have been designed to support low-latency
applications with minimal overhead to meet strict SLOs~\cite{peter2013arrakis,
belay2014ix, prekas2017zygos, shinjuku, shenango}. For example, Shinjuku~\cite{shinjuku} uses efficient mechanisms to
implement preemption and context switching, in order to realize
\emph{centralized} scheduling policies to avoid head-of-line blocking and
address temporal load imbalance between multiple cores.

However, as application demands continue to increase, \emph{scaling up} a single
server is not enough. Serving larger demands requires these systems
to \emph{scale out} to multiple servers in a rack, which is termed as rack-scale
computers, such as Berkeley Firebox~\cite{firebox}, Intel Rack Scale
Architecture~\cite{intel-rsd}, and HP ``The Machine''~\cite{hp-the-machine}.
Though previous efforts have not fully paned out yet, we believe it is
inevitable, as evidenced by the emerging TPU Pods that pack high-density
specialized hardware into a rack~\cite{tpu-pods}. Even a traditional rack
contains tens of servers and hundreds to thousands of cores, posing a challenge for
scheduling microsecond-scale requests.

While dataplane operating systems address intra-server scheduling between multiple
\emph{cores}, head-of-line blocking and temporal load imbalance between multiple
\emph{servers} arise when the systems scale out. Using a single core for
centralized scheduling and queue management is amenable for one server with a
few to tens of cores~\cite{shinjuku}. But the core would quickly become the
bottleneck if it were used to queue and schedule requests for a rack with
hundreds to thousands of cores.

In this paper, we present \sysname, the first rack-level microsecond-scale
scheduler that provides the abstraction of a rack-scale computer (i.e., a huge server with
hundreds to thousands of cores) to an external service with
network-system co-design. Serving microsecond-scale workloads is particularly
challenging because the scheduler needs to \emph{simultaneously} provide high
scheduling speed (i.e., high throughput and low latency of the scheduler) and
high scheduling quality (i.e., low tail latency to complete requests).
Given the scheduling latency of modern OSes on a single server being a few microseconds~\cite{cerqueira2013comparison, shinjuku}, our goal is to design a rack-level scheduler with comparable scheduling latency that can scale out to hundreds to thousands of cores in a rack.
While there
has been a long line of work on scheduling and load balancing, existing
solutions are not designed for microsecond-scale workloads: software-based
solutions~\cite{gog2016firmament,sparrow,ananta,maglev} suffer from low
scheduling throughput and high scheduling latency (at least millisecond-scale);
hardware-based ones~\cite{duet,miao2017silkroad} are coarse-grained (based on
five tuple) and server-agnostic, and thus suffer from long tail latency.

The core contribution of \sysname is a novel network-system co-design
that \emph{simultaneously} achieves high scheduling speed and high scheduling
quality (\S\ref{sec:overview}). We propose a two-layer scheduling framework that
integrates inter-server scheduling in the top-of-rack (ToR) switch with
intra-server scheduling in each server to approximate centralized scheduling
policies.
This two-layer design, and the
line-rate, on-path inter-server scheduling in the ToR switch, are critical for
the scheduler to achieve high speed.

\begin{figure}[t]
\centering
    \includegraphics[width=0.9\linewidth]{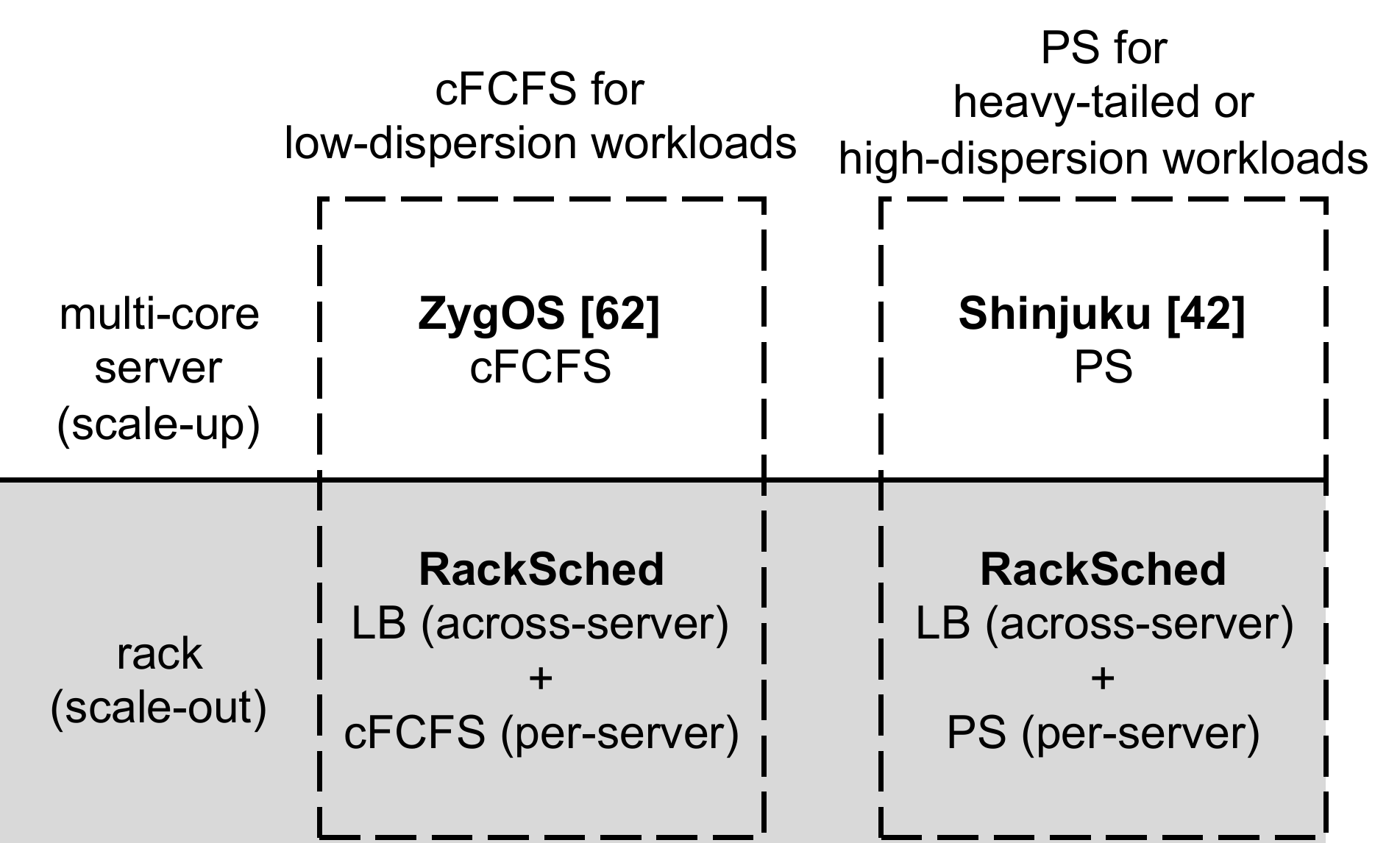}
\vspace{-0.1in}
\caption{Key idea of \sysname.}
\vspace{-0.1in}
\label{fig:intro_idea}
\end{figure}

To provide high quality scheduling decisions, our key insight is that the two
sources of long tail latency---load imbalance and head-of-line blocking---can be
decoupled and handled by separate mechanisms. The ToR switch tracks real-time
server loads, and schedules requests at \emph{per-request} granularity to
realize inter-server load balancing (LB). Each server keeps its
own queue, and uses intra-server scheduling to
preempt long requests that block pending short ones.
It is known that centralized first-come-first-serve (cFCFS) is near-optimal
for low-dispersion workloads, and processor sharing (PS)
is near-optimal for heavy-tailed workloads or light-tailed workloads with high dispersion~\cite{wierman2012tail, prekas2017zygos, shinjuku}.
We use a combination of
analytical results and simulations to show that our two-layer scheduling
framework provides \emph{near-optimal} performance as \emph{centralized}
policies, and is robust to different
workloads (Figure~\ref{fig:intro_idea}).

Realizing the inter-server scheduler in the
ToR switch requires the switch to schedule requests based on \emph{server
loads} on \emph{per-request} granularity (\S\ref{sec:design}). Today's stateful
network load balancers map connections to servers based the hash of the five
tuple~\cite{ananta, maglev, duet, miao2017silkroad, olteanu2018stateless}, which
is \emph{static}, and cannot dynamically adapt the server selection under
\emph{microsecond-scale} load imbalance. There are three aspects of our approach
to address this challenge. $(i)$ We leverage the switch on-chip memory to store
server loads, and use the multi-stage processing pipeline to implement
power-of-k-choices and to support a variety of
practical scheduling requirements, such as multi-queue policies. $(ii)$ We
design a \emph{request state table} for \emph{request affinity}, which
guarantees the packets of the same request are sent to the same server. To
maintain the dynamic mapping between requests and servers, the request state
table supports \emph{insert} (after scheduling the first packet), \emph{read}
(for sending following packets), and \emph{remove} (when the request is
completed) all in the \emph{data plane}. $(iii)$ We leverage \emph{in-network
telemetry} (INT) to monitor server loads with minimal overhead. Servers
\emph{piggyback} their load information in their \emph{normal} traffic, and the
switch tracks the latest reported load for each server.

Recent switch-based solution R2P2~\cite{kogias2019r2p2} relies on expensive
recirculation which does not scale for high request rate, and its scheduling
policy has long tail latency under heavy-tailed or high-dispersion workloads due to head-of-line
blocking (\S\ref{sec:evaluation:other}).
In addition, R2P2 needs an extra round trip for multi-packet requests to ensure request affinity,
while \sysname can finish in one round trip.
\sysname is also more general in supporting many practical policies (\S\ref{sec:design:extension}).

In summary, we make the following contributions.
\begin{itemize}[leftmargin=*]
    \item We propose \sysname, the first rack-level microsecond-scale scheduler
    that provides the abstraction of a rack-scale computer to an external service.

    \item We design a two-layer scheduling framework that integrates
    inter-server scheduling in the ToR switch and intra-server scheduling in
    each server. We use a combination of analytical results and simulations to
    show that it provides near-optimal performance as centralized policies.

    \item We design a custom switch data plane for the inter-server scheduler,
    which realizes power-of-k-choices for near-optimal load balancing, ensures
    request affinity, and tracks server loads accurately and efficiently.

    \item We implement a \sysname prototype on a cluster of twelve commodity servers
    with a Barefoot Tofino switch.
    End-to-end experiments show that \sysname improves the throughput by up to 1.44$\times$,
    and scales out the throughput
    near linearly, while maintaining the same tail latency as one server until
    the system is saturated.
\end{itemize}

\medskip\noindent
The code of \sysname is open-source and available at
\url{https://github.com/netx-repo/RackSched}.

\section{Design Decisions}
\label{sec:overview}

In this section, we navigate through the design space of building a
microsecond-scale scheduler for rack-scale computers, and highlight the design
rationale of \sysname.

\para{Scaling out to a rack.} Supporting large application
demands requires datacenter systems to scale out to multiple servers in a rack. While existing solutions like
Shinjuku~\cite{shinjuku} solve the problem of scheduling requests to multiple cores
(i.e., intra-server scheduling), they do not address the problem of scheduling
requests to different servers (i.e., inter-server scheduling). When requests are
simply scheduled to the servers randomly, the load imbalance and head-of-line
blocking can happen at the server level, causing long tail latency for the
entire system.

\begin{figure}[t]
    \centering
    \subfigure[Low-dispersion workload.]{
        \label{fig:sim_exp}
        \includegraphics[width=0.48\linewidth]{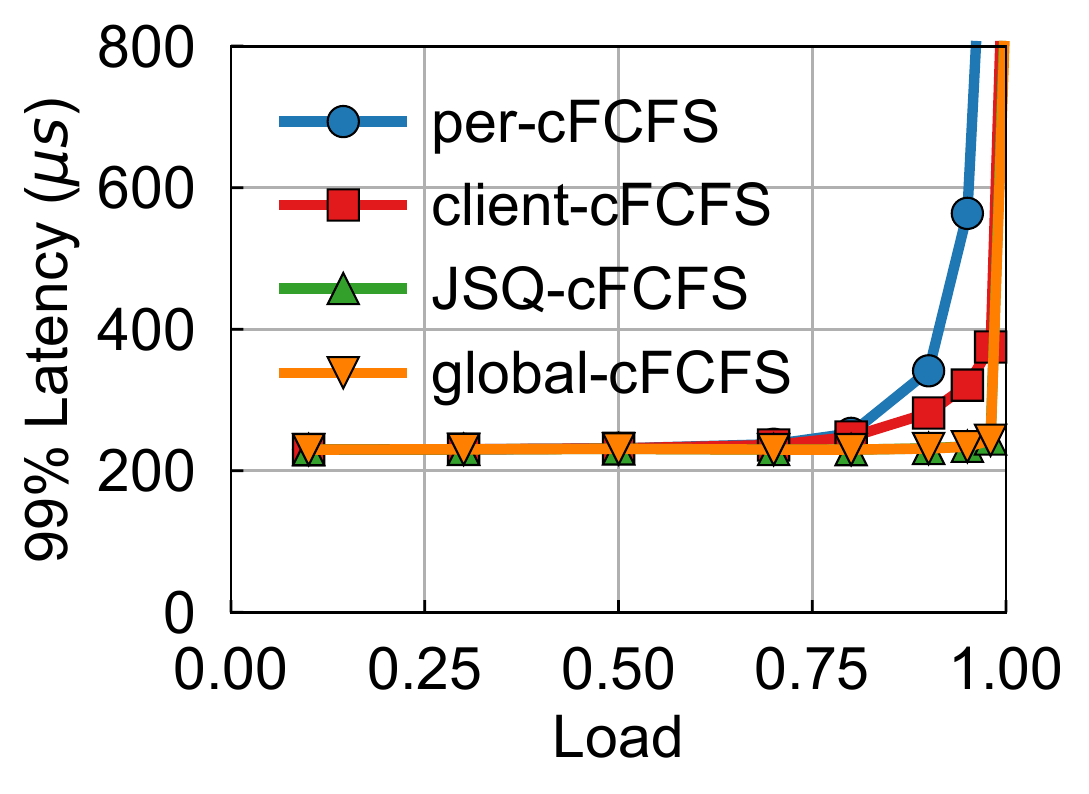}}
    \subfigure[High-dispersion workload.]{
        \label{fig:sim_bimodal}
        \includegraphics[width=0.48\linewidth]{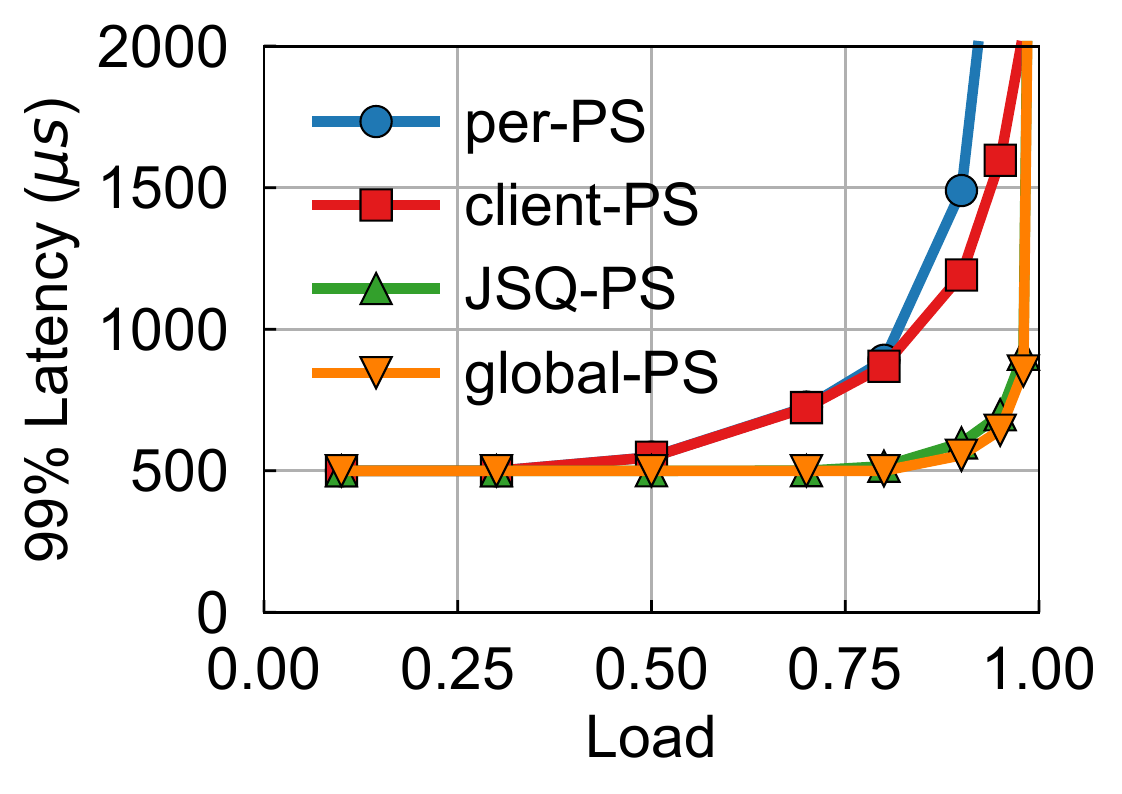}}
    \vspace{-0.15in}
    \caption{Simulation results.}
    \vspace{-0.1in}
    \label{fig:simulation}
\end{figure}

To motivate our work, we use simulations on representative workloads to show the
impact of ineffective scheduling policies. We use the following two request
service time distributions: $(i)$ Exp(50) is an exponential distribution with
mean = 50 $\mu s$, which is representative for low-dispersion workloads; $(ii)$
Trimodal(33.3\%-5, 33.3\%-50, 33.3\%-500) is a trimodal distribution with 33.3\% of requests
taking 5 $\mu s$, 33.3\% taking 50 $\mu s$ and 33.3\% taking 500 $\mu s$, which is representative for
high-dispersion workloads. The simulations assume eight servers and each server
has eight workers (cores). The PS time slice used in the simulations is 25 $\mu s$.

We first compare the baseline policies that randomly send requests to servers
and only use cFCFS or PS inside each server (per-cFCFS and per-PS) with the
ideal centralized policies across all workers (global-cFCFS and global-PS).
Figure~\ref{fig:simulation} shows that the centralized policies perform better
than the baseline policies. The tail latencies
of per-cFCFS and per-PS quickly go up when the system load exceeds 0.75 and 0.5 respectively, while
global-cFCFS and global-PS keep low tail latencies until the system is almost
saturated.

\para{Centralized scheduling cannot scale.} The policy comparison in Figure~\ref{fig:simulation} shows that
there is a substantial gap between the tail latencies of the centralized
policies (global-cFCFS and global-PS) and the baseline policies (per-cFCFS and
per-PS). However, directly implementing the centralized policies is challenging
because they would require a centralized scheduler for the entire rack. While a
single core is capable of running a centralized scheduler to handle the requests
for a multi-core server, it is unlikely to scale to a multi-server rack. Indeed,
a single scheduler in Shinjuku~\cite{shinjuku} can scale to up to 11 cores,
which falls well short of the demands of a rack with hundreds to thousands of
cores.

\begin{figure}[t]
\centering
    \includegraphics[width=0.95\linewidth]{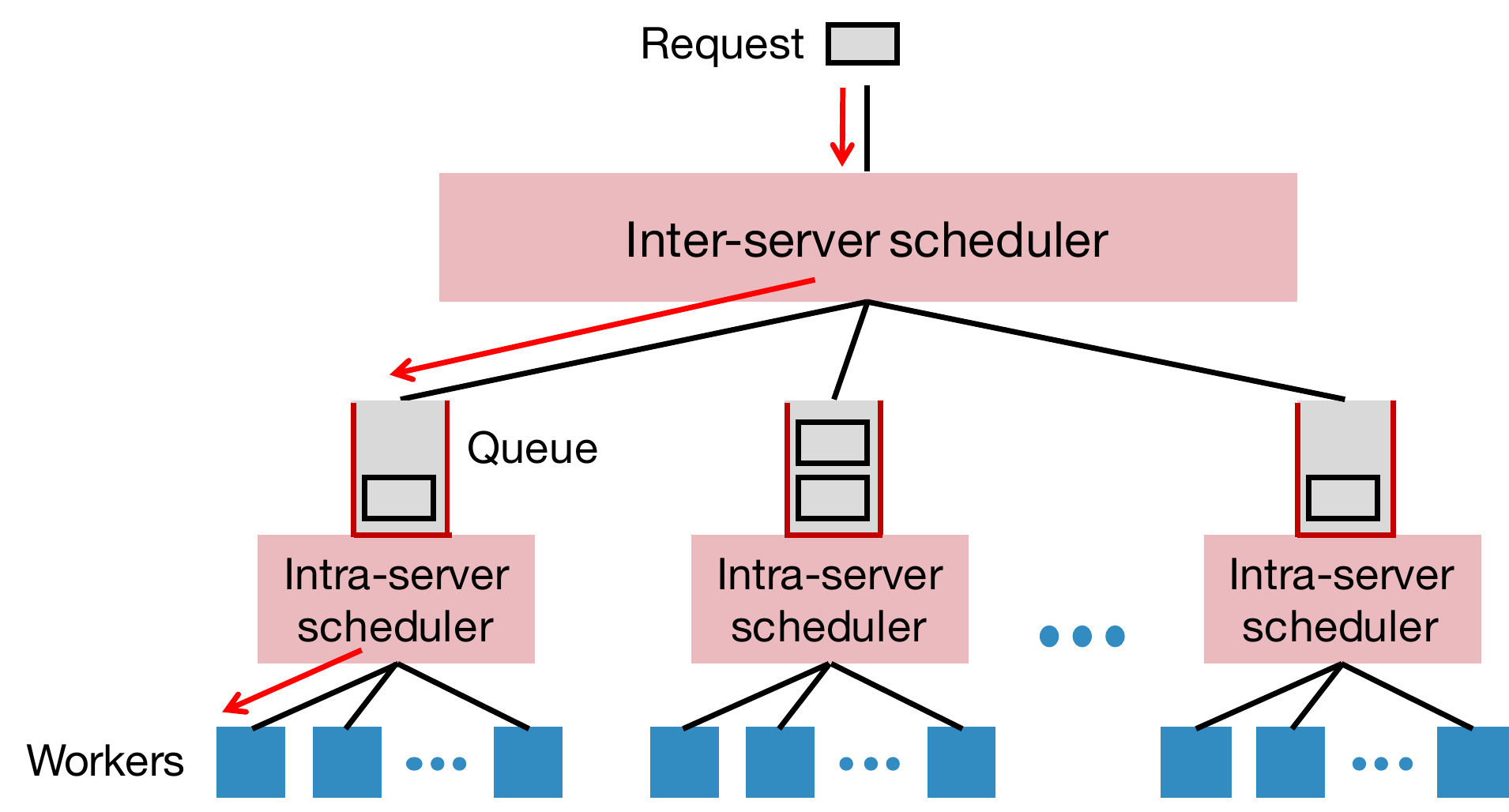}
\vspace{-0.15in}
\caption{Two-layer scheduling framework.}
\vspace{-0.1in}
\label{fig:design_concept}
\end{figure}

\para{Hierarchical scheduling.} One natural solution to scale up the rack-scale scheduler
is a two-layer hierarchical scheduler consisting of an inter-server scheduler at the high level,
and per-server schedulers at the low level (Figure~\ref{fig:design_concept}). This way, the
inter-server scheduler only needs to schedule requests across tens of servers, instead of
hundreds or thousands of cores. Each server employs its own intra-server scheduler to
schedule requests across its cores.

\para{Scaling the inter-server scheduler.} While the inter-server scheduler only needs
to schedule requests across the servers in the rack (instead of accross all cores),
it still needs to handle every  request. Assuming a rack with $1000$ cores and
10 ${\mu}s$ requests, the inter-server scheduler must handle up to 100 millions
requests per second (MRPS) to saturate the rack! Unfortunately, such a scheduler would
need to process a request every 10 $ns$, which exceeds the capability of a general-purpose computer.

To address this challenge, in this paper we propose to leverage emerging
\emph{programmable switches}, and have the ToR switch  implement the inter-server scheduler.
This design has the key benefit that the ToR switch is already on the path of the requests sent
to the rack, and thus can readily process all these requests at line rate.


\para{JSQ is near optimal and robust.} A natural way to approximate a
centralized scheduler with a two-layer scheduler is to implement the
same scheduler at both layers. For example, if the global scheduler is
cFCFS, in the corresponding hierarchical scheduler all the inter-server and
intra-server schedulers will be cFCFS as well.

Unfortunately, the cFCFS scheduler needs to maintain a queue, and existing
programmable switches have too limited memory to buffer requests, and
are not well equipped to maintain dynamic data structures,
such as queues.
The join-the-shortest-queue (JSQ) scheduler can address the challenge because it is a bufferless scheduler.
Upon a request arrival, it immediately forwards the request to
the server with the shortest queue. This way, JSQ achieves fine-grained load balancing
across the rack's servers. Figure~\ref{fig:simulation} confirms that JSQ-cFCFS and JSQ-PS can
deliver nearly the same performance as the centralized policies (global-cFCFS and global-PS).

Theoretically, the two-layer scheduling framework implements the A/S/K/JSQ/P models in
queueing theory, where A is the inter-arrival distribution, S is the service time distribution, K is the
number of servers, JSQ is the join-the-shortest-queue policy implemented by the
inter-server scheduler, and P is the intra-server scheduling policy which is
either cFCFS or PS in this case. In particular, it is known that JSQ provides
near-optimal load balancing, and importantly, is \emph{robust} against request service time
distributions.
An expanded discussion is in
Appendix~\ref{sec:appendix:analysis}.

\para{Approximating JSQ.} While conceptually simple, JSQ cannot be
implemented in its definite form in practice, because it requires the switch to
know the exact queue length of each server when scheduling a request. It takes a
round trip time for the switch to ask each server, during which the queue
lengths may have changed for microsecond-scale workloads.
Furthermore, imperfect JSQ based on delayed server status is prone to
\emph{herding}, where several consecutive requests are sent to the same server
before the server load is updated, and this can generate micro bursts and
degrade system performance. Note that herding here is not caused by multiple asynchronous load balancers as there is only one load balancer (the inter-server scheduler), but from stale server load information in the load balancer.
In addition, the switch can only do a limited number
of operations for each request, and finding the shortest queue cannot be
implemented for many tens of servers. Thus, we use power-of-k-choices to
approximate JSQ, which samples k servers for each request and chooses the one
with the minimum load. This approximation provides comparable
performance as JSQ in theory~\cite{bramson2010randomized}
(see Appendix~\ref{sec:appendix:analysis}),
and handles these practical
limitations well.

\para{Why not a distributed, client-based solution?} An alternative
solution is to implement distributed scheduling at
each client. The clients can use JSQ, power-of-k-choices or more complicated
solutions like C3~\cite{c3} to pick workers for their requests. Such a
client-based solution has two drawbacks. First, it needs client modification
and increases system complexity. The clients need to probe server
loads and make scheduling decisions. More importantly, the clients need to be
notified for \emph{every} system reconfiguration (e.g., adding or removing
servers), because they have to know which set of servers a request can be sent
to. Notifying a large number of clients of the latest system configuration
imposes both a consistency challenge and high system overhead. Putting these
functionalities in a scheduler, on the other hand, simplifies the clients and
avoids these system complexities.

Second, a distributed, client-based solution provides a worse trade-off between
performance and probing overhead than a centralized scheduler for
microsecond-scale workloads. This is because microsecond-scale workloads are
IO-intensive, and a probing request incurs comparable processing cost as a
normal request at the servers. Thus, probing needs to be minimized to improve
the throughput of processing the actual requests. No matter whether probing is
done proactively or piggybacked in reply packets, given $n$ clients with the same sending rate, a
centralized scheduler can utilize $n$ times as much probing data as that of one
client in a client-based solution, resulting in better scheduling quality.
Figure~\ref{fig:simulation} confirms the benefit of the centralized scheduler
over a client-based solution with a piggyback-based probing mechanism (client-cFCFS and client-PS).
The simulation does not model the client software delay and the network delay to
get the server loads, which favors the client-based solution. The client-based solution performs
worse in real experiments (\S\ref{sec:evaluation:other}).
In a multi-pipeline switch, though states are not shared across pipelines, \sysname
can approximate JSQ within each pipeline. It works better than the client-based solution
because the number of pipelines (e.g., 4) is far smaller than the number of clients (e.g., 1000 or more).

\para{Putting it all together.} We propose a two-layer scheduling framework that
integrates inter-server scheduling in the ToR switch and intra-server scheduling
in each server. The ToR switch uses power-of-k-choices to achieve inter-server
load balancing, and each server uses cFCFS or PS to minimize head-of-line
blocking. This solution approximates centralized scheduling for the entire rack,
and provides the abstraction of a rack-scale computer: the capacity (throughput)
of the rack-scale computer is the sum of that of its servers, and the tail
latency is maintained as that of one server.

\para{Challenges.} Translating the two-layer scheduling framework to a working
system implementation presents several technical challenges:

\begin{itemize}[leftmargin=*]
    \item What is the system architecture to realize this two-layer scheduling
    framework?

    \item How does the switch schedule requests based on the
    server loads, and handle practical scheduling requirements?

    \item How does the system ensure request affinity (i.e., the packets of the
    same request are sent to the same server), when the switch processes each
    packet independently?

    \item How do servers expose their states to the switch so that the switch
    can track the real-time loads on the servers efficiently and accurately?
\end{itemize}

\begin{figure}[t]
    \centering
    \subfigure[\sysname architecture.]{
        \label{fig:design_architecture}
        \includegraphics[width=0.95\linewidth]{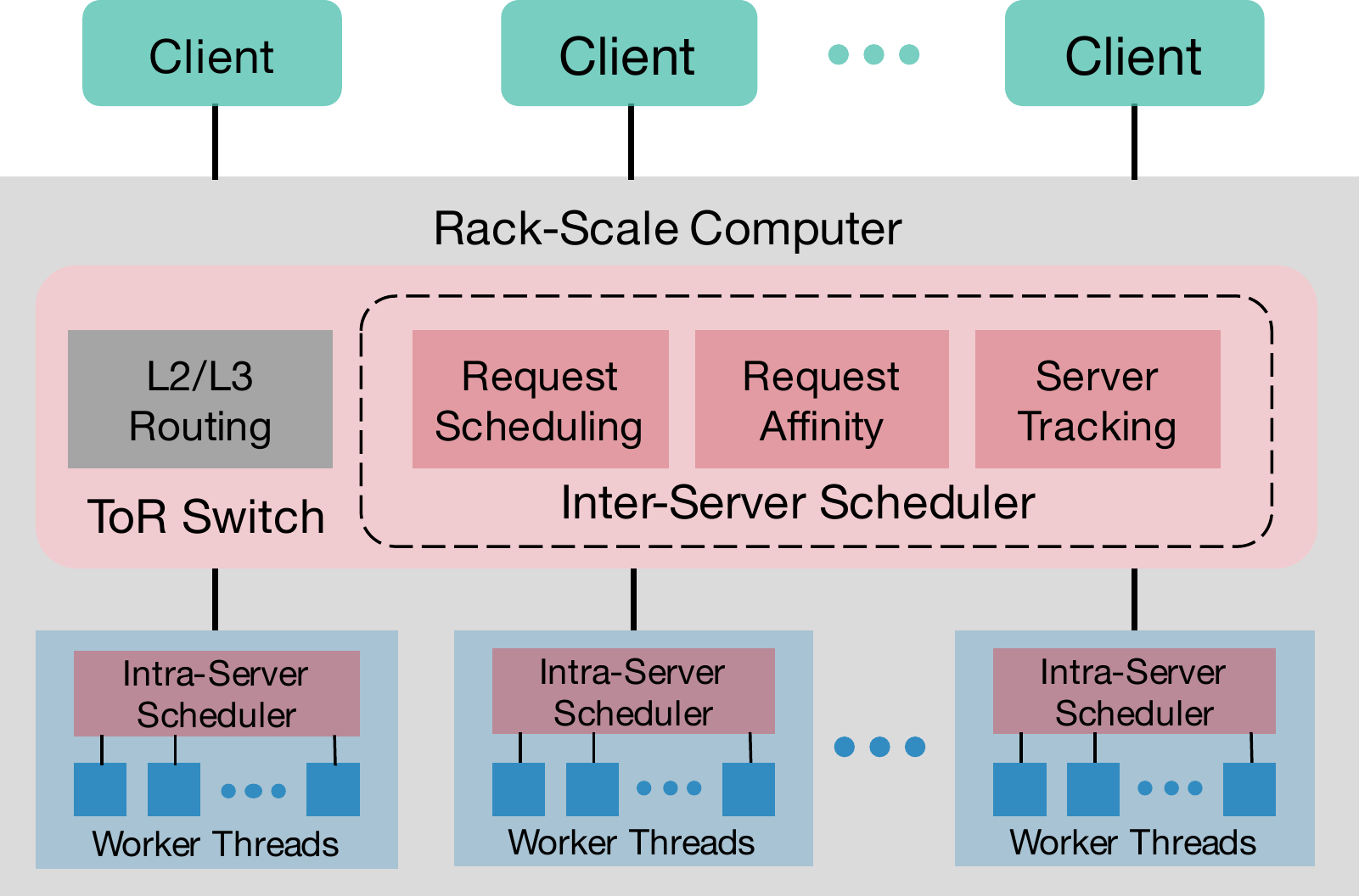}}
    \subfigure[\sysname packet format.]{
        \label{fig:design_format}
        \includegraphics[width=0.95\linewidth]{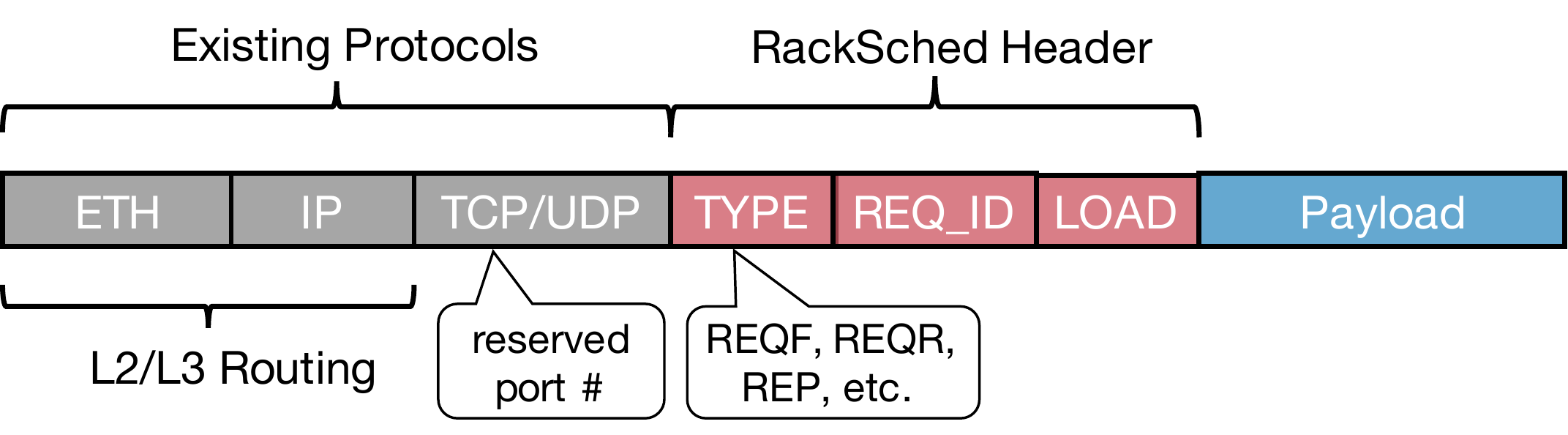}}
    \vspace{-0.1in}
    \caption{\sysname overview.}
    \vspace{-0.1in}
    \label{fig:design_overview}
\end{figure}

\section{\sysname Design}
\label{sec:design}

In this section, we present the design of \sysname to address the challenges. We
first give an overview of the system architecture, and then describe each
component in detail.

\subsection{System Architecture}
\label{sec:design:architecture}

The core of \sysname is a two-layer scheduling framework that combines
inter-server scheduling and intra-server scheduling.
Figure~\ref{fig:design_architecture} shows the \sysname architecture.

\begin{figure}[t]
    \centering
    \subfigure[The first packet of a request.]{
        \label{fig:design_request1}
        \includegraphics[width=0.95\linewidth]{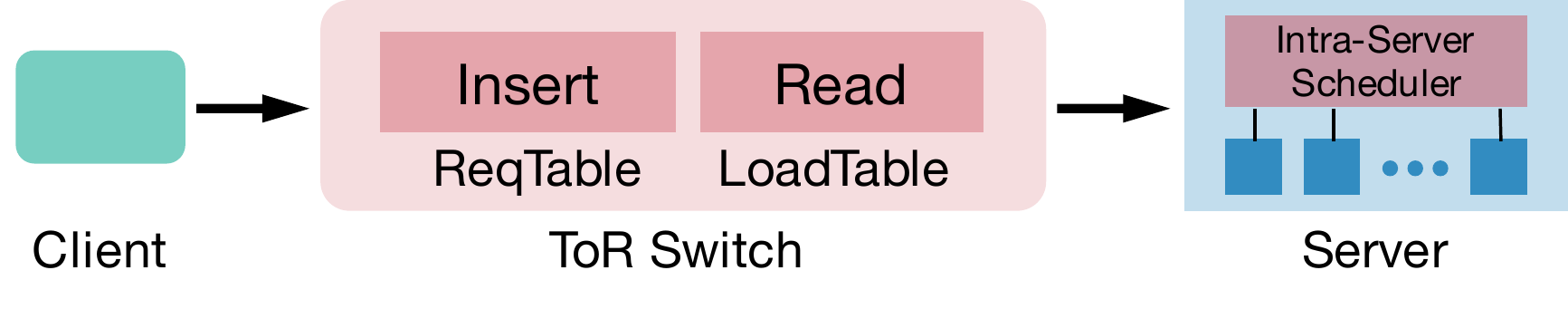}}
    \subfigure[The following packets of the request.]{
        \label{fig:design_request2}
        \includegraphics[width=0.95\linewidth]{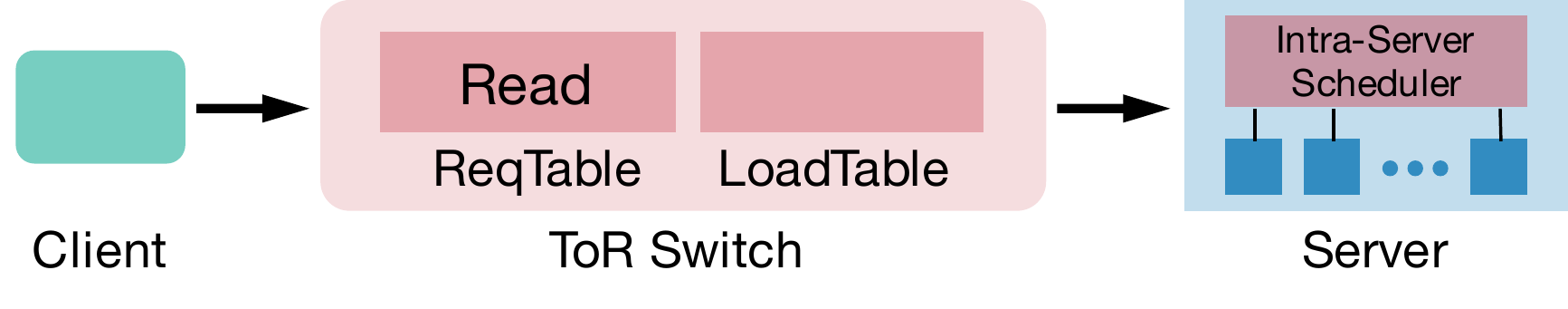}}
    \subfigure[The reply packets.]{
        \label{fig:design_request3}
        \includegraphics[width=0.95\linewidth]{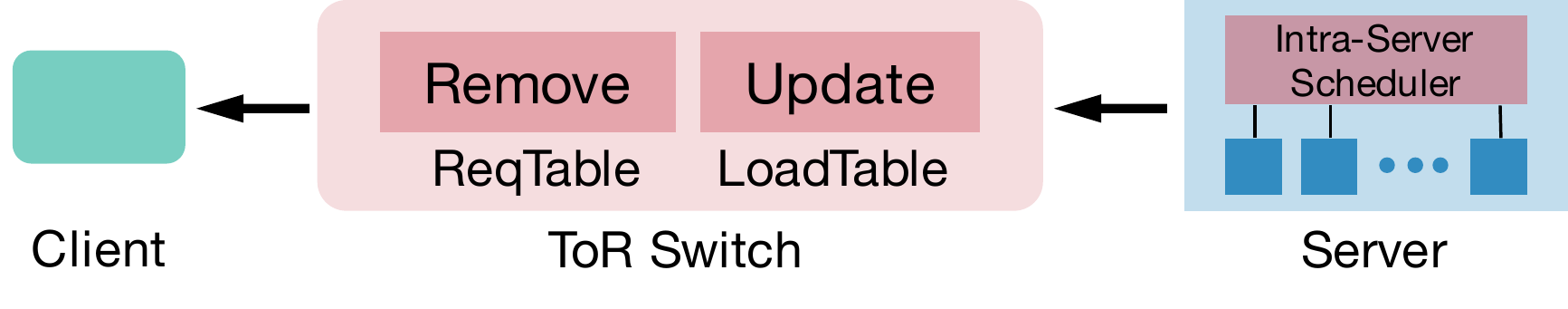}}
    \vspace{-0.15in}
    \caption{Request processing in \sysname.}
    \vspace{-0.1in}
    \label{fig:design_request}
\end{figure}

\para{Inter-server scheduling.} The ToR switch
performs inter-server scheduling at per-request
granularity via three modules: a request scheduling module that selects a server
for a new incoming request based on server loads (\S\ref{sec:design:scheduling})
and scheduling requirements (\S\ref{sec:design:extension}), a request affinity
module that forwards the packets of the same request to the same selected server
(\S\ref{sec:design:affinity}), and a server tracking module that tracks the
real-time load on each server (\S\ref{sec:design:tracking}). All three modules
are implemented in the switch data plane that enables the inter-server scheduler
to run at line rate.

\para{Intra-server scheduling.} Each multi-core server in the rack runs multiple
worker threads to process requests. Each server has a centralized scheduler to
queue and schedule requests to its own workers. The scheduler implements centralized scheduling policies
for intra-server scheduling.
Each server also piggybacks its load
information in reply messages to report its load to the ToR switch.

\para{Deployment options.} There are two deployment options for \sysname. $(i)$ The
first one is to integrate \sysname with the ToR switch of a rack-scale computer. This option adds additional functionalities to the
ToR switch, but does not change any other part of the datacenter network. $(ii)$
The second option is to treat the switch-based scheduler as a specialized server
with a programmable switching ASIC. This server can be connected to
the same ToR switch as worker servers. It owns the anycast
IP address so all requests would be first sent to it for scheduling. By properly updating the addresses, it can also
force the reply packets to pass through it before returning to the clients, in
order to clear request states and update server loads. This option does not even
modify the ToR switch, but has the latency cost of an extra hop by
the detour to the switch-based scheduler.

\subsection{Request Processing}
\label{sec:design:processing}

\paraf{Network protocol.} \sysname is designed for intra-datacenter scenarios.
It uses an application-layer protocol which is
embedded inside the L4 payload. We reserve an L4 port to distinguish \sysname
packets from other packets. The network uses existing L2/L3 routing protocols to
forward packets. There are no
modifications to the switches in the network other than the ToR switch. The ToR
switch uses the reserved L4 port to invoke the custom modules to process
\sysname packets. Other switches do not need to understand the \sysname
protocol, nor do they need to process \sysname packets.
\sysname is orthogonal to and compatible with other network functionalities, such as flow/congestion control,
which is typically implemented by the transport layer or the RPC layer (e.g., eRPC~\cite{erpc}).
\sysname ensures request affinity under packet loss and retransmission by maintaining connection state (\S\ref{sec:design:affinity}).

\sysname only requires the applications to add the
\sysname header between the TCP/UDP header and the payload
(Figure~\ref{fig:design_format}), so that it can make scheduling decisions based on the \sysname header.
Note that for TCP handshake packets that do not have any payload, the \sysname header
should be appended after the TCP header to expose necessary information to the
switch. We emphasize that \sysname focuses on microsecond-scale workloads. It is
not intended to support long-lived TCP connections, which impose unnecessary
system overhead to maintain connection states, especially under switch failures
(\S\ref{sec:design:affinity}), and restrict the scheduler from making
per-request scheduling decisions to address temporal load imbalance. \sysname
does support request dependency for tasks with multiple
requests (\S\ref{sec:design:extension}).

The \sysname header contains three major
fields, which are \texttt{TYPE}, \texttt{REQ\_ID}, and \texttt{LOAD}.
\texttt{TYPE} indicates the type of the packet, e.g., \texttt{REQF} (the first
packet of a request), \texttt{REQR} (a remaining packet of a request), and
\texttt{REP} (a reply packet). \texttt{REQ\_ID} is a unique ID for each
request. All packets of the same request and the corresponding reply use the same
\texttt{REQ\_ID}. To ensure a \texttt{REQ\_ID} is globally unique, each client
appends its client ID in front of its locally generated unique request ID, i.e.,
a tuple of $<$client ID, local request ID$>$. \texttt{LOAD} indicates the load
of the server. The server piggybacks its current queue length in the
\texttt{LOAD} field in reply packets. \texttt{LOAD} is not used in
request packets.

\begin{algorithm}[t!]
\caption{ProcessPacket(pkt)}
\begin{itemize}[noitemsep,nolistsep]
    \item[--] $ReqTable$: on-chip memory for request-server mapping
    \item[--] $LoadTable$: on-chip memory for server loads
\end{itemize}
\begin{algorithmic}[1]
\Statex \textbf{// first packet of a request}
\If{$pkt.type == REQF$}
    \State $server\_ip \gets LoadTable.select\_server()$
    \State $ReqTable.insert(pkt.req\_id, server\_ip)$
\Statex \textbf{// remaining packets of a request}
\ElsIf{$pkt.type == REQR$}
    \State $server\_ip \gets ReqTable.read(pkt.req\_id)$
\Statex \textbf{// reply packets}
\ElsIf{$pkt.type == REP$}
    \State $ReqTable.remove(pkt.req\_id)$
    \State $LoadTable.update(pkt.srcip, pkt.load)$
\EndIf
\State Update packet header and forward
\end{algorithmic}
\label{alg:switch}
\end{algorithm}

\para{Processing request packets.} Clients use an \emph{anycast} IP address as
the destination IP to send requests to the rack-scale computer, and are unaware
of the number of servers behind the ToR switch.
Figure~\ref{fig:design_request} illustrates how \sysname processes packets, and
Algorithm~\ref{alg:switch} shows the high-level pseudo code of the switch. The switch
keeps two essential sets of state in the switch on-chip memory. One is
$ReqTable$ which stores the mapping from the request IDs to the servers, and the
other is $LoadTable$ which stores the queue lengths of the servers. As shown in
Figure~\ref{fig:design_request1}, when the first packet of a request arrives at the switch, the
switch selects a server based on $LoadTable$, and remembers this selection by
inserting an entry to  $ReqTable$ (line 1-3). Then in Figure~\ref{fig:design_request2}, when remaining packets of the
request arrive, the switch checks the $ReqTable$ to get the selected server
(line 4-5), which ensures request affinity. The switch uses the
selected server IP to update the destination IP in the packet header and sends
the packet to the corresponding server (line 9).

\para{Processing reply packets.} After a server receives a request, it uses its
local scheduler to schedule and processes the request.
Then the server generates a reply, and sets the \texttt{LOAD} field with its
current queue length. As shown in Figure~\ref{fig:design_request3}, the switch deletes the mapping
from $ReqTable$, because the request has completed and the memory space can be
freed for other requests (line 7). The switch also updates the server load in
$LoadTable$ based on the \texttt{LOAD} field (line 8). We do not
distinguish the first and following packets for a reply even if the reply
contains \emph{multiple} packets (also equivalent to \emph{multiple} replies).
Because $ReqTable$ checks $req\_id$ for deletion, if a slot is reused by another request, the following reply packets of the previous request would not be applied.
The updates of $LoadTable$ only affect server selection of new requests. Note that this is
compatible with TCP even if the client initiates the termination of the
connection, as the mapping of this request is removed from $ReqTable$ when the
server receives the request and sends the first reply packet back to the client.
The switch control plane periodically checks the data plane to remove
\emph{stale} mappings from $ReqTable$, which can be caused by server failures or
lost reply packets. In the end, the switch updates the source IP to the anycast
IP in the packet header, and sends the packet back to the client (line 9).

\subsection{Request Scheduling}
\label{sec:design:scheduling}

\begin{figure}[t]
\centering
    \includegraphics[width=0.95\linewidth]{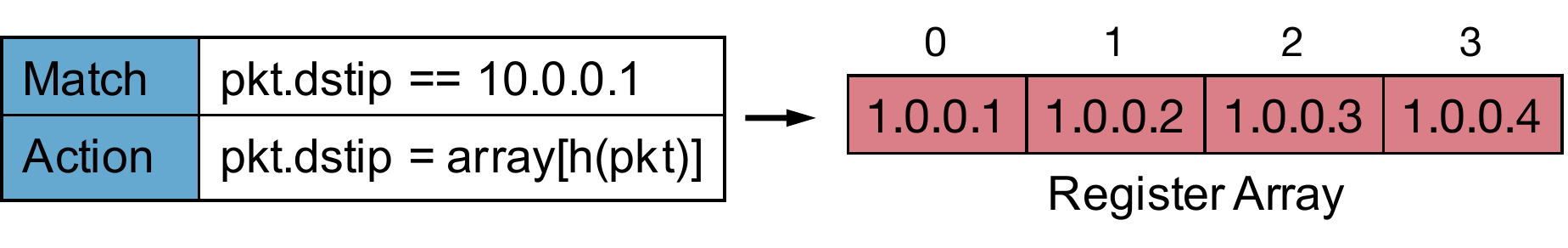}
\vspace{-0.2in}
\caption{Traditional hash-based random dispatching.}
\vspace{-0.15in}
\label{fig:design_schedule1}
\end{figure}

The request scheduling module dynamically schedules
requests based on server loads. Unfortunately, this is not supported
in today's switches. Today's switch-based load balancers such as
SilkRoad~\cite{miao2017silkroad} only support ECMP-like random dispatching based on the five tuple.
Figure~\ref{fig:design_schedule1} shows how
hash-based random selection is implemented in the data plane. The register array stores a
set of server IPs for the anycast IP 10.0.0.1. The rule in the
match-action table matches packets with their destination IP being the anycast
IP 10.0.0.1, and the action rewrites the destination IP to an IP in the register
array, which is selected by computing a hash on the packet header (usually the
five tuple). Because the selection is static, and is purely based on the hash,
it can cause load imbalance and long tail latency as discussed in
\S\ref{sec:overview}. We now describe how to realize dynamic request scheduling
based on server loads. Handling practical scheduling requirements is
in \S\ref{sec:design:extension}.

\begin{figure}[t]
\centering
    \includegraphics[width=0.95\linewidth]{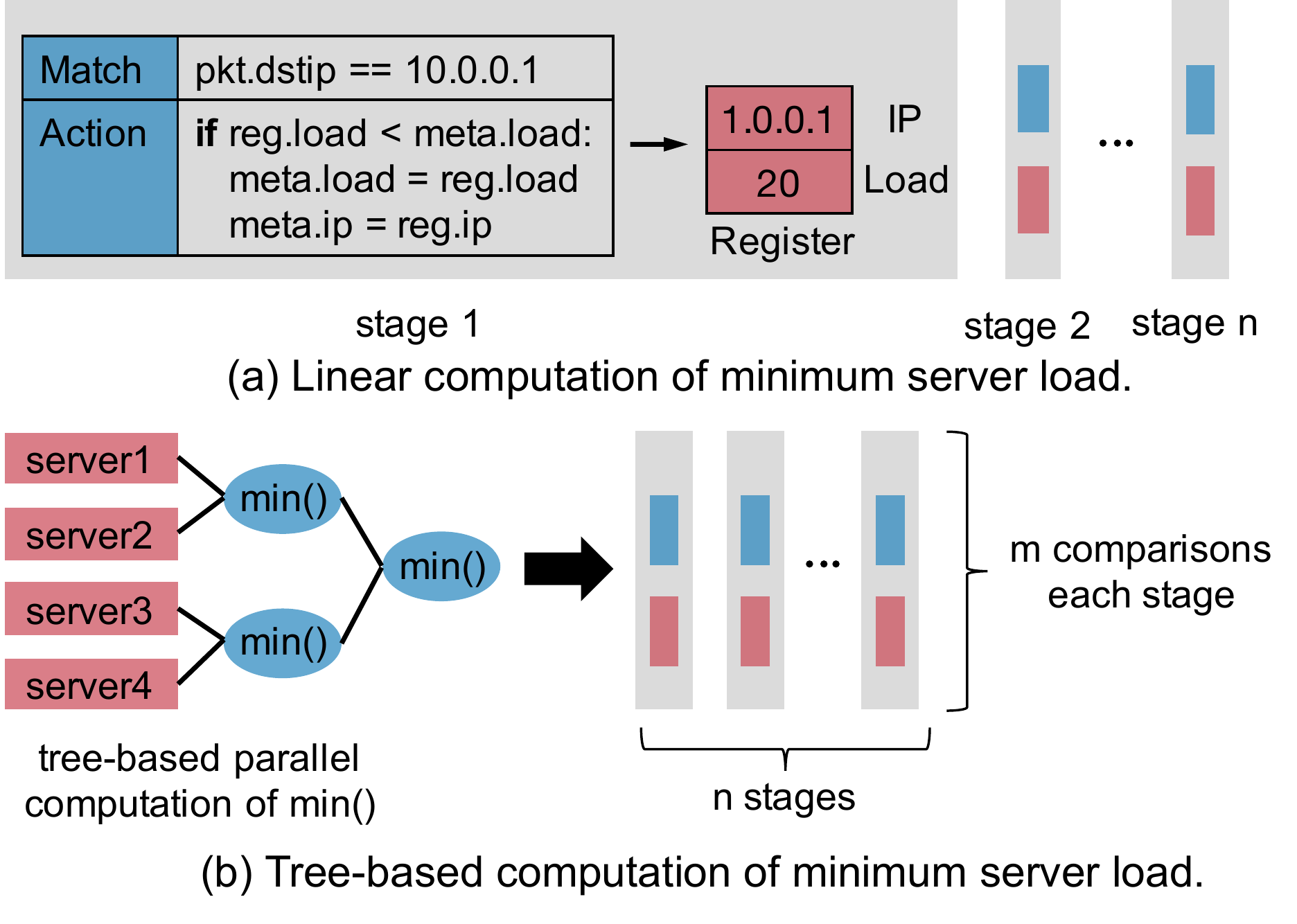}
\vspace{-0.2in}
\caption{Optimal server selection.}
\vspace{-0.15in}
\label{fig:design_schedule2}
\end{figure}
\begin{figure}[t]
\centering
    \includegraphics[width=0.95\linewidth]{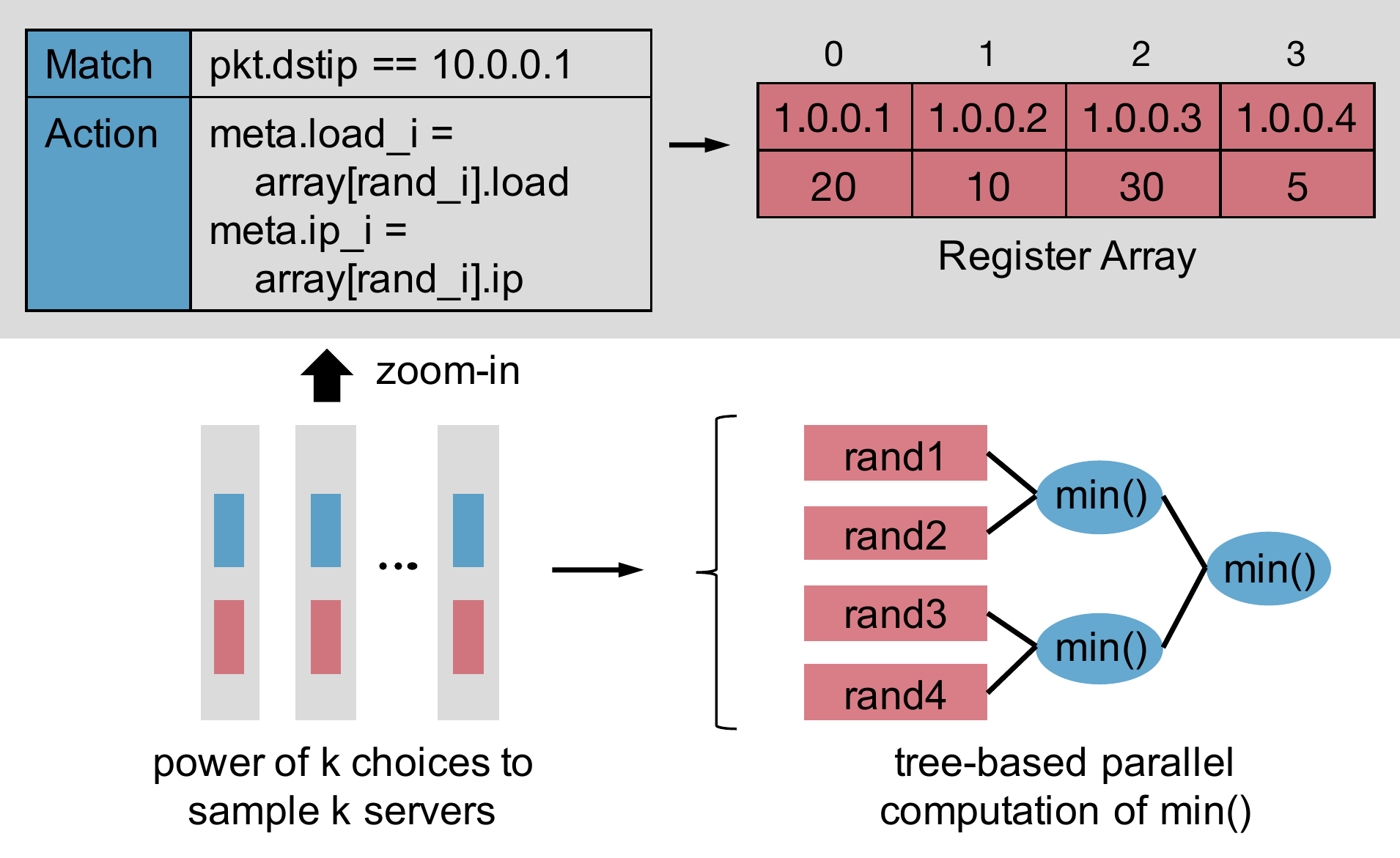}
\vspace{-0.15in}
\caption{Approximate server selection.}
\vspace{-0.1in}
\label{fig:design_schedule3}
\end{figure}

\para{Optimal server selection.} We leverage the register arrays to store the
server loads together with the server IPs and use the multi-stage packet processing
pipeline to compute the minimum. A naive way is to use multiple stages to
scan the server loads linearly, as shown in
Figure~\ref{fig:design_schedule2}(a). The number of servers this solution can
support is limited to the number of stages. As a switch typically has 10--20
stages and many stages need to be used by other switch functionalities, this
solution is not scalable. It can be optimized by computing the minimum in a tree
structure as shown in Figure~\ref{fig:design_schedule2}(b). The comparisons
between two servers in each layer of the tree have no dependencies, and thus can
be done in parallel in the same stage. In the ideal case, given $n$ servers,
this tree-based solution requires $log(n)$ stages, while the naive solution
requires $n$ stages. Let each stage support up to $m$ comparisons. The
comparisons in the first few layers need to be distributed to multiple stages,
if they are larger than $m$.

\para{Approximate server selection.} As discussed in \S\ref{sec:overview}, always choosing
the server with the shortest queue is prone to herding, and due to the limited stages and the
need to support other functionalities, the tree-based approach cannot scale to
many tens of servers. We design an approximate server selection mechanism based
on power-of-k-choices~\cite{bramson2010randomized}, i.e., the switch samples $k$ servers and chooses
the one with the shortest queue from them.
As shown in Figure~\ref{fig:design_schedule3}, the sampling can be done via multiple stages
if $k$ is bigger than the number of register read operations supported by one
stage. After the $k$ servers are sampled, the tree-based mechanism can be applied to get
the one with the shortest queue.

\subsection{Request Affinity}
\label{sec:design:affinity}

Request affinity ensures all packets of the same request are sent to the same
server. This is challenging because the switch processes each packet
independently. In traditional network load balancers~\cite{ananta, maglev, duet,
miao2017silkroad, olteanu2018stateless}, the server selection is solely based on
the hash of the packet header, and the switch does not need to keep any state for request affinity. But
in \sysname, the selection is dynamic. If the switch performs a server selection
for every packet,
the packets of the same request might be sent to
different servers.

\begin{figure}[t]
\centering
    \includegraphics[width=0.85\linewidth]{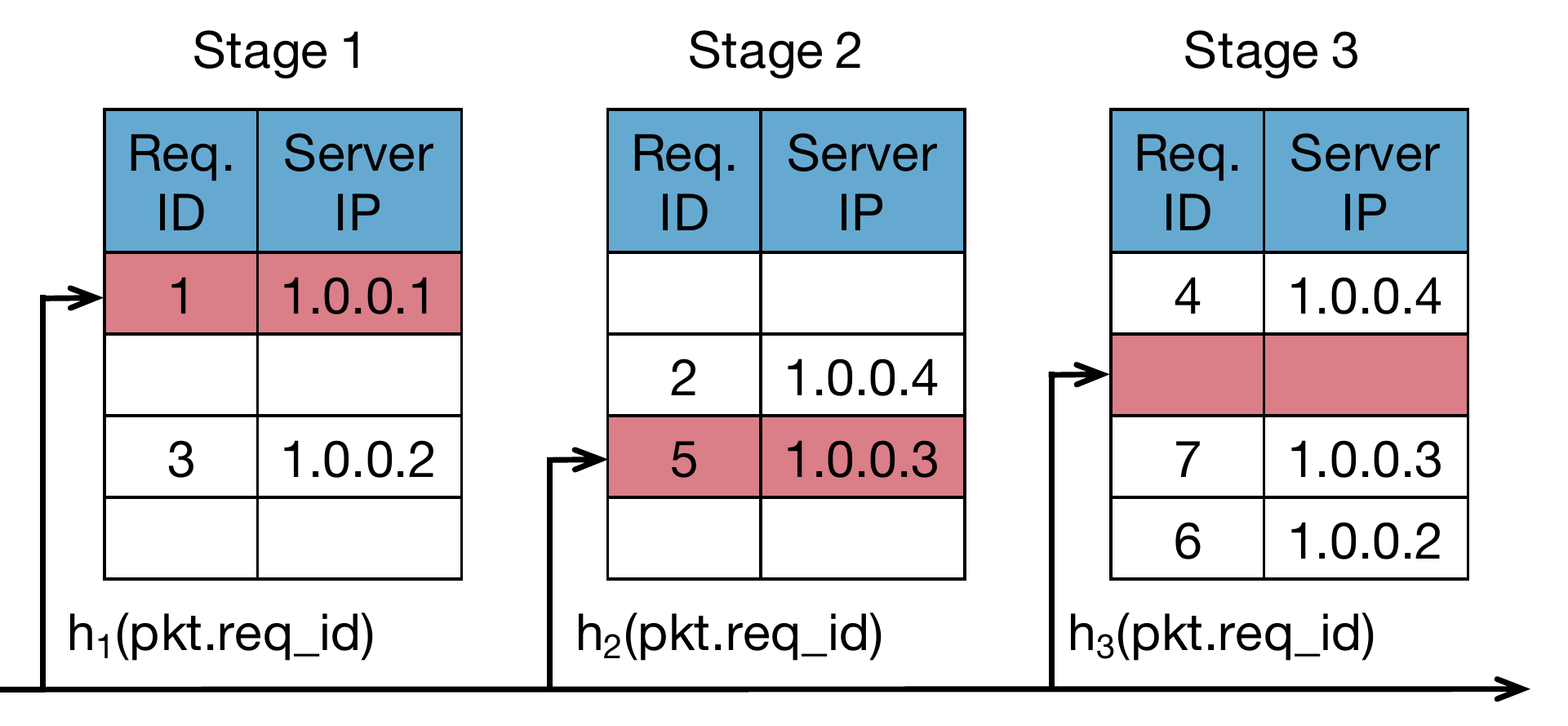}
\vspace{-0.1in}
\caption{Multi-stage hash table for request affinity.}
\vspace{-0.1in}
\label{fig:design_affinity}
\end{figure}

Realizing request affinity requires the switch to keep states. Abstractly, the
switch should maintain a request state table to store the mapping from request
IDs to server IPs (i.e., $ReqTable$ in Algorithm~\ref{alg:switch}). One option
is to use a match-action table, where the request IDs are stored in the match,
and the servers IPs are stored in the action to update the destination IP of the
packets (e.g., used by SilkRoad~\cite{miao2017silkroad}). This option, however,
does not work for microsecond-scale requests at million RPS throughput, because
updating the match-action table (e.g., adding or removing a request) requires
the control plane, which can only do about 10K updates per
second~\cite{switchkv, netcache, noviswitch}. To address this challenge, our design leverages
register arrays to realize a multi-stage hash table that implements all
necessary operations (i.e., $insert$, $read$ and $remove$) for $ReqTable$ in the
data plane, as shown in Figure~\ref{fig:design_affinity}. Unlike match-action
tables, register arrays can only be accessed via an index. We use the hash of
the request ID to find the slot for a request, and the slot stores the request
state, i.e., the request ID and server IP. To handle hash collisions and the
limited array size in each stage, we leverage multiple stages to build a
multi-stage hash table. Algorithm~\ref{alg:affinity} shows the pseudo code to
implement the three operations on $ReqTable$ in Algorithm~\ref{alg:switch}. The
switch iterates over the stages to find an empty slot to insert a new request
(line 1-5), and to find a matched slot to read the server IP (line 6-9) or
remove a completed request (line 10-14).
\sysname does not decrease the capability of the system to defend against DoS attacks. The switch has sufficient memory for $ReqTable$ to support high throughput (\S\ref{sec:evaluation:methodology}), and a DoS attack that overwhelms $ReqTable$ could have overwhelmed the servers first.

\para{Handling switch failure.} There is a relevant notion
to request affinity called Per-Connection Consistency (PCC) for stateful layer-4
load balancers, which requires a TCP connection to be kept across load balancer
failures and system reconfigurations~\cite{ananta, maglev, duet,
miao2017silkroad, olteanu2018stateless}. We emphasize that \sysname focuses on
microsecond-scale requests with strict deadlines (e.g., a couple of the request
execution time). Rebooting a failed switch or replacing it with a backup switch
takes a few minutes, by the time of which the requests have already
\emph{missed} their deadlines. Therefore, different from PCC, maintaining
request affinity across switch failures is a non-goal for \sysname.
Because of the \emph{fate sharing} between the ToR switch and the rack, it is
safe to disregard the $ReqTable$ upon a switch failure, and the new switch
starts with an empty $ReqTable$. Note that \sysname does not increase the chance
of switch failures as a normal ToR switch without \sysname can still fail and make the
rack disconnected.

\begin{algorithm}[t]
\footnotesize
\caption{Request affinity}
\begin{itemize}[noitemsep,nolistsep]
    \item[--] $ReqTable[n][m]$: register arrays to store request state, which
    spans $n$ stages and has $m$ slots in each stage
\end{itemize}
\begin{algorithmic}[1]
\Function{Insert}{$req\_id, server\_ip$}
\For{stage $i$ in all stages}
    \If{$ReqTable[i][h(req\_id)] == None$}
        \State $ReqTable[i][h(req\_id)] \gets (req\_id, server\_ip)$
        \State \Return
    \EndIf
\EndFor
\EndFunction

\Function{Read}{$req\_id$}
\For{stage $i$ in all stages}
    \If{$ReqTable[i][h(req\_id)].req\_id == req\_id$}
        \State \Return $ReqTable[i][h(req\_id)].server\_ip$
    \EndIf
\EndFor
\EndFunction

\Function{Remove}{$req\_id$}
\For{stage $i$ in all stages}
    \If{$ReqTable[i][h(req\_id)].req\_id == req\_id$}
        \State $ReqTable[i][h(req\_id)] \gets None$
    \EndIf
\EndFor
\State \Return
\EndFunction
\end{algorithmic}
\label{alg:affinity}
\end{algorithm}

\para{Handling system reconfiguration.} Unlike switch failures, there is
no fate sharing between each server and the rack, and \sysname does maintain
request affinity across system reconfigurations such as adding or removing servers
for an application. Because \sysname uses $ReqTable$ to store the mapping,
ongoing requests simply check $ReqTable$ to go to the correct servers. Only the
request scheduling module (\S~\ref{sec:design:scheduling}) needs to be updated
to have the right set of servers to choose from for new requests. We pre-allocate a
large number of registers for $LoadTable$ at compilation time, and use another
register to indicate the number of \emph{active} servers, which is dynamically
updated for system reconfigurations. For an unplanned server removal (e.g., a
server failure), \sysname uses the switch control plane to update the $ReqTable$ and delete the stale
entries related to the removed server.

\subsection{Server Tracking}
\label{sec:design:tracking}

The server tracking module updates the server loads (i.e., $LoadTable$ in Algorithm~\ref{alg:switch}) for the
switch to make scheduling decisions. The challenge is to
accurately track the server loads at real-time with low overhead. A
straightforward solution is to let the switch control plane periodically poll
the queue lengths at each server and update the data plane. However, due to the
millisecond-scale delay and the limited rate of control plane updates~\cite{noviswitch, dionysus}, this
solution does not apply to the microsecond-scale workloads targeted by \sysname.
To do this in the data plane, a possible solution is to let the switch
\emph{proactively} track the server loads, i.e., incrementing and decrementing the
counters for queue lengths when processing request and reply packets. This
solution suffers from estimation errors due to packet loss and retransmissions,
and fixes like decreasing the counters based on the server processing
rate~\cite{pegasus} cannot handle temporal load imbalance in high-dispersion
microsecond-scale workloads.

\sysname leverages in-network telemetry to accurately track server loads
with minimal overhead. In-network telemetry is widely used in
network monitoring and diagnosis where switches put relevant measurement data
into packet headers in the data plane. \sysname applies this mechanism to track
server loads.
Then the servers piggyback their loads in the reply
packets to update the counters in the switch, which does not introduce new
packets and thus minimizes system overhead. A potential problem is the feedback
loop delay as stale information can cause herding, which can degrade the
scheduling performance and make the system unstable (i.e., swing between
overloading different servers). In \sysname, the switch and the servers are
directly connected in the same rack, and the server-side data plane
implementation bypasses traditional TCP/IP stack to report its queue length to
the switch quickly, making the feedback loop delay minimal. And together with
power-of-k-choices scheduling, \sysname can effectively avoid herding. An
alternative solution that only keeps the server with the minimum load in the
switch and updates it based on in-network telemetry cannot leverage
power-of-k-choices to avoid herding. We show the impact of different ways to
track and
represent server loads in \S\ref{sec:evaluation:analysis}.

\subsection{Handling Scheduling Requirements}
\label{sec:design:extension}

\paraf{Multi-queue support.} By default, \sysname uses a single-queue policy,
i.e., the system does not differentiate a priori between request types and aims
to meet a single SLO for tail latency (e.g., a photo caching workload
with only $get$ requests). \sysname also supports multiple
queues if the workload has multiple request types that have distinct service
time distributions (e.g., a key-value store workload with both $get$ and $range$
requests). Applications indicate the request type in the \texttt{TYPE} field of the
packet header. Each server maintains a
separate queue for each type for intra-server scheduling. The switch maintains the
counters for each type in $LoadTable$, and schedules requests based on the
queue lengths of the request type.
We remark that there is no fundamental limit on the number of queues on each server,
since the queues are implemented in software and the switch only needs to keep a counter for each queue on each server.

\para{Locality and placement constraints.} \sysname handles two types of common
locality and placement constraints, which are data locality and request
dependency. $(i)$ Data locality requires a request
to be processed on a subset of servers that hold the input data.
To support data locality,
applications set different \texttt{LOCALITY} values to represent
different locality requirements (i.e., different sets of servers that can process this request),
and the switch maintains different mappings, which map a server ID to a server
for different \texttt{LOCALITY} values.
$(ii)$ Request
dependency requires multiple requests to be scheduled to the same server, e.g.,
a task consists of multiple requests and the input of one request is the output
of one or more other tasks. Request dependency is supported using request
affinity: relevant requests carry the same \texttt{REQ\_ID} in their headers, so
that they will be sent to the same server.
Additional information is included in the \sysname header for the number of subsequent requests to expect.
The server can send replies to each request separately and independently.
 \sysname only requires the server to set \texttt{TYPE} to be reply in replies after it has received all requests in the set
 in order to safely delete the state in the switch.
 Note that applications can
still use different request IDs for different requests and receive replies as
soon as some requests are completed, by adding application-specific metadata in
the payload.

\para{Resource allocation policies.} The scheduler is responsible for allocating
resources when the demand exceeds the capacity of the rack-scale computer.
\sysname supports two types of common resource allocation policies, which are
strict priority and weighted fair sharing. $(i)$ To support strict priority, each
server maintains a separate queue for each priority. Similar to the multi-queue
support, the switch tracks the queue lengths, and balances the server loads for
each priority. Each server uses intra-server
scheduling to preempt low-priority requests when high-priority requests
arrive, which can be done in 5 $\mu s$ in our implementation based on
Shinjuku~\cite{shinjuku}. $(ii)$ Supporting weighted fair sharing is similar. Each
server maintains a separate queue for each client, and performs weighted fair
queueing~\cite{demers1989analysis} for intra-server scheduling on the granularity of slice in PS. The switch tracks the queue
lengths and balances the server loads for each client.

\section{Evaluation}
\label{sec:evaluation}

In this section, we evaluate \sysname with a variety of synthetic and real
application workloads. We provide additional experiment results, including locality
constraints, priority policies and multiple applications, in
Appendix~\ref{sec:appendix:add_eval}.

\subsection{Methodology}
\label{sec:evaluation:methodology}

\paraf{Testbed.} The experiments are conducted on a testbed of twelve server machines connected by a 6.5Tbps Barefoot
Tofino switch. Each server has an 8-core CPU (Intel Xeon E5-2620 @ 2.1GHz), 64GB memory, and one 40G NIC (Intel XL710).
Eight servers are used as workers to process requests, and they run Shinjuku~\cite{shinjuku}
with our extension. Four servers are used as clients to generate requests.
The bottleneck of the system is at the workers.

\begin{figure*}[t]
    \centering
    \subfigure[Exp(50).]{
        \label{fig:eval_exp}
        \includegraphics[width=0.24\linewidth]{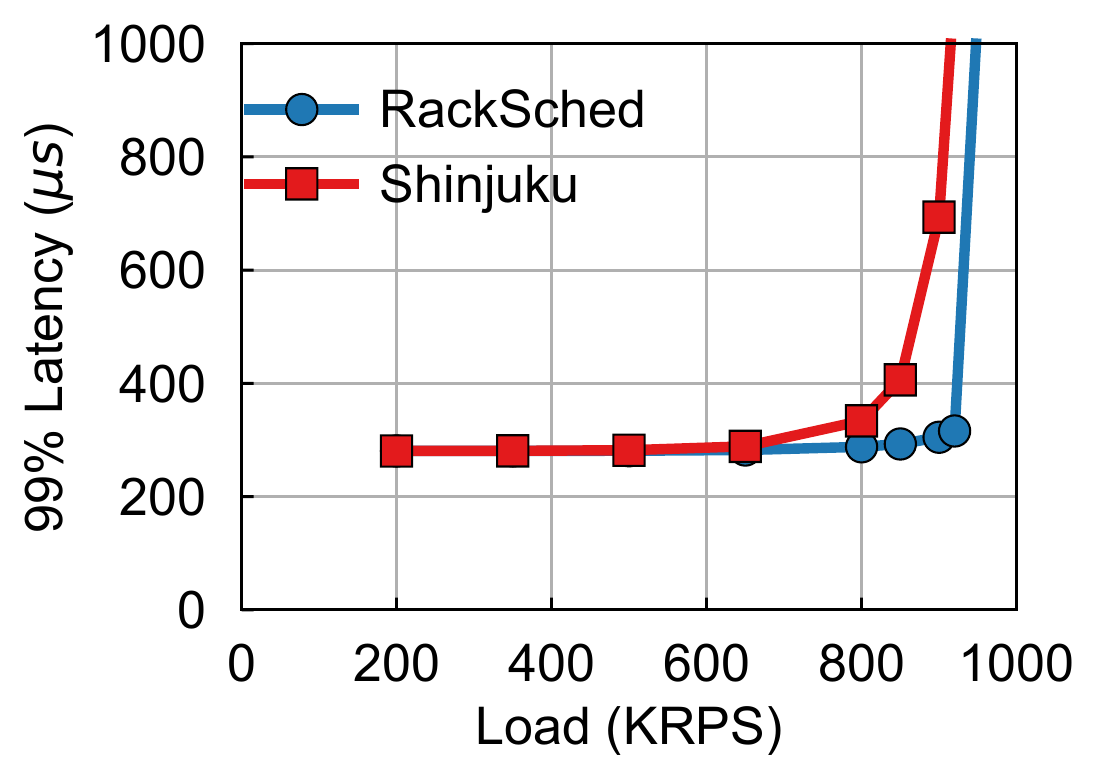}}
    \subfigure[Bimodal(90\%-50, 10\%-500).]{
        \label{fig:eval_bimodal}
        \includegraphics[width=0.24\linewidth]{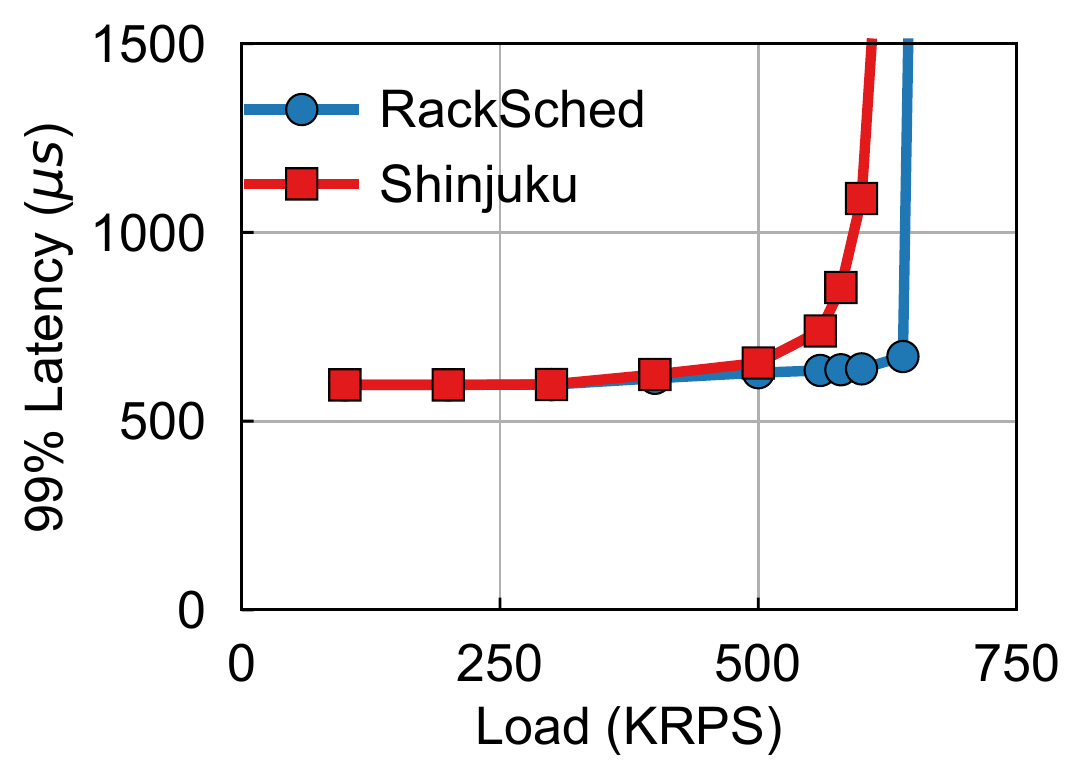}}
    \subfigure[Bimodal(50\%-50, 50\%-500).]{
        \label{fig:eval_port_bimodal}
        \includegraphics[width=0.24\linewidth]{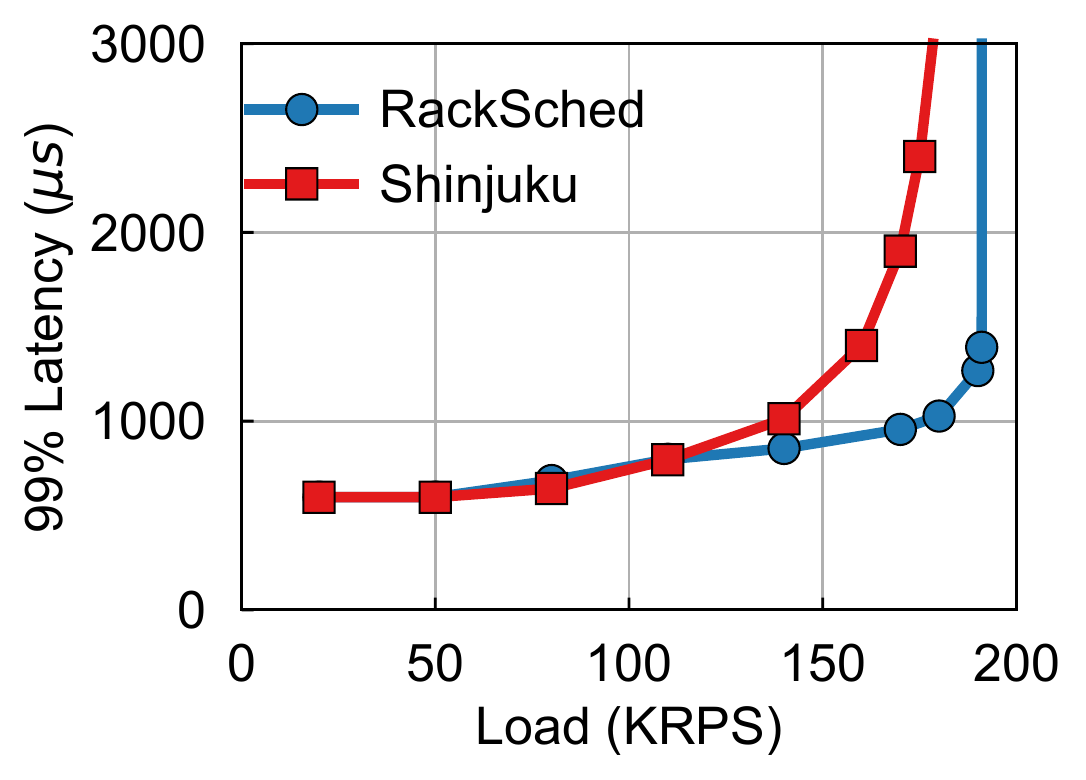}}
    \subfigure[Trimodal(33.3\%-50, 33.3\%-500, 33.3\%-5000).]{
        \label{fig:eval_trimodal}
        \includegraphics[width=0.24\linewidth]{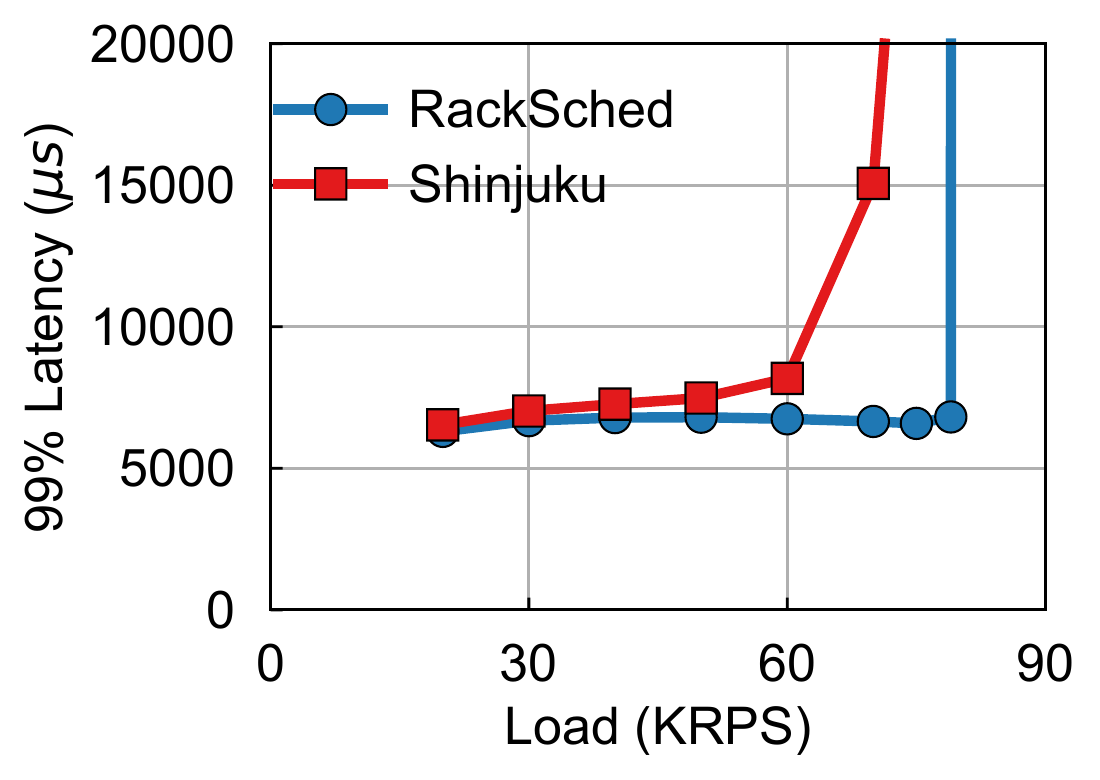}}
    \vspace{-0.2in}
    \caption{Experimental results for synthetic workloads with homogeneous servers.}
    \vspace{-0.2in}
    \label{fig:eval_synthetic}
\end{figure*}

\begin{figure*}[t]
    \centering
    \subfigure[Exp(50).]{
        \label{fig:eval_hetero_exp}
        \includegraphics[width=0.24\linewidth]{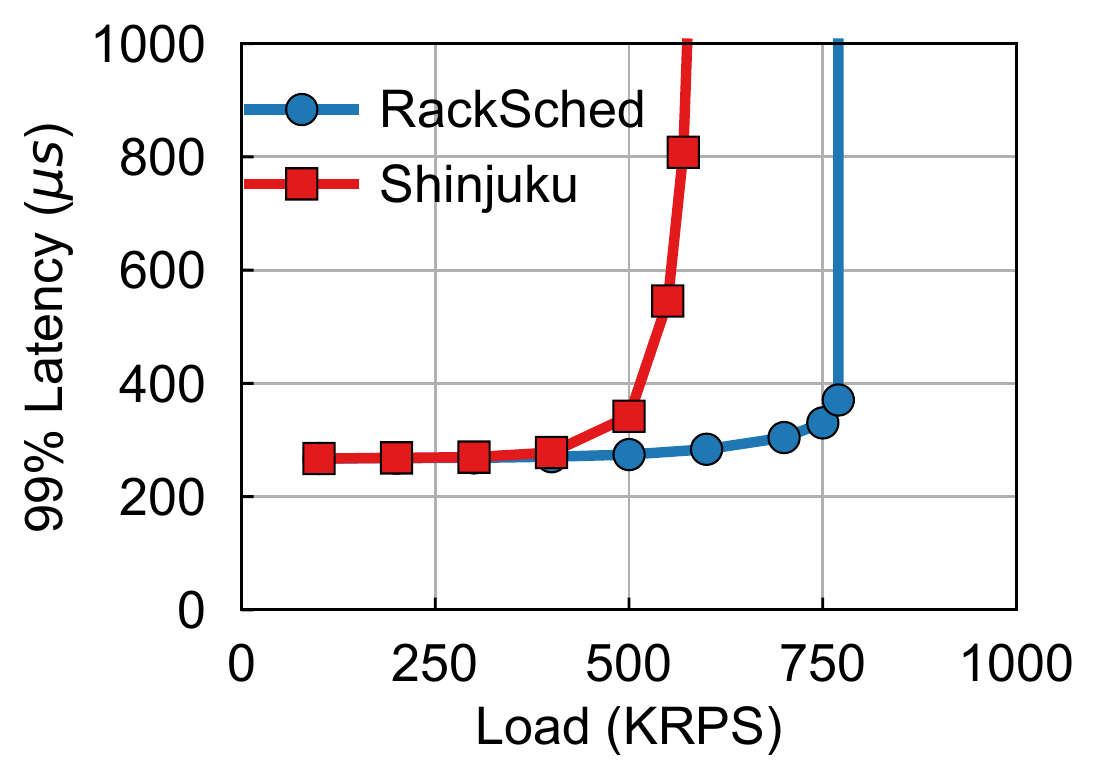}}
    \subfigure[Bimodal(90\%-50, 10\%-500).]{
        \label{fig:eval_hetero_bimodal}
        \includegraphics[width=0.24\linewidth]{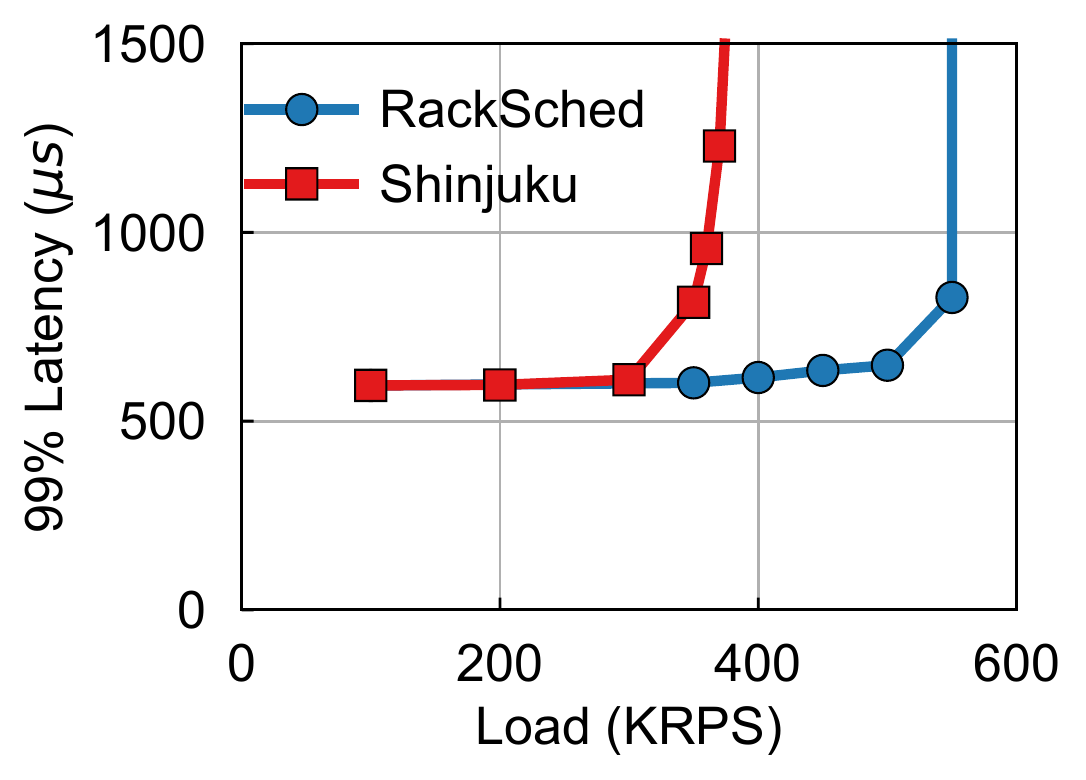}}
    \subfigure[Bimodal(50\%-50, 50\%-500).]{
        \label{fig:eval_hetero_port_bimodal}
        \includegraphics[width=0.24\linewidth]{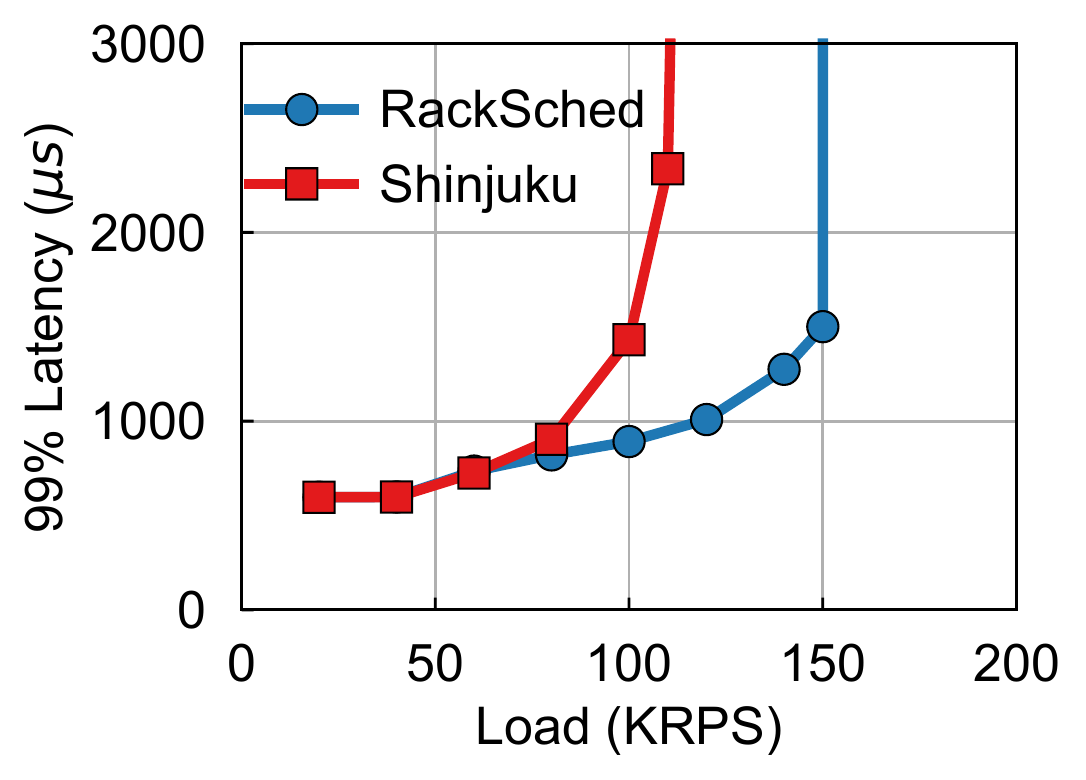}}
    \subfigure[Trimodal(33.3\%-50, 33.3\%-500, 33.3\%-5000).]{
        \label{fig:eval_hetero_trimodal}
        \includegraphics[width=0.24\linewidth]{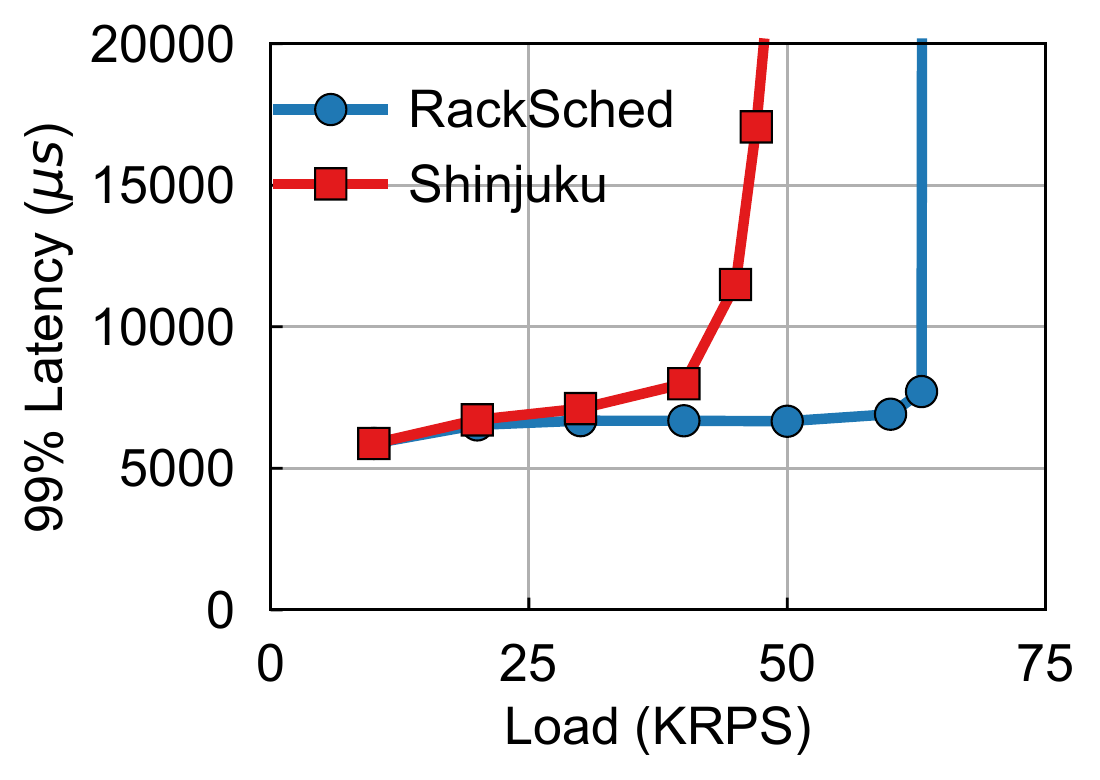}}
    \vspace{-0.2in}
    \caption{Experimental results for synthetic workloads with heterogeneous servers.}
    \vspace{-0.2in}
    \label{fig:eval_hetero}
\end{figure*}

\para{Implementation.} We have implemented a \sysname prototype and integrated it with Shinjuku~\cite{shinjuku}. $(i)$ The switch data plane is
written in P4~\cite{p4-ccr} and compiled to Barefoot Tofino ASIC~\cite{tofino} with P4
Studio~\cite{p4studio}. The request state table contains a hash table with 64K slots.
The default implementation uses power-of-2-choices.
$(ii)$ The worker server is based on Shinjuku~\cite{shinjuku}. We have extended
Shinjuku to support the \sysname packet header, maintain a counter
and update the counter upon request arrival and reply departure
to track the queue length and append the queue
length in reply packets.
Both \sysname and Shinjuku preempt requests that exceed 250 $\mu s$ in our experiments.
$(iii)$ The client is open-loop and implemented in C using Intel
DPDK 16.11.1~\cite{dpdk}. It can generate requests at high request rate based on synthetic
workloads and the RocksDB application, and measure the throughput and latency of
\sysname.

\para{Resource consumption.} 
Our prototype uses 13.12\% SRAM, 9.96\% Match Input Crossbar, 12.5\% Hash Unit and 25\% Stateful ALUs of the Tofino ASIC resources.
We provide an analysis and a back-of-the-envelop
calculation to show that \sysname consumes little switch memory.
\sysname has two sets of state on the switch,
i.e., $LoadTable$ and $ReqTable$. $(i)$ $LoadTable$ maintains a counter for each
queue of each server. Let the counter be 4 bytes, the number of queues in each
server be 3 and the number of servers be 32. It only consumes 384 bytes. $(ii)$
$ReqTable$ maintains the selected server IPs for the \emph{ongoing} requests,
not \emph{all} requests the system have received and processed. Each slot can be
\emph{reused} by many times each second because the requests are
microsecond-scale. Given an average request processing latency of 50 $\mu s$, a
slot can support 20 KRPS throughput, and a table with 64K slots can support 1.28
BRPS throughput. Let the request ID and server IP both be 4 bytes. A table with
64K slots (our implementation) consumes 256 KB, which is only a few percent of the on-chip memory
(tens of MB). Overflowed requests can fall back to hash-based random dispatching
which preserves request affinity.

\para{Workloads.} We use a combination of synthetic and application workloads.
They include the following workloads. By default, the workloads use one-packet requests.
\begin{itemize}[leftmargin=*]
    \item Exp(50) is an exponential distribution with mean = 50 $\mu s$, which represents common query and storage workloads, such as $get$ requests in photo caching.

    \item Bimodal(90\%-50, 10\%-500) is a bimodal distribution
    with 90\% of requests taking 50 $\mu s$ and 10\% taking 500 $\mu
    s$, which represents workloads with a mix of simple requests and complex requests, such as $get$ and $range$ requests in key-value stores.

    \item Bimodal(50\%-50, 50\%-500) is a bimodal distribution
    with 50\% of requests taking 50 $\mu s$ and 50\% taking 500 $\mu
    s$, which represents workloads with half simple requests and half complex requests.

    \item Trimodal(33.3\%-50, 33.3\%-500, 33.3\%-5000) is a trimodal distribution
    with a third of requests taking 50 $\mu s$, 500 $\mu s$ and 5000 $\mu s$, respectively, which represents workloads with more diverse request types,
    such as $point$, $range$ and complex $join$ requests in databases.
\end{itemize}
We also use RocksDB 5.13~\cite{rocksdb}, an open-source production-quality key-value
store, as a real application workload to evaluate \sysname. RocksDB is
configured to store data in DRAM 
to avoid blocking behavior and achieve low latency.

\subsection{Synthetic Workloads}
\label{sec:evaluation:synthetic}

We evaluate the system on synthetic workloads that cover large application
space. We compare \sysname with that directly runs Shinjuku in the cluster,
i.e., the requests are randomly sent to the servers.

Figure~\ref{fig:eval_exp} and Figure~\ref{fig:eval_bimodal} compare \sysname and
Shinjuku under Exp(50) and Bimodal(90\%-50, 10\%-500) workloads, respectively.
In these two figures, both \sysname and Shinjuku use a single-queue policy.
Under Exp(50), the 99\% latencies of \sysname and Shinjuku are similar at low load.
But the 99\% latency of Shinjuku quickly goes up after 800 KRPS, while that of \sysname
can support the system load up to 950 KRPS. Under Bimodal(90\%-50, 10\%-500),
the 99\% latency of Shinjuku quickly increases after 500 KRPS, while that of \sysname
stays stable until 650 KRPS. In both workloads, \sysname supports larger request
load with lower tail latency, because its inter-server scheduling addresses
temporal load imbalance between servers, while Shinjuku experiences short bursts
and long queues in individual servers under high request load.

Figure~\ref{fig:eval_port_bimodal} and Figure~\ref{fig:eval_trimodal} show the results for Bimodal(50\%-50, 50\%-500) and
Trimodal(33.3\%-50, 33.3\%-500, 33.3\%-5000) workloads, respectively. In these two figures, both \sysname and Shinjuku have a separate queue for each request
type. Again, \sysname significantly outperforms Shinjuku. The improvement of \sysname is larger in Trimodal(33.3\%-50, 33.3\%-500, 33.3\%-5000) than Bimodal(50\%-50,
50\%-500), because Trimodal(33.3\%-50, 33.3\%-500, 33.3\%-5000) has more diverse service times and can benefit more from effective inter-server scheduling.

Figure~\ref{fig:eval_hetero} shows the results with heterogeneous servers.
In this case, four servers have four workers and the other four servers have seven workers (one core used by the scheduler).
This evaluates the cases when some servers are slower or some cores of these servers are grabbed for other purposes~\cite{robinhood, shenango}.
We compare \sysname with Shinjuku under the same four distributions in Figure~\ref{fig:eval_synthetic}.
\sysname is load-aware and tends to send requests to the servers with shorter queue lengths, while Shinjuku
distributes the requests to the servers uniformly, disregarding the heterogeneity.
\sysname can improve the performance further with heterogeneous servers.

\subsection{Scalability}
\label{sec:evaluation:scalability}
\begin{figure}[t]
\centering
    \includegraphics[width=0.9\linewidth]{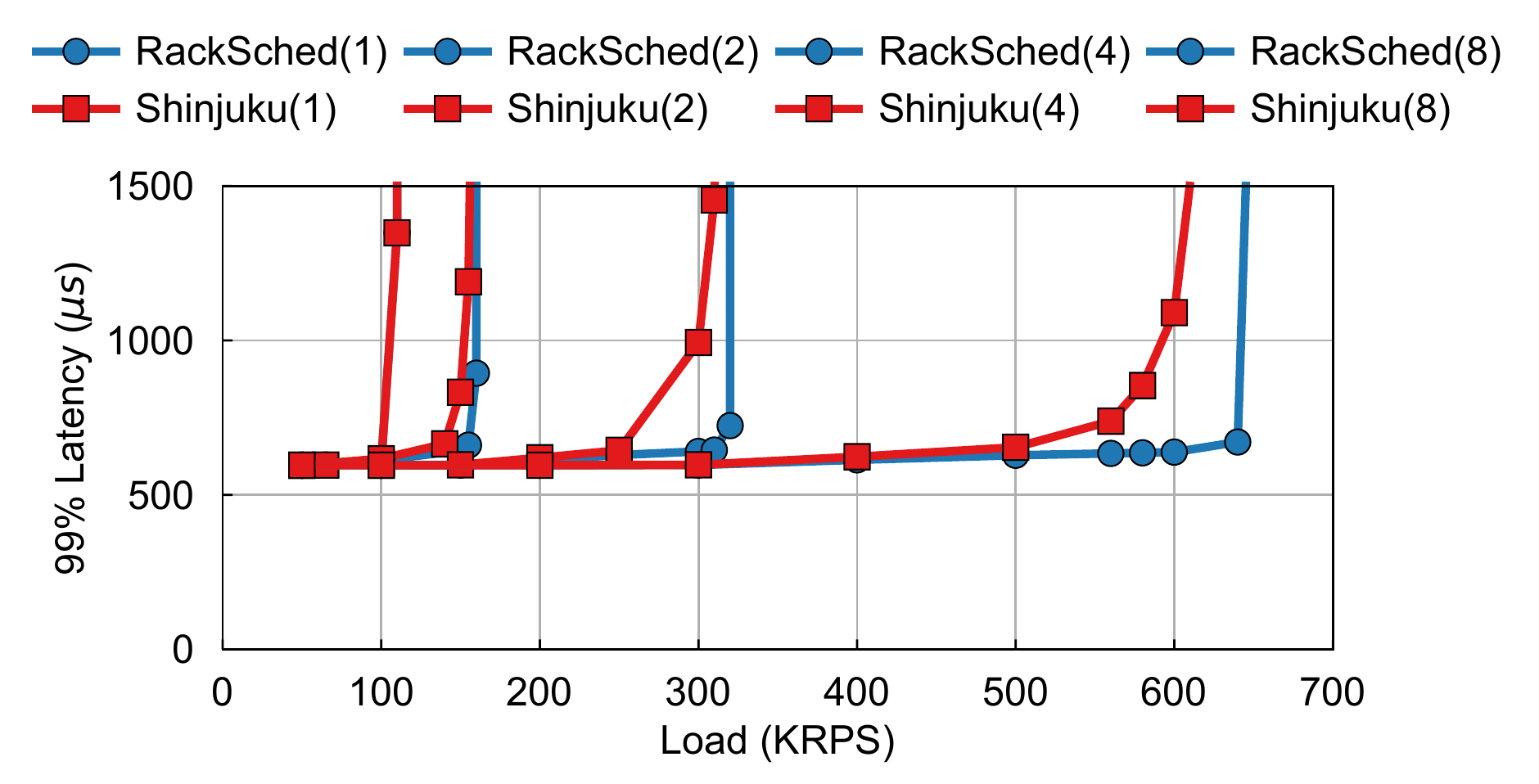}
\vspace{-0.1in}
\caption{Scalability results.}
\vspace{-0.05in}
\label{fig:eval_scalability}
\end{figure}

The key benefit of \sysname is that it enables the system to scale out by adding
servers, while achieving low tail latency at high throughput.
Figure~\ref{fig:eval_scalability} shows the 99\% latency under different request
load with one, two, four and eight servers, respectively. This figure uses
Bimodal(90\%-50, 10\%-500) workload, and the results for other workloads are
similar. With one server, the two systems, i.e., \sysname(1) and Shinjuku(1), have
the same performance, as there is no need for inter-server scheduling. With two
servers, load imbalance can happen, but the variability is small. With four servers, micro bursts can cause
bigger temporal load imbalance, and the improvement of inter-server scheduling
is also bigger. When there are
eight servers, there is more variability between the loads on the servers, and
inter-server scheduling has more opportunities to improve performance.
Shinjuku(8) can only maintain low tail latency until 500 KRPS, while \sysname(8)
can maintain low tail latency until 650 KRPS. We expect the improvement of
\sysname over Shinjuku would be larger with more servers, because there would be
more variabilities with more servers.

Overall, \sysname scales out the
total throughput of the system near linearly with the number of servers in the rack. And the
throughput improvement is achieved without increasing the tail latency. Even
with more servers, \sysname is still able to maintain the same tail latency as one
server until the system is saturated.

\subsection{Application: RocksDB}
\label{sec:evaluation:rocksdb}

We use RocksDB~\cite{rocksdb} to demonstrate the benefits of \sysname on real
applications. RocksDB is an open-source production-quality storage system that
is widely deployed to support many online services such as Facebook. In the
experiments, RocksDB is configured with an in-memory file
system (/tmpfs/) for microsecond-scale request processing. We use two request
types. One is $GET$ which gets 60 objects with a median request service time
of 50 $\mu s$. The other is $SCAN$ which scans 5000 objects with a median service
processing time of 740 $\mu s$. Only 326 lines of code are needed to port RocksDB to \sysname and Shinjuku.
\begin{figure}[t]
    \centering
    \subfigure[90\%-GET, 10\%-SCAN.]{
        \label{fig:eval_rocksdb_bimodal}
        \includegraphics[width=0.48\linewidth]{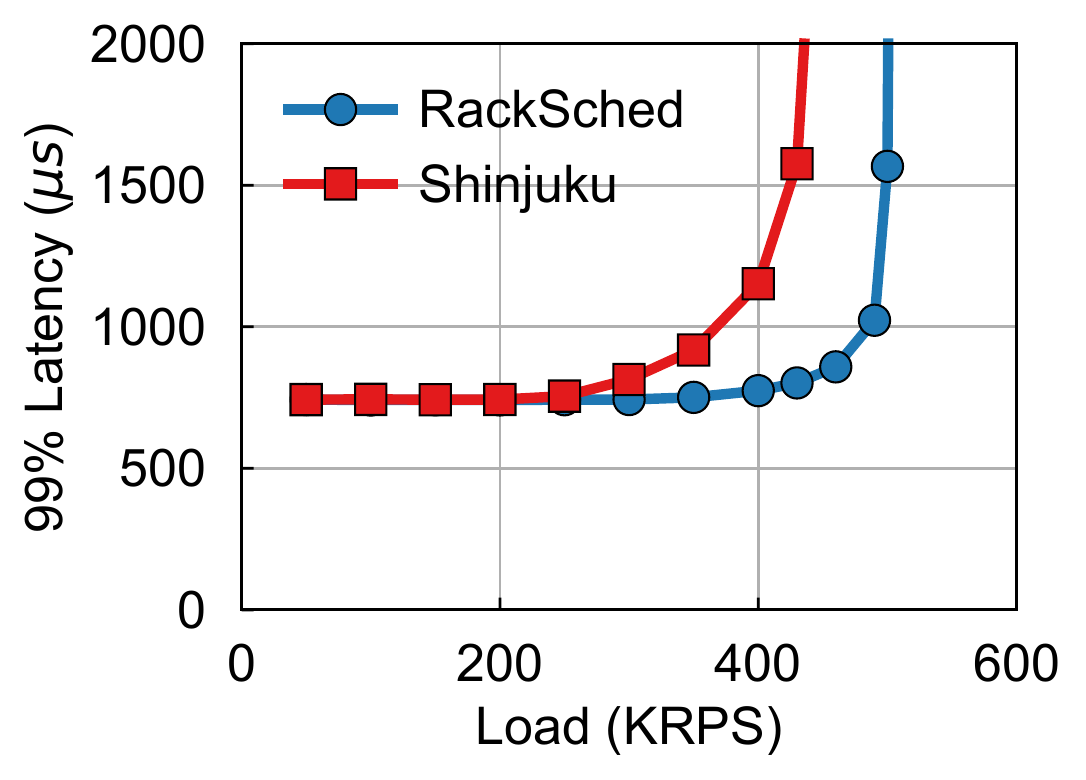}}
    \subfigure[50\%-GET, 50\%-SCAN.]{
        \label{fig:eval_rocksdb_port_bimodal}
        \includegraphics[width=0.48\linewidth]{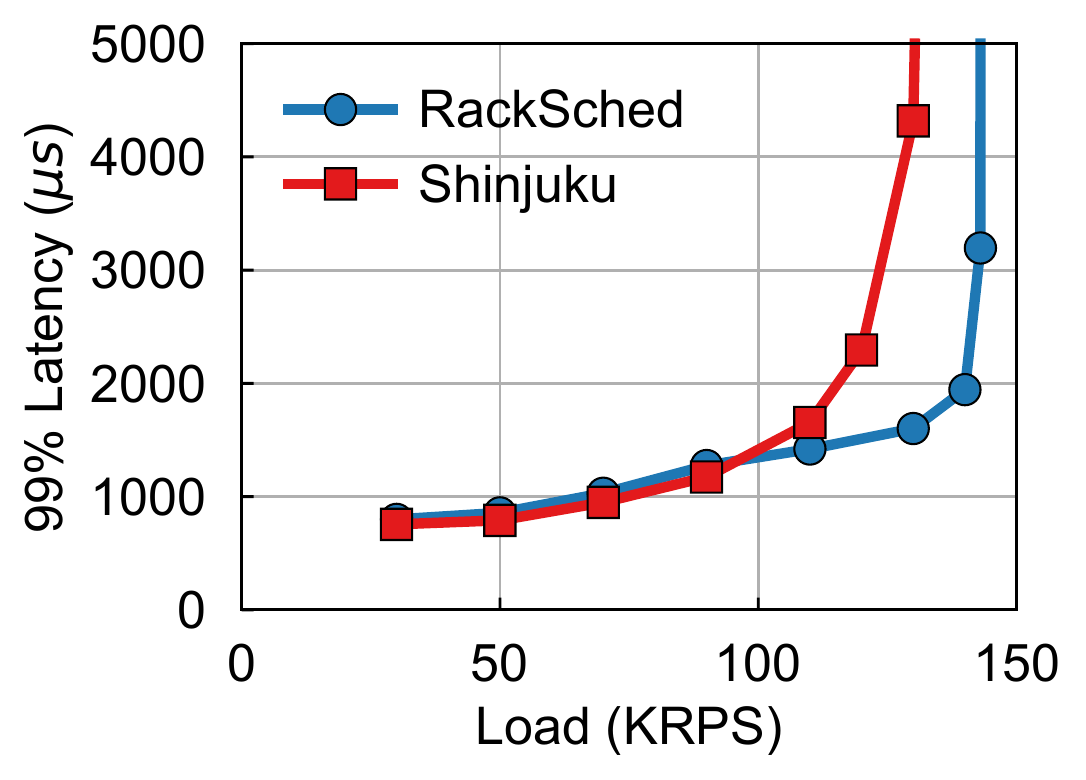}}
    \subfigure[GET in 50\%-GET, 50\%-SCAN.]{
        \label{fig:eval_rocksdb_port_bimodal_short}
        \includegraphics[width=0.48\linewidth]{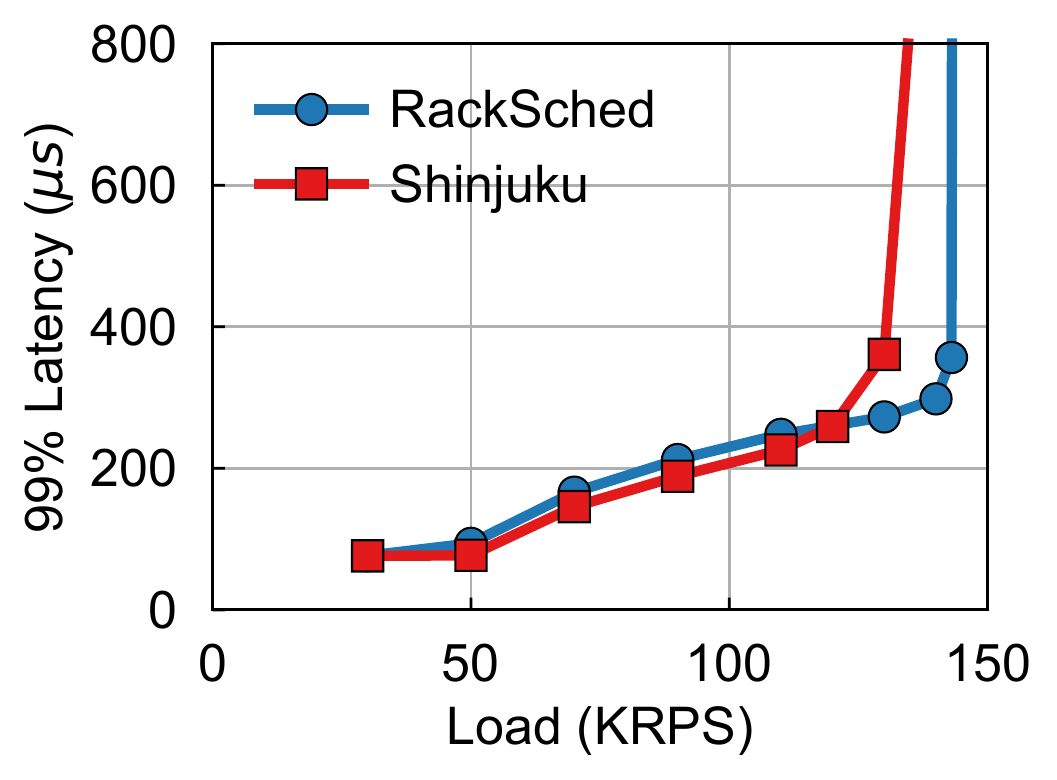}}
    \subfigure[SCAN in 50\%-GET, 50\%-SCAN.]{
        \label{fig:eval_rocksdb_port_bimodal_long}
        \includegraphics[width=0.48\linewidth]{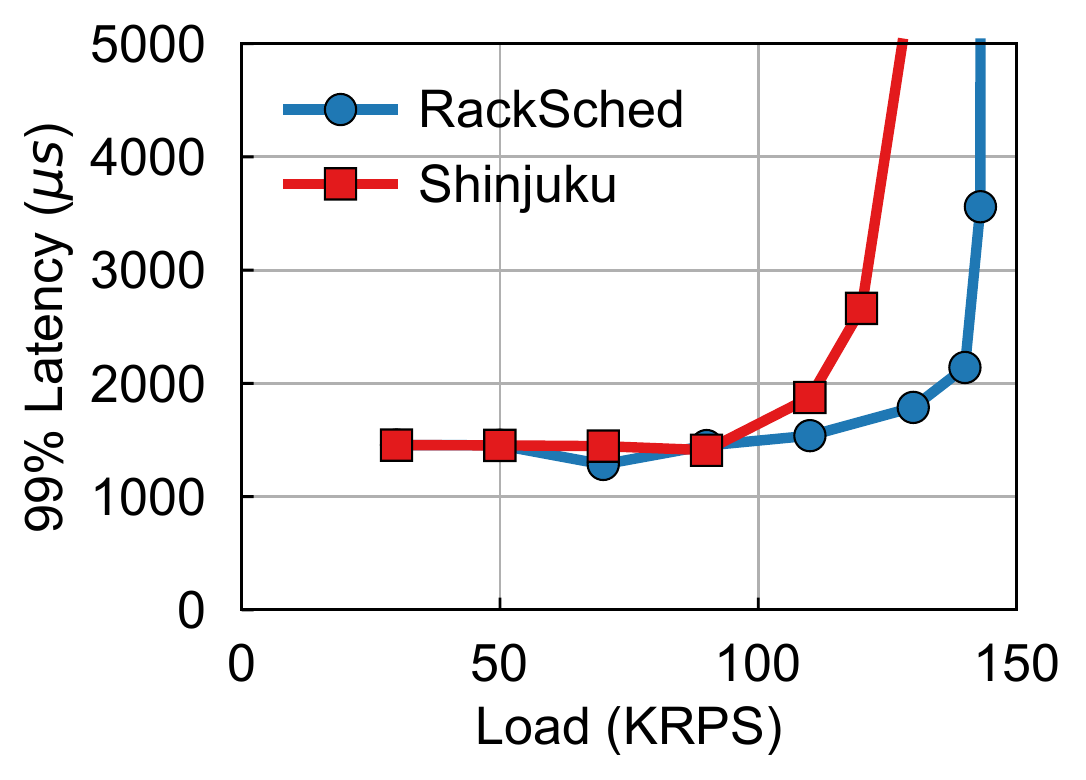}}
    \vspace{-0.1in}
    \caption{Experimental results for RocksDB.}
    \vspace{-0.1in}
    \label{fig:eval_rocksdb}
\end{figure}
Figure~\ref{fig:eval_rocksdb_bimodal} shows the results for the
workload that contains 90\% $GET$ requests and 10\% $SCAN$ requests. In this
experiment, the system uses a single-queue policy. At low request load, \sysname
and Shinjuku have comparable 99\% latency. But
Shinjuku can only maintain low tail latency until 300 KRPS, while
\sysname is able to keep low tail latency until 500 KRPS.

Figure~\ref{fig:eval_rocksdb_port_bimodal} shows the results for the
workload that contains 50\% $GET$ requests and 50\% $SCAN$
requests. In this experiment, the system uses a multi-queue policy. \sysname is able
to maintain low tail latency at a higher request load than Shinjuku.
\begin{figure}[t]
    \centering
    \subfigure[Bimodal(90\%-50,10\%-500).]{
        \label{fig:eval_client_bimodal}
        \includegraphics[width=0.48\linewidth]{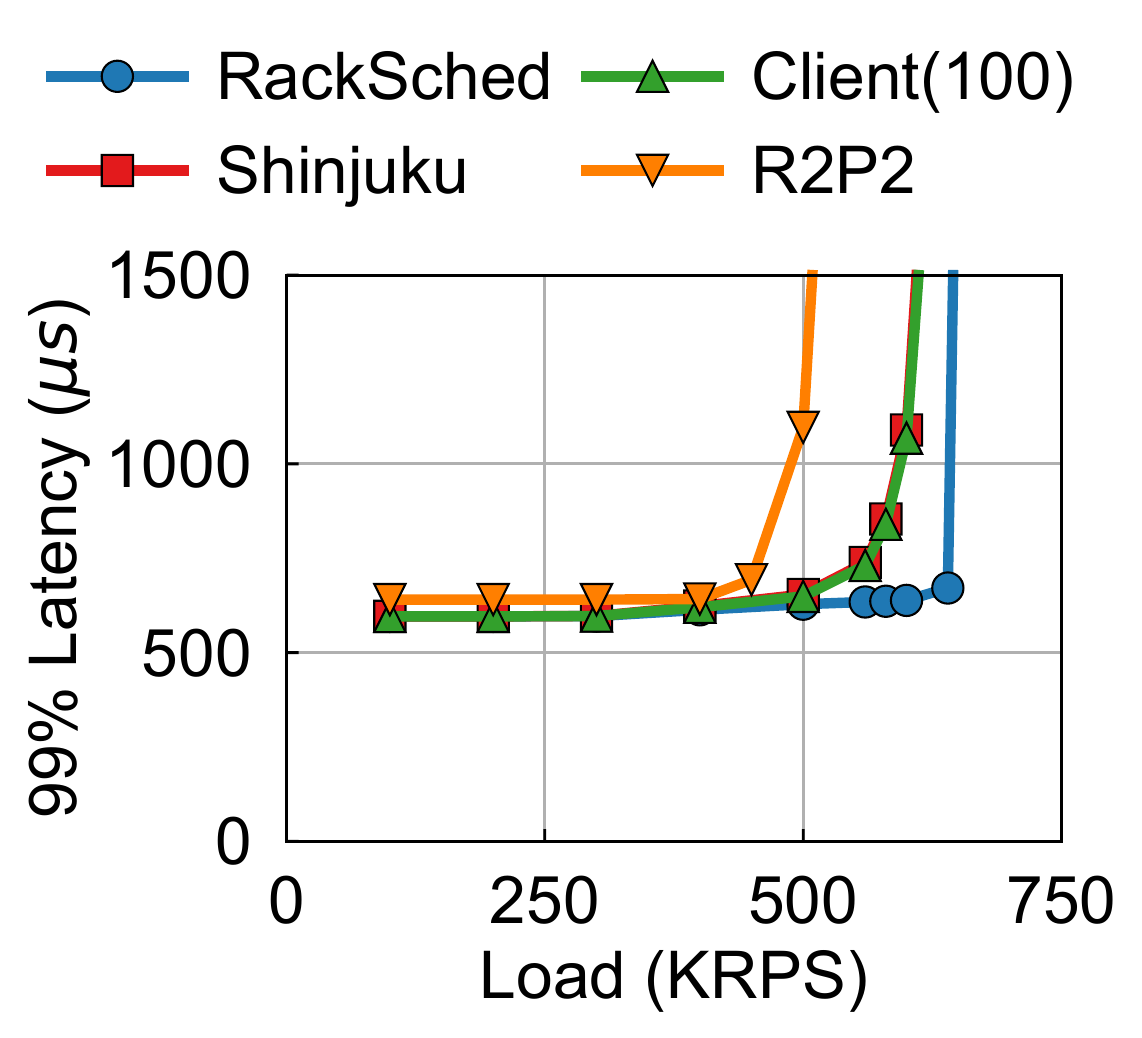}}
    \subfigure[Bimodal(50\%-50,50\%-500).]{
        \label{fig:eval_client_port_bimodal}
        \includegraphics[width=0.48\linewidth]{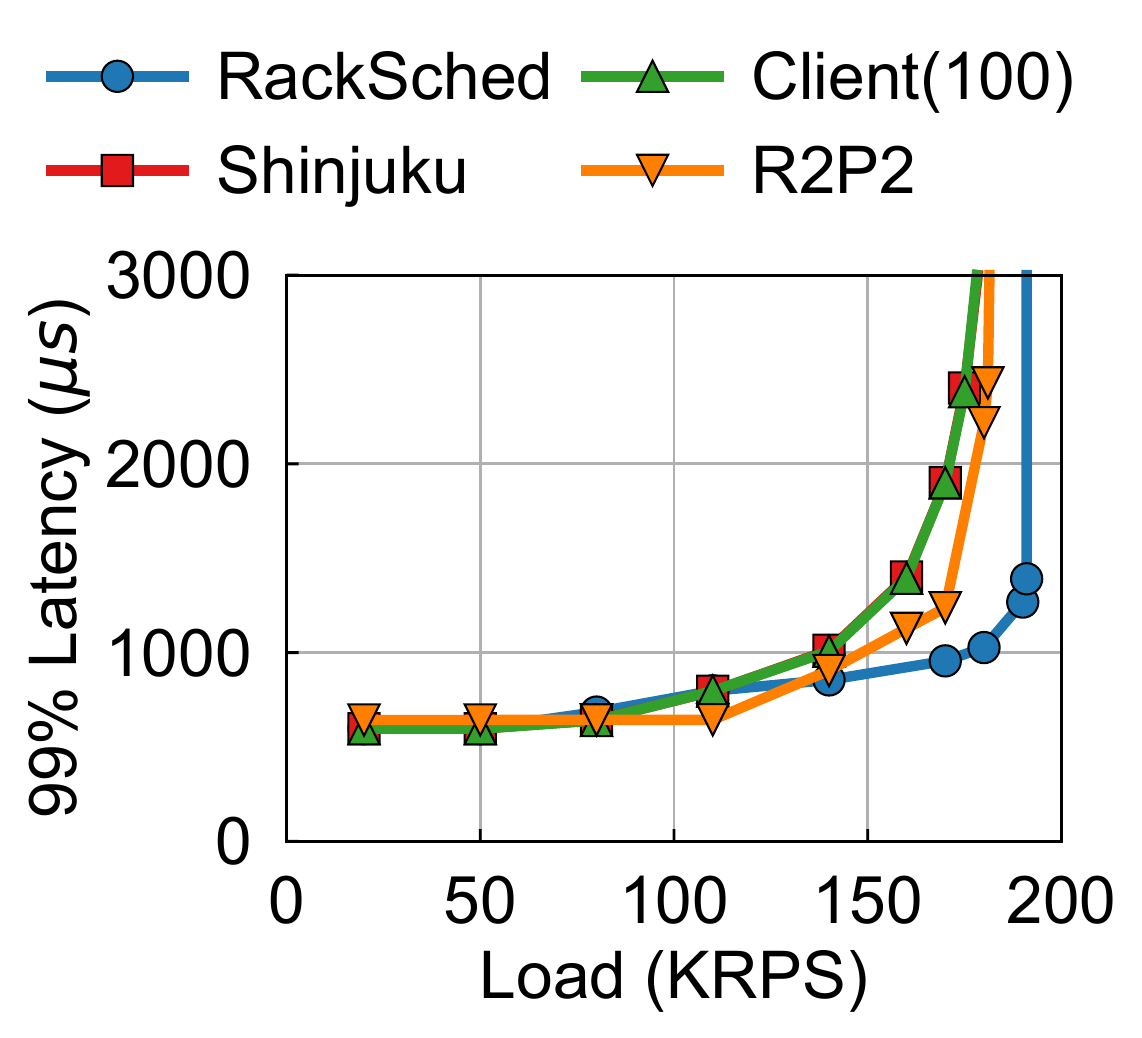}}
    \vspace{-0.1in}
    \caption{Comparison with other solutions.}
    \vspace{-0.1in}
    \label{fig:eval_client}
\end{figure}
We further break down the results for each request type for this
workload. Figure~\ref{fig:eval_rocksdb_port_bimodal_short} and
Figure~\ref{fig:eval_rocksdb_port_bimodal_long} show the 99\% latency for $GET$
and $SCAN$ under different total request load, respectively. Because \sysname
uses the switch to balance the load of each request type between the servers,
the improvement of \sysname over all requests does not come at the cost of
sacrificing any individual request type. For both request types, \sysname is
able to deliver comparable tail latency at low load, achieve significantly lower
tail latency at high load, and support higher total request load.

\subsection{Comparison with Other Solutions}
\label{sec:evaluation:other}

R2P2~\cite{kogias2019r2p2} is a recent solution that proposes a
join-bounded-shortest-queue (JBSQ) policy for request scheduling, and the
solution can be implemented on programmable switches. R2P2 does not have preemptive intra-server scheduling
and has head-of-line blocking. Thus, it suffers from long tail latency,
especially under high-dispersion workloads. Client-based solutions are lack of global view and
use power-of-k-choices scheduling based on stale server load information, and thus they suffer from inaccurate scheduling decisions.
We emulate 100 clients that generate requests with the same rate in the machines. Each client
performs the same policy as \sysname and tracks server queue lengths via piggybacking by its own.
The performance of Client(10) (which emulates 10 clients) and Client(1000) (which
emulates 1000 clients) are nearly the same as that of Client(100).
Figure~\ref{fig:eval_client} shows the performance of \sysname, Shinjuku, the client-based solution and R2P2
under Bimodal(90\%-50, 10\%-500) and Bimodal(50\%-50, 50\%-500) workloads. In both
workloads, \sysname outperforms others by maintaining low latency at higher
request rate, and Client(100) has nearly the same performance as Shinjuku. More
importantly, R2P2 is not robust to service time distributions.
It is close to \sysname under Bimodal(50\%-50, 50\%-500), and the gap between R2P2
and \sysname is
significantly larger under Bimodal(90\%-50, 10\%-500).

\subsection{Analysis of \sysname}
\label{sec:evaluation:analysis}

We analyze \sysname and show the impact of different
design choices, including different scheduling
policies of the switch-based inter-server scheduler and different
mechanisms to track the server loads.
\begin{figure}[t]
    \centering
    \subfigure[Bimodal(90\%-50,10\%-500).]{
        \label{fig:eval_policy_bimodal}
        \includegraphics[width=0.48\linewidth]{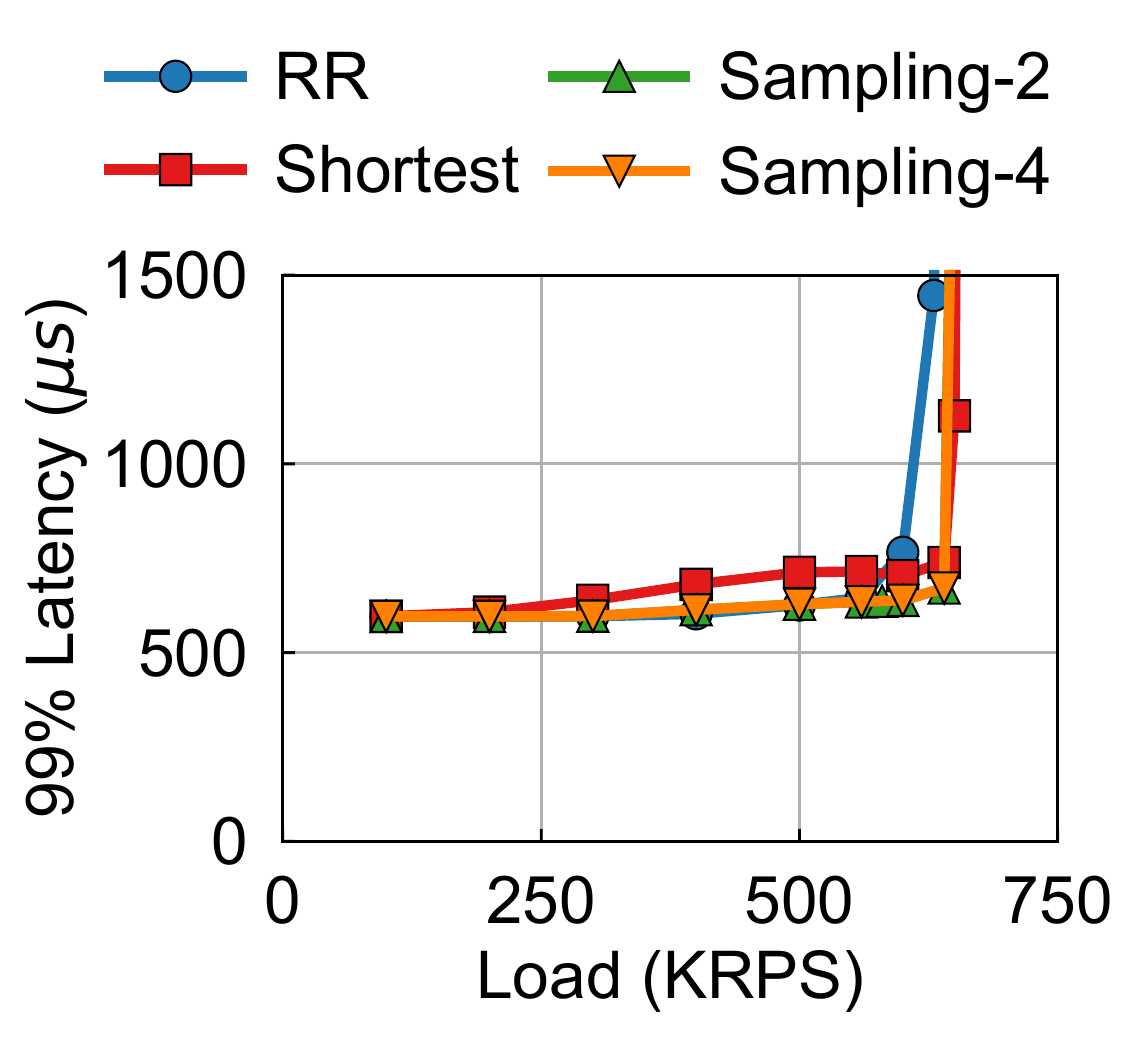}}
    \subfigure[Bimodal(50\%-50,50\%-500).]{
        \label{fig:eval_policy_port_bimodal}
        \includegraphics[width=0.48\linewidth]{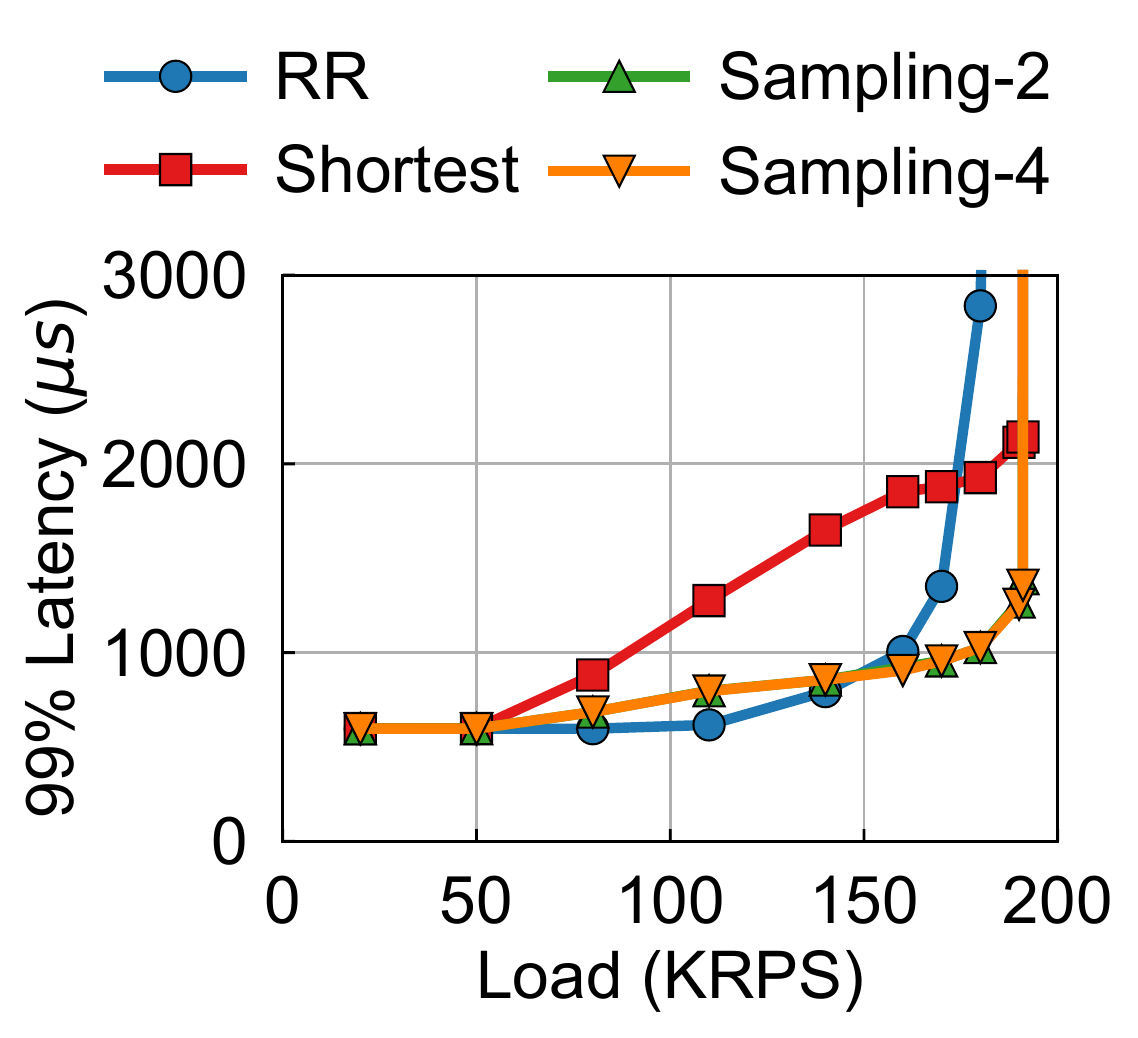}}
    \vspace{-0.1in}
    \caption{Impact of switch scheduling policies.}
    \vspace{-0.1in}
    \label{fig:eval_policy}
\end{figure}

\begin{figure}[t]
    \centering
    \subfigure[Bimodal(90\%-50,10\%-500).]{
        \label{fig:eval_tracking_bimodal}
        \includegraphics[width=0.48\linewidth]{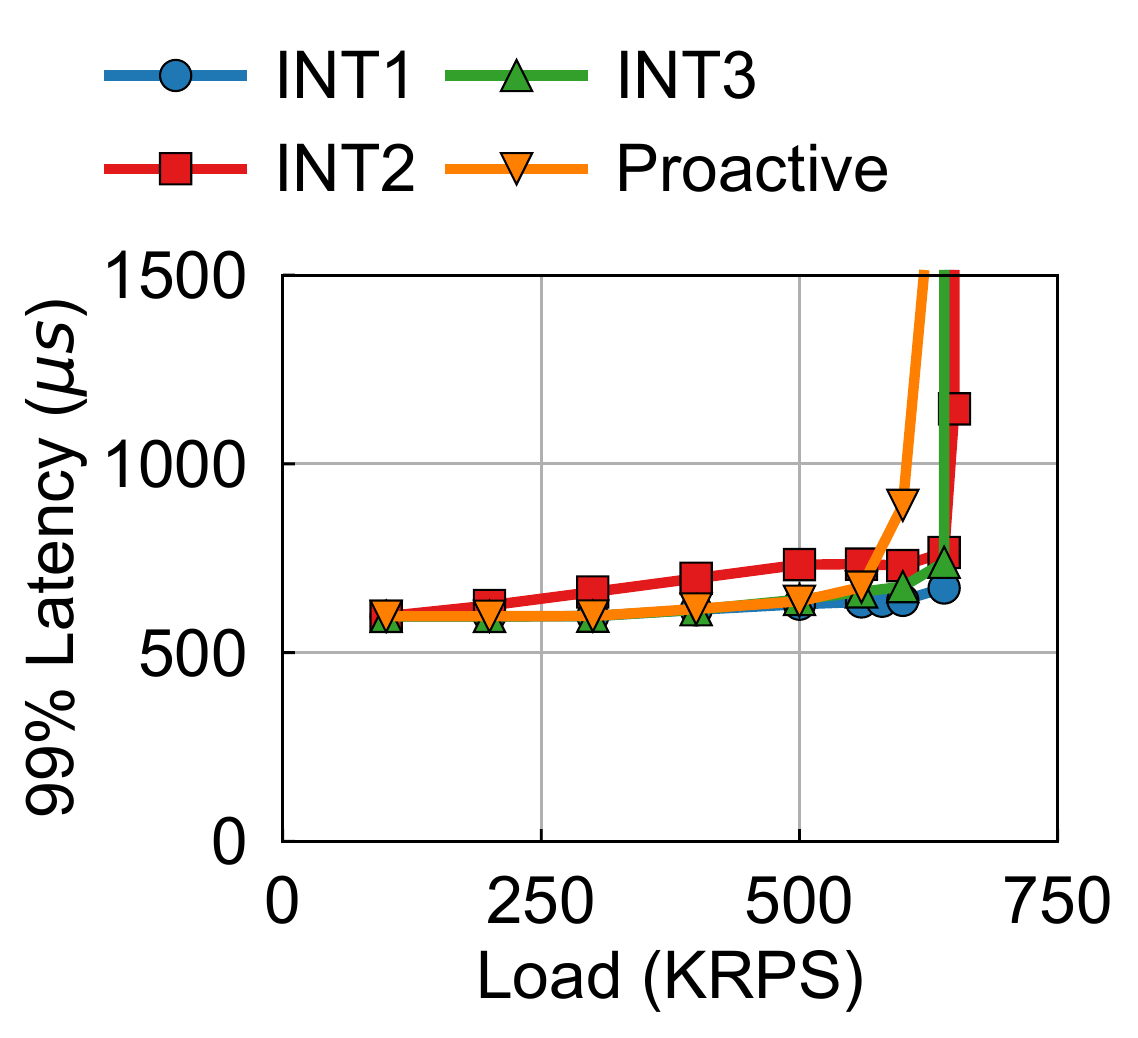}}
    \subfigure[Bimodal(50\%-50,50\%-500).]{
        \label{fig:eval_tracking_port_bimodal}
        \includegraphics[width=0.48\linewidth]{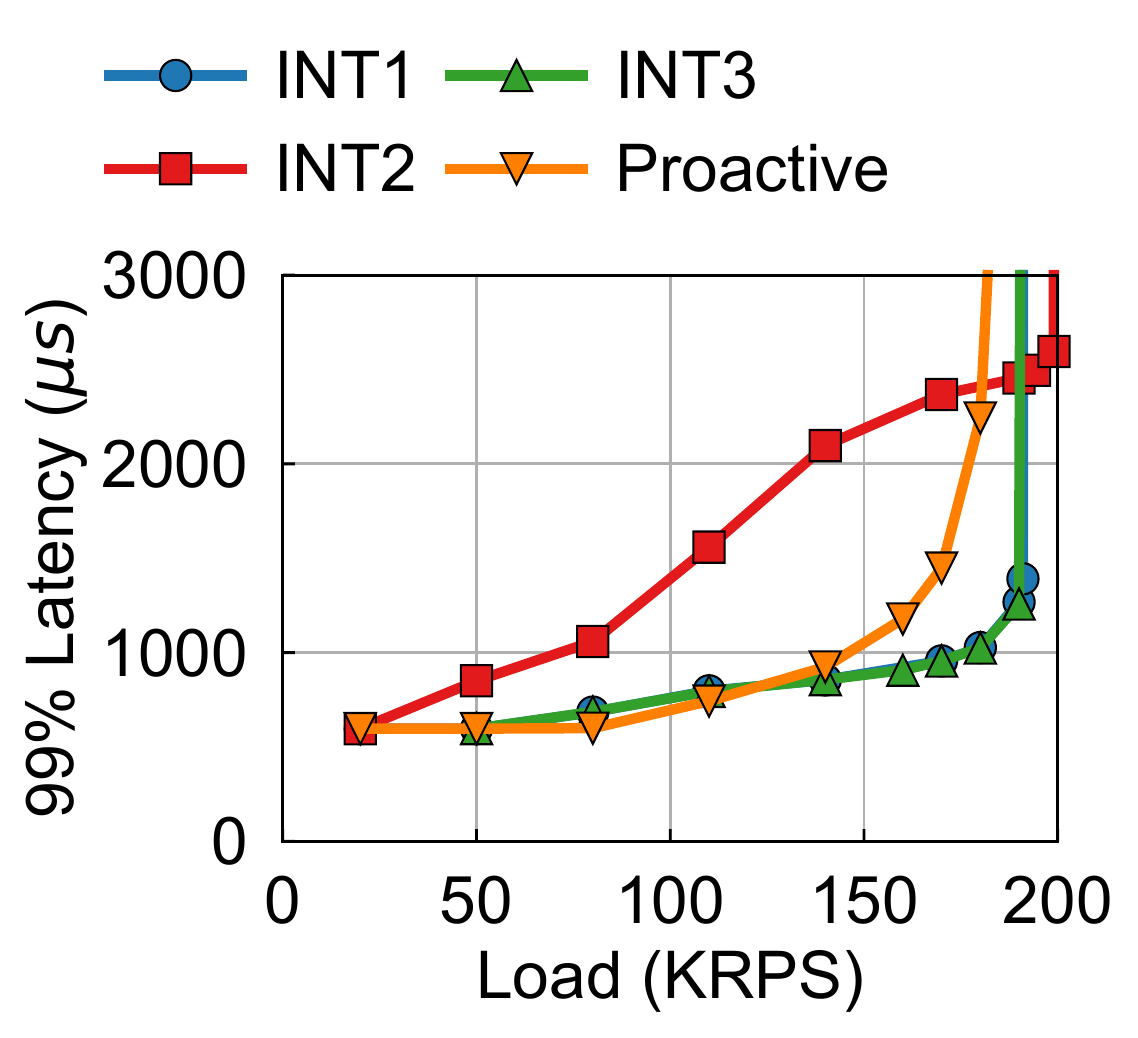}}
    \vspace{-0.1in}
    \caption{Impact of server load tracking mechanisms.}
    \vspace{-0.1in}
    \label{fig:eval_tracking}
\end{figure}

\para{Impact of switch scheduling policies.} Figure~\ref{fig:eval_policy}
evaluates the impact of different scheduling policies under Bimodal(90\%-50, 10\%-500) and
Bimodal(50\%-50, 50\%-500) workloads. We compare four scheduling
policies: $RR$ (which schedules requests to server with round-robin), $Shortest$
(which chooses the server with the smallest queue length), $Sampling$-2 (which
samples two servers and chooses the one with the smaller queue length), and
$Sampling$-4 (which samples four servers and chooses the one with the smallest
queue length). RR sends an even number of requests to each server, without
considering the variability of request service times. Thus, it suffers from long
tail latency at high request load. Theoretically, $Shortest$ can provide
effective load balancing, but it incurs high tail latency in practice, even at
low request load. As discussed in \S\ref{sec:overview}, the reason is that there is a delay to update the queue
lengths in the switch from the servers. When a server becomes the one
with the smallest queue length, multiple consecutive requests would all choose
this server, causing a micro \emph{herding} behavior. And the queue length of
this server has to wait to be updated until the new queue length is piggybacked in the first reply packet to update in the
switch. As discussed in \S\ref{sec:overview}, this herding behavior can be
handled by adding randomization to the scheduling process. The results in the
figure confirm the effectiveness of sampling. For the scale of the evaluated
scenario, sampling two and four servers have similar performance, because sampling two servers already provides enough choices to avoid hotspots and
enough randomization to avoid herding.

\para{Impact of server load tracking mechanisms.} Figure~\ref{fig:eval_tracking}
evaluates the impact of different mechanisms to track server loads, under both Bimodal(90\%-50, 10\%-500) and
Bimodal(50\%-50, 50\%-500) workloads. We compare three tracking
mechanisms discussed in \S\ref{sec:design:tracking}: $INT1$ (which tracks the number of outstanding requests for each server and computes the minimum), $INT2$ (which only tracks the minimum number of outstanding requests and updates on reply packets), $INT3$ (which tracks the total service time of outstanding requests for each server)
and $Proactive$ (which increments and
decrements the counters by the switch). $Proactive$ cannot precisely maintain
the queue length for each server as packet loss and retransmissions can introduce errors on the
counters, and as a result, it does not work well as others. $INT2$ performs
worse than $INT1$ because it only keeps one server with the minimum load, resulting in herding.
$INT3$ is comparable to $INT1$. However, it presumes that the service times are known as a priori, which is normally not the case in practice.
$INT1$ works the best because it accurately tracks server loads, enables
randomization to avoid herding for effective load balancing, and does not require any priori knowledge.

\begin{figure}[t]
    \centering
    \subfigure[\sysname handles a switch failure.]{
        \label{fig:eval_switch_failure}
        \includegraphics[width=0.75\linewidth]{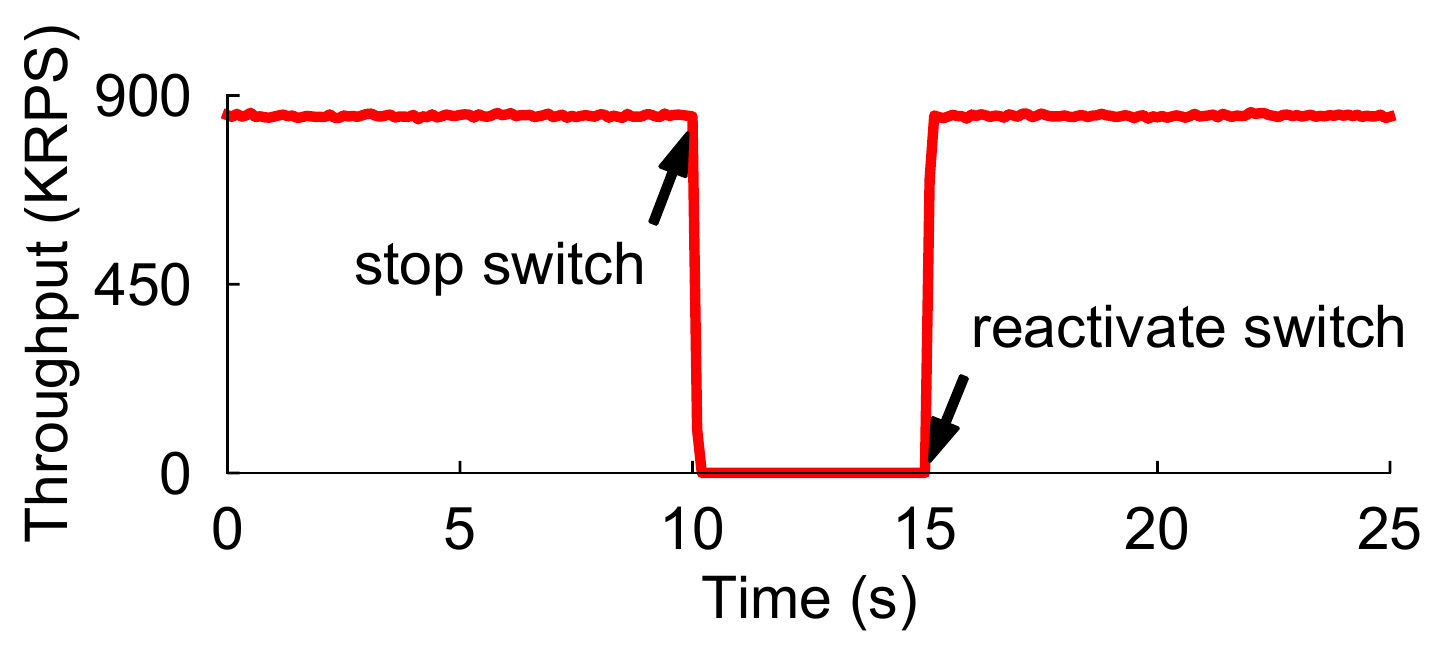}}
        \newline
    \subfigure[\sysname handles server reconfigurations.]{
        \label{fig:eval_server_reconfig}
        \includegraphics[width=0.75\linewidth]{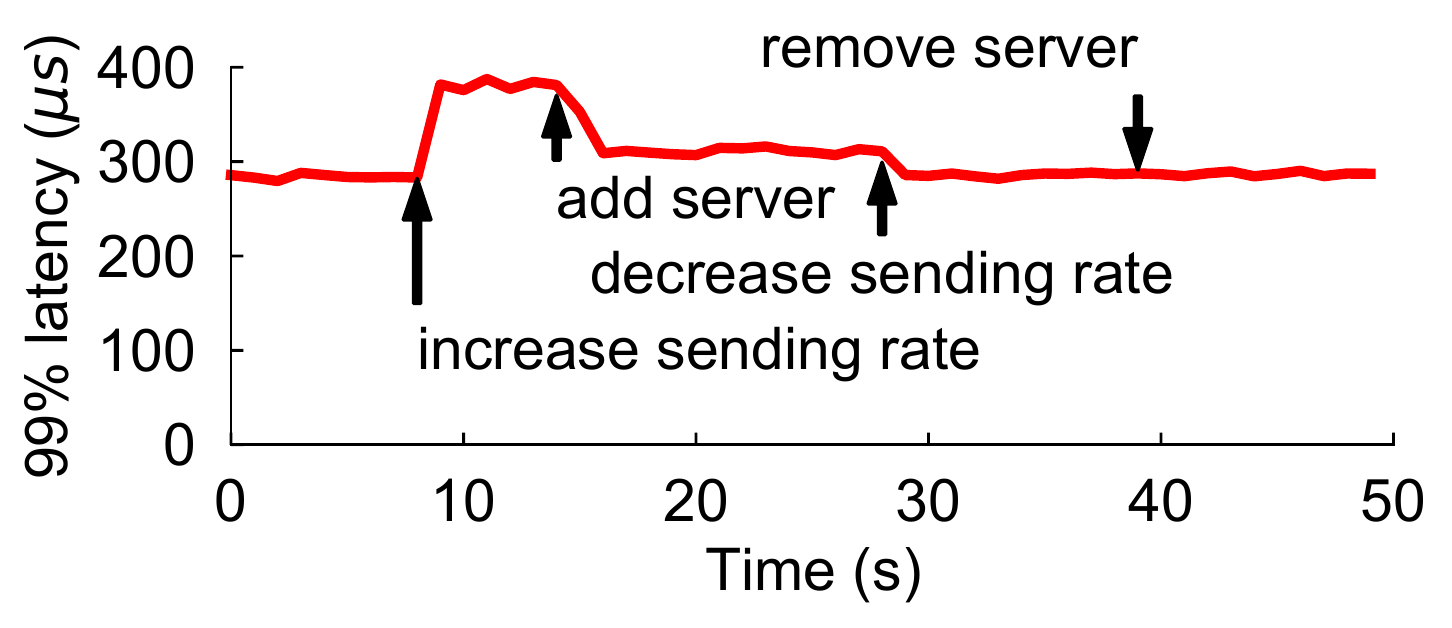}}
    \vspace{-0.1in}
    \caption{Handling failures and reconfigurations.}
    \vspace{-0.1in}
    \label{fig:eval_reconfig}
\end{figure}

\subsection{Request Affinity}
\label{sec:evaluation:affinity}

\paraf{Handling switch failures.} To simulate a switch failure, we first stop the
switch manually, then reactivate the switch after several seconds.
Figure~\ref{fig:eval_switch_failure} shows the total throughput during this
period under Exp(50) workload. At 10 s, the switch is stopped and the total
throughput drops to 0. We reactivate the switch after 5 seconds and the total
throughput recovers to the initial level. The
microsecond-scale requests have already timed out after 5 seconds. So it is safe to start with an empty
$ReqTable$ after the reactivation, as discussed in
\S\ref{sec:design:affinity}.

\para{Handling system reconfigurations.}
\sysname maintains request affinity during system reconfigurations.
Figure~\ref{fig:eval_server_reconfig} shows the 99\% latency under system reconfigurations.
We use two-packet requests under Exp(50) workload, and start with 500 KRPS load
and seven machines as the servers. At time 8 s, we increase the request sending
rate, and the 99\% latency goes up to around 380 $\mu s$. At time 14 s, we add
another server to serve the requests, and the 99\% latency drops to 310 $\mu s$.
At time 28 s, we set the request sending rate back to 500 KRPS. And the 99\%
latency drops to 280 $\mu s$ further. At time 39 s, we remove a server from the
rack. Since seven servers are enough for such workload, the 99\% latency remains
the same. As discussed in \S~\ref{sec:design:affinity}, the request affinity is
maintained by the $ReqTable$ in the above process.

\section{Discussion}
\label{sec:discussion}

\paraf{Target workloads.}
\sysname supports both stateless and replicated stateful services.
Examples include microservices, function-as-a-service, stream processing, replicated caches and storage, and replicated machine learning models for high-throughput inference.
It is unlikely for a stateful service to be replicated to all servers in a rack.
We expect a more practical scenario is that a rack would run multiple such services,
where each service is provided by a subset of (overlapping) servers, i.e., locality constraints.
The evaluation shows that \sysname can provide significant improvements for a service hosted on just 8 servers (\S\ref{sec:evaluation}).
And we provide additional results to show the benefits for multiple services with locality constraints in Appendix~\ref{sec:appendix:add_eval}.
These results demonstrate that \sysname provides significant benefits even when the service is replicated on just a couple of servers.

\para{Going beyond a rack.}
We focus on a single rack in this paper. A modern
rack can already pack hundreds of cores, and a future rack is expected to pack
thousands of cores~\cite{firebox, intel-rsd, hp-the-machine}, which is sufficient
for many services.
For planetary-scale services, a single microservice may span multiple racks. In
this scenario, there is no central place like the ToR switch in a rack that can
see and process all traffic. Yet, the abstraction of a rack-scale computer
provided by \sysname provides a useful building block for distributed inter-rack
scheduling. This would be an interesting direction for future work.

\section{Related Work}
\label{sec:related}

\paraf{Dataplane designs for low latency.} Conventional networking stacks and
operating systems usually sacrifice low latency for generality. To address the
need for low latency, various dataplane designs have been proposed, including optimized networking stacks~\cite{dpdk, jeong2014mtcp, Andromeda}
 and dataplane operating systems~\cite{peter2013arrakis,belay2014ix,prekas2017zygos,rpcvalet,shenango,shinjuku,humphries2019mind}. \sysname leverages such
dataplane designs and enhances them with an inter-server scheduler, realizing
low latency in a rack-scale computer.

\para{Scheduling and resource management.} There is a long line of
research on job scheduling and resource management~\cite{ghodsi2011dominant, grandl2016altruistic, grandl2016graphene, jyothi2016morpheus, peng2018optimus, tumanov2016tetrisched, vavilapalli2013apache, zaharia2008improving,
sparrow, kogias2019r2p2,kaffes2019centralized, humphries2019mind}. Many systems focus on large jobs that can run from seconds to hours, and they can afford running
sophisticated scheduling algorithms to make effective decisions. \sysname works at microsecond scale
and optimizes the tail latency with network-system co-design.

\para{Programmable switches.} Programmable switches bring new opportunities to
improve datacenter networks and systems, such as key-value stores~\cite{switchkv, incbricks-asplos17, netcache, distcache},
 coordination and consensus~\cite{netchain, nopaxos, eris, harmonia, netlock},
network telemetry~\cite{int, qpipe},
 machine learning acceleration~\cite{daiet, sapio2019scaling}
 and query processing offload~\cite{lerner2019case}.
There are also proposals for managing systems built with programmable switches~\cite{p4multitenancy, zheng2018p4visor, hyper4}.
 \sysname is a new solution that leverages the programmable switch
as an inter-rack scheduler to optimize microsecond-scale tail latency for
rack-scale computers.


\section{Conclusion}
\label{sec:conclusion}
We present \sysname, a rack-level microsecond-scale scheduler that provides the
abstraction of a rack-scale computer to an external service.
\sysname leverages a two-layer scheduling framework
to achieve scalability and low tail latency.
We hope that with the end
of Moore's law and Dennard's scaling, \sysname will inspire a new
generation of datacenter systems enabled by
domain-specific hardware and hardware-software co-design.

\vspace{0.1mm}

\para{Acknowledgments}
We thank our shepherd Ryan Stutsman and the anonymous reviewers for their
valuable feedback. We thank Jack Humphries for helping debug Shinjuku issues.
This work is supported in part by NSF grants CNS-1813487, CCF-1918757 and
CNS-1955487, a Facebook Communications \& Networking Research Award, and a
Google Faculty Research Award.

{
\bibliographystyle{abbrv}
\bibliography{paper}}

\begin{thebibliography}{10}

\bibitem{firebox}
{FireBox: A Hardware Building Block for 2020 Warehouse-Scale Computers}.
\newblock \url{https://www.usenix.org/node/179918}.

\bibitem{hp-the-machine}
{HP The Machine}.
\newblock \url{https://www.labs.hpe.com/the-machine}.

\bibitem{int}
{In-band Network Telemetry (INT) Dataplane Specification}.
\newblock \url{https://p4.org/specs/}.

\bibitem{dpdk}
{Intel Data Plane Development Kit (DPDK)}.
\newblock \url{http://dpdk.org/}.

\bibitem{intel-rsd}
{Intel Rack Scale Design}.
\newblock
  \url{https://www.intel.com/content/www/us/en/architecture-and-technology/rack-scale-design-overview.html}.

\bibitem{memcached}
Memcached key-value store.
\newblock \url{https://memcached.org/}.

\bibitem{redis}
Redis data structure store.
\newblock \url{https://redis.io/}.

\bibitem{tpu-pods}
{TPU Pods}.
\newblock \url{https://cloud.google.com/tpu/}.

\bibitem{voldtb}
Voltdb in-memory database.
\newblock \url{https://www.voltdb.com}.

\bibitem{p4studio}
{Barefoot P4 Studio}.
\newblock \url{https://www.barefootnetworks.com/products/brief-p4-studio/}.

\bibitem{tofino}
{Barefoot Tofino}.
\newblock \url{https://www.barefootnetworks.com/technology/}.

\bibitem{microseconds}
L.~Barroso, M.~Marty, D.~Patterson, and P.~Ranganathan.
\newblock Attack of the killer microseconds.
\newblock {\em Communications of the ACM}, 2017.

\bibitem{barroso2003web}
L.~A. Barroso, J.~Dean, and U.~H{\"o}lzle.
\newblock Web search for a planet: The {Google} cluster architecture.
\newblock {\em IEEE Micro}, 2003.

\bibitem{belay2014ix}
A.~Belay, G.~Prekas, A.~Klimovic, S.~Grossman, C.~Kozyrakis, and E.~Bugnion.
\newblock {IX}: A protected dataplane operating system for high throughput and
  low latency.
\newblock In {\em USENIX OSDI}, 2014.

\bibitem{robinhood}
D.~S. Berger, B.~Berg, T.~Zhu, S.~Sen, and M.~Harchol-Balter.
\newblock Robinhood: Tail latency aware caching--dynamic reallocation from
  cache-rich to cache-poor.
\newblock In {\em USENIX OSDI}, 2018.

\bibitem{p4-ccr}
P.~Bosshart, D.~Daly, G.~Gibb, M.~Izzard, N.~McKeown, J.~Rexford,
  C.~Schlesinger, D.~Talayco, A.~Vahdat, G.~Varghese, and D.~Walker.
\newblock P4: Programming protocol-independent packet processors.
\newblock {\em SIGCOMM CCR}, July 2014.

\bibitem{boucher2018putting}
S.~Boucher, A.~Kalia, D.~G. Andersen, and M.~Kaminsky.
\newblock Putting the ``micro'' back in microservice.
\newblock In {\em USENIX ATC}, 2018.

\bibitem{bramson2010randomized}
M.~Bramson, Y.~Lu, and B.~Prabhakar.
\newblock Randomized load balancing with general service time distributions.
\newblock {\em ACM SIGMETRICS performance evaluation review}, 38(1):275--286,
  2010.

\bibitem{cerqueira2013comparison}
F.~Cerqueira and B.~Brandenburg.
\newblock A comparison of scheduling latency in linux, preempt-rt, and litmus
  rt.
\newblock In {\em Annual Workshop on Operating Systems Platforms for Embedded
  Real-Time Applications}, 2013.

\bibitem{rpcvalet}
A.~Daglis, M.~Sutherland, and B.~Falsafi.
\newblock {RPCValet}: {NI}-driven tail-aware balancing of $\mu$s-scale {RPCs}.
\newblock In {\em ACM ASPLOS}, 2019.

\bibitem{Andromeda}
M.~Dalton, D.~Schultz, J.~Adriaens, A.~Arefin, A.~Gupta, B.~Fahs,
  D.~Rubinstein, E.~C. Zermeno, E.~Rubow, J.~A. Docauer, J.~Alpert, J.~Ai,
  J.~Olson, K.~DeCabooter, M.~de~Kruijf, N.~Hua, N.~Lewis, N.~Kasinadhuni,
  R.~Crepaldi, S.~Krishnan, S.~Venkata, Y.~Richter, U.~Naik, and A.~Vahdat.
\newblock Andromeda: Performance, isolation, and velocity at scale in cloud
  network virtualization.
\newblock In {\em USENIX NSDI}, 2018.

\bibitem{Dean:tail}
J.~Dean and L.~A. Barroso.
\newblock The tail at scale.
\newblock {\em Communications of the ACM}, February 2013.

\bibitem{demers1989analysis}
A.~Demers, S.~Keshav, and S.~Shenker.
\newblock Analysis and simulation of a fair queueing algorithm.
\newblock In {\em ACM SIGCOMM}, 1989.

\bibitem{maglev}
D.~E. Eisenbud, C.~Yi, C.~Contavalli, C.~Smith, R.~Kononov, E.~Mann-Hielscher,
  A.~Cilingiroglu, B.~Cheyney, W.~Shang, and J.~D. Hosein.
\newblock Maglev: A fast and reliable software network load balancer.
\newblock In {\em USENIX NSDI}, 2016.

\bibitem{ephremides1980simple}
A.~Ephremides, P.~Varaiya, and J.~Walrand.
\newblock A simple dynamic routing problem.
\newblock {\em IEEE transactions on Automatic Control}, 25(4):690--693, 1980.

\bibitem{duet}
R.~Gandhi, H.~H. Liu, Y.~C. Hu, G.~Lu, J.~Padhye, L.~Yuan, and M.~Zhang.
\newblock Duet: Cloud scale load balancing with hardware and software.
\newblock In {\em ACM SIGCOMM}, August 2015.

\bibitem{ghodsi2011dominant}
A.~Ghodsi, M.~Zaharia, B.~Hindman, A.~Konwinski, S.~Shenker, and I.~Stoica.
\newblock Dominant resource fairness: Fair allocation of multiple resource
  types.
\newblock In {\em USENIX NSDI}, 2011.

\bibitem{gog2016firmament}
I.~Gog, M.~Schwarzkopf, A.~Gleave, R.~N. Watson, and S.~Hand.
\newblock Firmament: Fast, centralized cluster scheduling at scale.
\newblock In {\em USENIX OSDI}, 2016.

\bibitem{grandl2016altruistic}
R.~Grandl, M.~Chowdhury, A.~Akella, and G.~Ananthanarayanan.
\newblock Altruistic scheduling in multi-resource clusters.
\newblock In {\em USENIX OSDI}, 2016.

\bibitem{grandl2016graphene}
R.~Grandl, S.~Kandula, S.~Rao, A.~Akella, and J.~Kulkarni.
\newblock Graphene: Packing and dependency-aware scheduling for data-parallel
  clusters.
\newblock In {\em USENIX OSDI}, 2016.

\bibitem{gupta2007analysis}
V.~Gupta, M.~H. Balter, K.~Sigman, and W.~Whitt.
\newblock Analysis of join-the-shortest-queue routing for web server farms.
\newblock {\em Performance Evaluation}, 64(9-12):1062--1081, 2007.

\bibitem{gupta2007insensitivity}
V.~Gupta, K.~Sigman, M.~Harchol-Balter, and W.~Whitt.
\newblock Insensitivity for ps server farms with jsq routing.
\newblock {\em ACM SIGMETRICS Performance Evaluation Review}, 35(2):24--26,
  2007.

\bibitem{hyper4}
D.~Hancock and J.~Van~der Merwe.
\newblock Hyper4: Using p4 to virtualize the programmable data plane.
\newblock In {\em ACM CoNEXT}, 2016.

\bibitem{hennessy2019new}
J.~L. Hennessy and D.~A. Patterson.
\newblock A new golden age for computer architecture.
\newblock {\em Communications of the ACM}, 2019.

\bibitem{humphries2019mind}
J.~T. Humphries, K.~Kaffes, D.~Mazi{\`e}res, and C.~Kozyrakis.
\newblock Mind the gap: A case for informed request scheduling at the nic.
\newblock In {\em ACM SIGCOMM HotNets Workshop}, 2019.

\bibitem{qpipe}
N.~Ivkin, Z.~Yu, V.~Braverman, and X.~Jin.
\newblock {QPipe}: Quantiles sketch fully in the data plane.
\newblock In {\em ACM CoNEXT}, December 2019.

\bibitem{jeong2014mtcp}
E.~Jeong, S.~Wood, M.~Jamshed, H.~Jeong, S.~Ihm, D.~Han, and K.~Park.
\newblock {mTCP}: a highly scalable user-level {TCP} stack for multicore
  systems.
\newblock In {\em USENIX NSDI}, 2014.

\bibitem{netchain}
X.~Jin, X.~Li, H.~Zhang, N.~Foster, J.~Lee, R.~Soul{\'e}, C.~Kim, and
  I.~Stoica.
\newblock {NetChain}: Scale-free sub-{RTT} coordination.
\newblock In {\em USENIX NSDI}, 2018.

\bibitem{netcache}
X.~Jin, X.~Li, H.~Zhang, R.~Soul{\'e}, J.~Lee, N.~Foster, C.~Kim, and
  I.~Stoica.
\newblock {NetCache}: Balancing key-value stores with fast in-network caching.
\newblock In {\em ACM SOSP}, 2017.

\bibitem{dionysus}
X.~Jin, H.~H. Liu, R.~Gandhi, S.~Kandula, R.~Mahajan, M.~Zhang, J.~Rexford, and
  R.~Wattenhofer.
\newblock Dynamic scheduling of network updates.
\newblock In {\em ACM SIGCOMM}, 2014.

\bibitem{jyothi2016morpheus}
S.~A. Jyothi, C.~Curino, I.~Menache, S.~M. Narayanamurthy, A.~Tumanov,
  J.~Yaniv, R.~Mavlyutov, {\'I}.~Goiri, S.~Krishnan, J.~Kulkarni, et~al.
\newblock Morpheus: Towards automated slos for enterprise clusters.
\newblock In {\em USENIX OSDI}, 2016.

\bibitem{shinjuku}
K.~Kaffes, T.~Chong, J.~T. Humphries, A.~Belay, D.~Mazi{\`e}res, and
  C.~Kozyrakis.
\newblock Shinjuku: Preemptive scheduling for $\mu$second-scale tail latency.
\newblock In {\em USENIX NSDI}, 2019.

\bibitem{kaffes2019centralized}
K.~Kaffes, N.~J. Yadwadkar, and C.~Kozyrakis.
\newblock Centralized core-granular scheduling for serverless functions.
\newblock In {\em ACM Symposium on Cloud Computing}, 2019.

\bibitem{erpc}
A.~Kalia, M.~Kaminsky, and D.~Andersen.
\newblock Datacenter rpcs can be general and fast.
\newblock In {\em USENIX NSDI}, 2019.

\bibitem{kogias2019r2p2}
M.~Kogias, G.~Prekas, A.~Ghosn, J.~Fietz, and E.~Bugnion.
\newblock {R2P2}: Making {RPCs} first-class datacenter citizens.
\newblock In {\em USENIX ATC}, 2019.

\bibitem{splinter}
C.~Kulkarni, S.~Moore, M.~Naqvi, T.~Zhang, R.~Ricci, and R.~Stutsman.
\newblock Splinter: Bare-metal extensions for multi-tenant low-latency storage.
\newblock In {\em USENIX OSDI}, 2018.

\bibitem{lerner2019case}
A.~Lerner, R.~Hussein, P.~Cudre-Mauroux, and U.~eXascale Infolab.
\newblock The case for network accelerated query processing.
\newblock In {\em CIDR}, 2019.

\bibitem{eris}
J.~Li, E.~Michael, and D.~R.~K. Ports.
\newblock Eris: Coordination-free consistent transactions using in-network
  concurrency control.
\newblock In {\em ACM SOSP}, October 2017.

\bibitem{nopaxos}
J.~Li, E.~Michael, N.~K. Sharma, A.~Szekeres, and D.~R. Ports.
\newblock Just say {NO} to {Paxos} overhead: Replacing consensus with network
  ordering.
\newblock In {\em USENIX OSDI}, November 2016.

\bibitem{pegasus}
J.~Li, J.~Nelson, X.~Jin, and D.~R. Ports.
\newblock Pegasus: Load-aware selective replication with an in-network
  coherence directory.
\newblock {\em Technical Report UW-CSE-18-12-01}, 2018.

\bibitem{switchkv}
X.~Li, R.~Sethi, M.~Kaminsky, D.~G. Andersen, and M.~J. Freedman.
\newblock Be fast, cheap and in control with {SwitchKV}.
\newblock In {\em USENIX NSDI}, March 2016.

\bibitem{incbricks-asplos17}
M.~Liu, L.~Luo, J.~Nelson, L.~Ceze, A.~Krishnamurthy, and K.~Atreya.
\newblock {IncBricks}: Toward in-network computation with an in-network cache.
\newblock In {\em ACM ASPLOS}, April 2017.

\bibitem{distcache}
Z.~Liu, Z.~Bai, Z.~Liu, X.~Li, C.~Kim, V.~Braverman, X.~Jin, and I.~Stoica.
\newblock Distcache: Provable load balancing for large-scale storage systems
  with distributed caching.
\newblock In {\em USENIX FAST}, 2019.

\bibitem{miao2017silkroad}
R.~Miao, H.~Zeng, C.~Kim, J.~Lee, and M.~Yu.
\newblock Silkroad: Making stateful layer-4 load balancing fast and cheap using
  switching asics.
\newblock In {\em ACM SIGCOMM}, 2017.

\bibitem{noviswitch}
{NoviSwitch}.
\newblock \url{http://noviflow.com/products/noviswitch/}.

\bibitem{olteanu2018stateless}
V.~Olteanu, A.~Agache, A.~Voinescu, and C.~Raiciu.
\newblock Stateless datacenter load-balancing with {Beamer}.
\newblock In {\em USENIX NSDI}, 2018.

\bibitem{shenango}
A.~Ousterhout, J.~Fried, J.~Behrens, A.~Belay, and H.~Balakrishnan.
\newblock Shenango: Achieving high {CPU} efficiency for latency-sensitive
  datacenter workloads.
\newblock In {\em USENIX NSDI}, 2019.

\bibitem{sparrow}
K.~Ousterhout, P.~Wendell, M.~Zaharia, and I.~Stoica.
\newblock Sparrow: Distributed, low latency scheduling.
\newblock In {\em ACM SOSP}, 2013.

\bibitem{ananta}
P.~Patel, D.~Bansal, L.~Yuan, A.~Murthy, A.~Greenberg, D.~A. Maltz, R.~Kern,
  H.~Kumar, M.~Zikos, H.~Wu, et~al.
\newblock Ananta: Cloud scale load balancing.
\newblock In {\em SIGCOMM CCR}, 2013.

\bibitem{peng2018optimus}
Y.~Peng, Y.~Bao, Y.~Chen, C.~Wu, and C.~Guo.
\newblock Optimus: An efficient dynamic resource scheduler for deep learning
  clusters.
\newblock In {\em EuroSys}, 2018.

\bibitem{peter2013arrakis}
S.~Peter, J.~Li, I.~Zhang, D.~R. Ports, A.~Krishnamurthy, T.~Anderson, and
  T.~Roscoe.
\newblock Arrakis: The operating system is the control plane.
\newblock In {\em USENIX OSDI}, 2013.

\bibitem{prekas2017zygos}
G.~Prekas, M.~Kogias, and E.~Bugnion.
\newblock {ZygOS}: Achieving low tail latency for microsecond-scale networked
  tasks.
\newblock In {\em ACM SOSP}, 2017.

\bibitem{richa2001power}
A.~W. Richa, M.~Mitzenmacher, and R.~Sitaraman.
\newblock The power of two random choices: A survey of techniques and results.
\newblock {\em Combinatorial Optimization}, 9:255--304, 2001.

\bibitem{rocksdb}
RocksDB.
\newblock {RocksDB}.
\newblock \url{https://rocksdb.org/}.

\bibitem{daiet}
A.~Sapio, I.~Abdelaziz, A.~Aldilaijan, M.~Canini, and P.~Kalnis.
\newblock In-network computation is a dumb idea whose time has come.
\newblock In {\em ACM SIGCOMM HotNets Workshop}, November 2017.

\bibitem{sapio2019scaling}
A.~Sapio, M.~Canini, C.-Y. Ho, J.~Nelson, P.~Kalnis, C.~Kim, A.~Krishnamurthy,
  M.~Moshref, D.~R. Ports, and P.~Richt{\'a}rik.
\newblock Scaling distributed machine learning with in-network aggregation.
\newblock {\em arXiv}, 2019.

\bibitem{c3}
L.~Suresh, M.~Canini, S.~Schmid, and A.~Feldmann.
\newblock C3: Cutting tail latency in cloud data stores via adaptive replica
  selection.
\newblock In {\em USENIX NSDI}, 2015.

\bibitem{silo}
S.~Tu, W.~Zheng, E.~Kohler, B.~Liskov, and S.~Madden.
\newblock Speedy transactions in multicore in-memory databases.
\newblock In {\em ACM SOSP}, 2013.

\bibitem{tumanov2016tetrisched}
A.~Tumanov, T.~Zhu, J.~W. Park, M.~A. Kozuch, M.~Harchol-Balter, and G.~R.
  Ganger.
\newblock Tetrisched: global rescheduling with adaptive plan-ahead in dynamic
  heterogeneous clusters.
\newblock In {\em EuroSys}, 2016.

\bibitem{vavilapalli2013apache}
V.~K. Vavilapalli, A.~C. Murthy, C.~Douglas, S.~Agarwal, M.~Konar, R.~Evans,
  T.~Graves, J.~Lowe, H.~Shah, S.~Seth, et~al.
\newblock {Apache Hadoop YARN}: Yet another resource negotiator.
\newblock In {\em ACM Symposium on Cloud Computing}, 2013.

\bibitem{facebook-tao-sigmod12}
V.~Venkataramani, Z.~Amsden, N.~Bronson, G.~Cabrera~III, P.~Chakka, P.~Dimov,
  H.~Ding, J.~Ferris, A.~Giardullo, J.~Hoon, S.~Kulkarni, N.~Lawrence,
  M.~Marchukov, D.~Petrov, and L.~Puzar.
\newblock {TAO}: How {Facebook} serves the social graph.
\newblock In {\em ACM SIGMOD}, May 2012.

\bibitem{p4multitenancy}
T.~Wang, H.~Zhu, F.~Ruffy, X.~Jin, A.~Sivaraman, D.~R. Ports, and A.~Panda.
\newblock Multitenancy for fast and programmable networks in the cloud.
\newblock In {\em USENIX HotCloud Workshop}, July 2020.

\bibitem{weber1978optimal}
R.~R. Weber.
\newblock On the optimal assignment of customers to parallel servers.
\newblock {\em Journal of Applied Probability}, 15(2):406--413, 1978.

\bibitem{wierman2012tail}
A.~Wierman and B.~Zwart.
\newblock Is tail-optimal scheduling possible?
\newblock {\em Operations research}, 60(5):1249--1257, 2012.

\bibitem{winston1977optimality}
W.~Winston.
\newblock Optimality of the shortest line discipline.
\newblock {\em Journal of Applied Probability}, 14(1):181--189, 1977.

\bibitem{netlock}
Z.~Yu, Y.~Zhang, V.~Braverman, M.~Chowdhury, and X.~Jin.
\newblock Netlock: Fast, centralized lock management using programmable
  switches.
\newblock In {\em ACM SIGCOMM}, 2020.

\bibitem{zaharia2008improving}
M.~Zaharia, A.~Konwinski, A.~D. Joseph, R.~H. Katz, and I.~Stoica.
\newblock Improving mapreduce performance in heterogeneous environments.
\newblock In {\em USENIX OSDI}, 2008.

\bibitem{zheng2018p4visor}
P.~Zheng, T.~Benson, and C.~Hu.
\newblock {P4Visor}: Lightweight virtualization and composition primitives for
  building and testing modular programs.
\newblock In {\em ACM CoNEXT}, 2018.

\bibitem{harmonia}
H.~Zhu, Z.~Bai, J.~Li, E.~Michael, D.~R. Ports, I.~Stoica, and X.~Jin.
\newblock Harmonia: Near-linear scalability for replicated storage with
  in-network conflict detection.
\newblock {\em Proceedings of the VLDB Endowment}, 2019.

\end{thebibliography}

\newpage
\begin{appendix}

\section{Analysis}
\label{sec:appendix:analysis}

This section examines the policies $M/G/K/JSQ/FCFS$ for low-dispersion jobs and
$M/G/K/JSQ/PS$ for high-dispersion or heavy-tailed jobs.  Theoretical properties of
$M/G/K/JSQ/FCFS$ have been extensively
examined~\cite{weber1978optimal,winston1977optimality,ephremides1980simple,richa2001power,gupta2007insensitivity},
whereas  queuing theorists tackle $M/G/K/JSQ/PS$  systems by using a combination
of rigorous and heuristic analysis. For completeness, this section outlines
analysis and key intuitions developed in the literature
(see~\cite{gupta2007analysis,bramson2010randomized,wierman2012tail} for
details).

\para{Major conclusions.} We have three major conclusions.

\noindent{\emph{C1. The load distribution resembles that in a standard
Power-of-Two-Choices model (PoT).}} Our algorithm here can be viewed as a
power-of-$k$-choices algorithm, where $k$ is the number of servers. Thus, it is
more powerful than PoT. Each server's load distribution resembles that in PoT
systems but has ``better parameters.''

\noindent{\emph{C2. JSQ is the only known policy that's robust to the service
distribution.}} The load distribution is \emph{insensitive} to service
distribution and therefore, our algorithm can handle both heavy and light tails.
Note that other policies, such as round-robin, random, or least-load-left are
\emph{not} distribution insensitive. Therefore, JSQ is the most robust among all
known policies.

\noindent{\emph{C3. When job distributions are known to have low-dispersion, we
can trade robustness for higher speed.}} The insensitivity results in C2 is an
asymptotic result, characterizing the limiting behavior of a system (in which
both servers and jobs grow to infinite). When the jobs are known to have
low-dispersion, the $M/G/K/JSQ/FCFS$ policy can further optimize the
performance. This effectively trades robustness for speed when distributional
information about the jobs is available.

We remark that for high-dispsersion or heavy-tailed jobs, it is theoretically
possible to improve the tail of the processing time (e.g., make the average
processing time for the 1\% slowest jobs smaller) when the service distribution
is known~\cite{wierman2012tail}. But the performance gain is limited and it can
be difficult to exactly characterize service distribution in practice. Thus,
optimizing policies based on the service distribution is not always feasible.

\subsection{Analysis for C1 and C2}

This section shows C1 and C2.

\para{Techniques.} Our analysis consists of two steps (see
e.g.,~\cite{gupta2007analysis}). In step 1, we show that the queues are
approximately independent when the service times follow an exponential
distribution. Therefore, a balanced condition technique can be used to
characterize the equilibrium state. In step 2, one analyzes special cases and
performs extensive simulations to confirm that equilibrium state in our
algorithm is insensitive to service time distribution.

\para{Notation.} There are a total number of $K$ servers. The arrival of the
jobs follows a Poisson process with arrival rate $\lambda$ and the service time
follows a general distribution with mean $\mu^{-1}$. Let us also define $\rho =
\lambda/(K\mu)$.

We say a queue is $M_n/M/1$ if the arrival process into queue $Q$ is a
stochastic point process with state-dependent rates. In particular, the arrival
process into queue $Q$ has stochastic intensity $\lambda(n)$, where $n$ is the
current length of the queue.

\para{Step 1. Independence of queues under exponential service time.} We first
consider a special case, in which the service time follows an exponential
distribution (i.e., $M/M/K/JSQ/PS$). There is the following theorem.

\begin{theorem}\label{thm:independent}~\cite{gupta2007analysis} Consider an
$M/M/K/JSQ/PS$ model. Let $Q$ be any particular server in the model. There
exists an $M_n(\lambda(\cdot))/M/1$ such that this queue's steady-state
queue-length distribution is the same as $Q$.
\end{theorem}

\noindent{\emph{Interpretation and an example.}} Consider two queues $Q_1$ and
$Q_2$ in a system of interest. $Q_1$ has 10 jobs and $Q_2$ has 100 jobs.
Theorem~\ref{thm:independent} states that the arrival rate of $Q_1$ is
$\lambda(10)$ and that of $Q_2$ is $\lambda(100)$. We shall also see soon that
$\lambda(100) \ll \lambda(10)$. These arrival rates are \emph{independent} to
the states of other servers at the limit. The fact that $\lambda(100) \ll
\lambda(10)$ is also quite intuitive: when a queue has 100 jobs, it is much less
likely to be the shortest one (than when it has 10 jobs).

\noindent{\emph{Balanced condition.}} The function $\lambda(n)$ controls the
efficacy of our system, and can be computed by using a balanced condition
technique. This technique is widely used in analyzing PoT
systems~\cite{richa2001power}. Let $x_n$ denotes the limiting probability that
there are $n$ jobs in a single queue. We have \begin{equation} x_n \lambda(n) =
x_{n + 1}\mu. \end{equation} The left hand side represents the fraction of
servers that will move from state $n$ to state $n + 1$ (state $n$ means that the
queue has $n$ jobs). The right hand side represents the fraction of servers that
will move from state $n + 1$ to state $n$. At the equilibrium, these two masses
should be the same. This also implies $x_n = \rho^{nK}$. One can check that this
policy is strictly better than the PoT policy. For example, the probability that
there exists a server that has $c \log \log n$ jobs is exponentially small in
our system but is only polynomial small in a PoT system.

\noindent{\emph{Quality of approximation.}} Our system needs to use
power-of-$k$-choices to choose the server for performance and latency reasons
(Section~\ref{sec:overview}). One can use results
from~\cite{bramson2010randomized} to obtain a result that is similar to
Theorem~\ref{thm:independent} (the queues can be treated as being independent).
In addition, the balanced condition technique is still applicable. In this case,
the distribution of queue length decays in a doubly exponential manner.

\begin{figure*}[t]
    \centering
    \subfigure[Total requests.]{
        \label{fig:eval_locality_total}
        \includegraphics[width=0.3\linewidth]{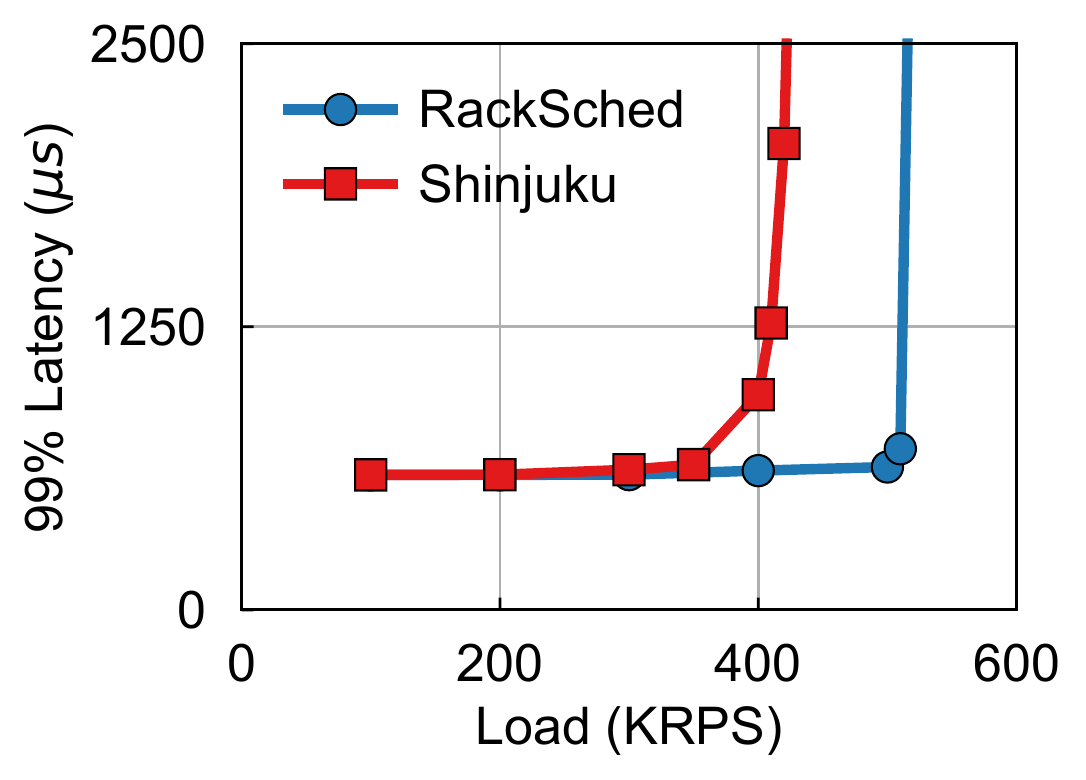}}
    \subfigure[Type-I requests (with locality constraint).]{
        \label{fig:eval_locality_1}
        \includegraphics[width=0.3\linewidth]{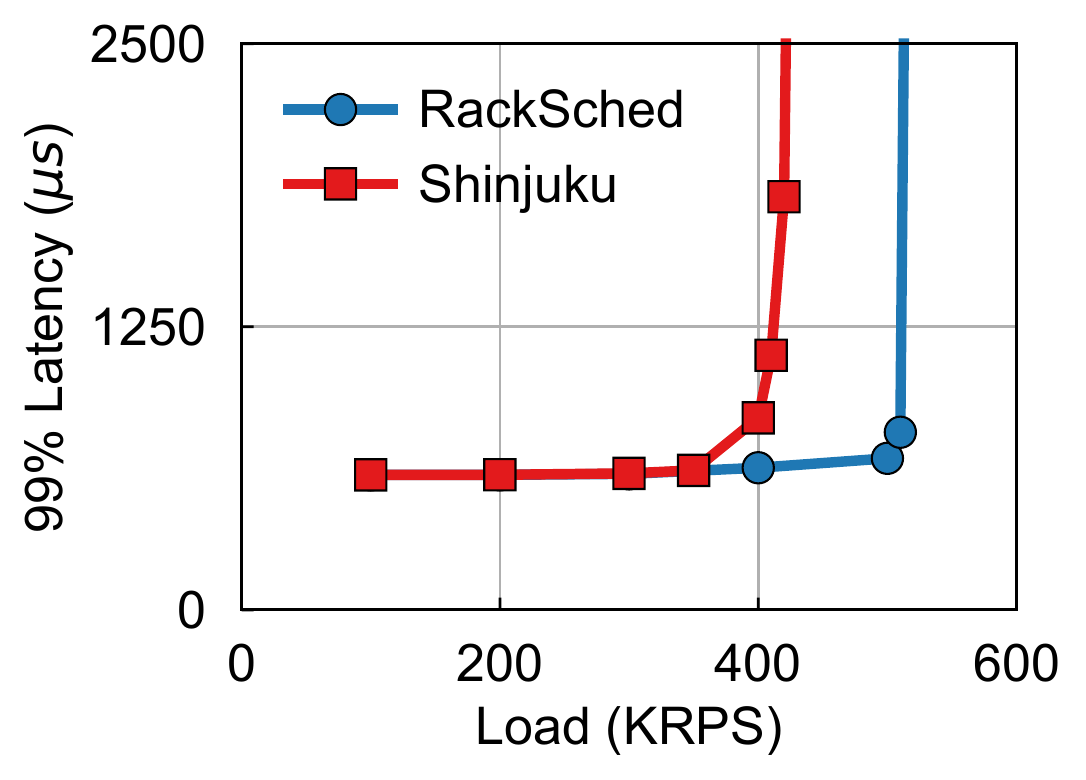}}
    \subfigure[Type-II requests (w/o locality constraint).]{
        \label{fig:eval_locality_2}
        \includegraphics[width=0.3\linewidth]{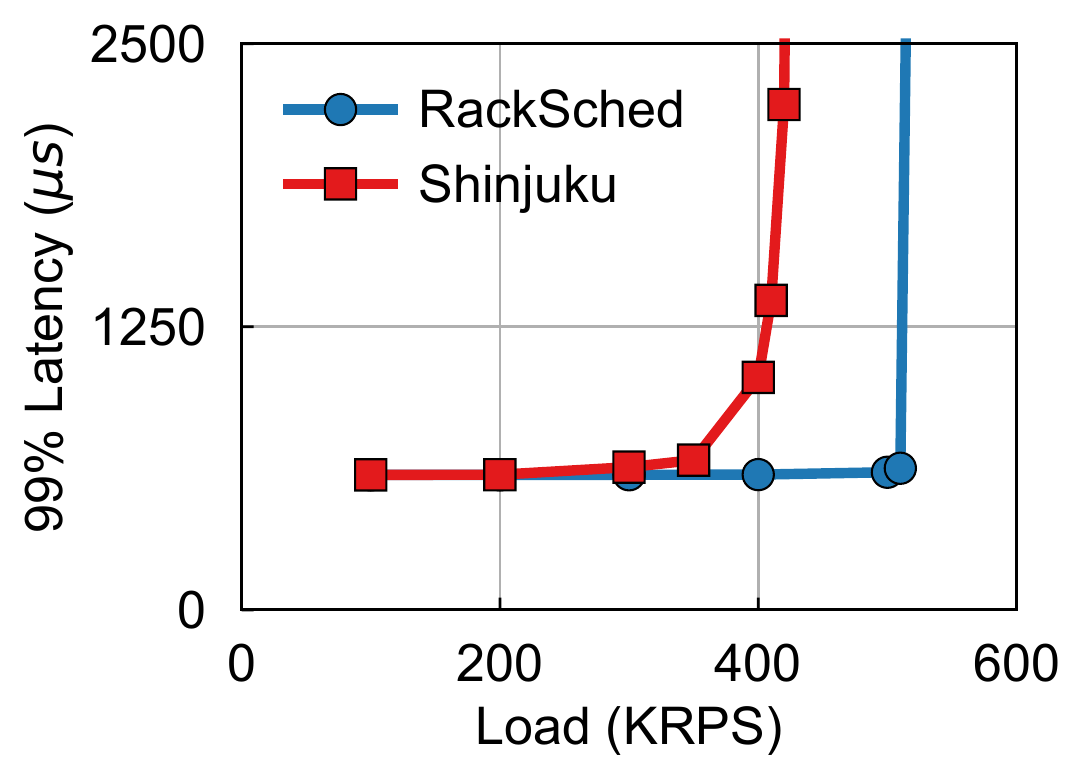}}
    \vspace{-0.1in}
    \caption{Data locality results with Bimodal(90\%-50, 10\%-500).}
    \vspace{-0.1in}
    \label{fig:eval_locality}
\end{figure*}

\begin{figure}[t]
    \centering
    \subfigure[Without priority policy.]{
        \label{fig:eval_priority_wo}
        \includegraphics[width=0.48\linewidth]{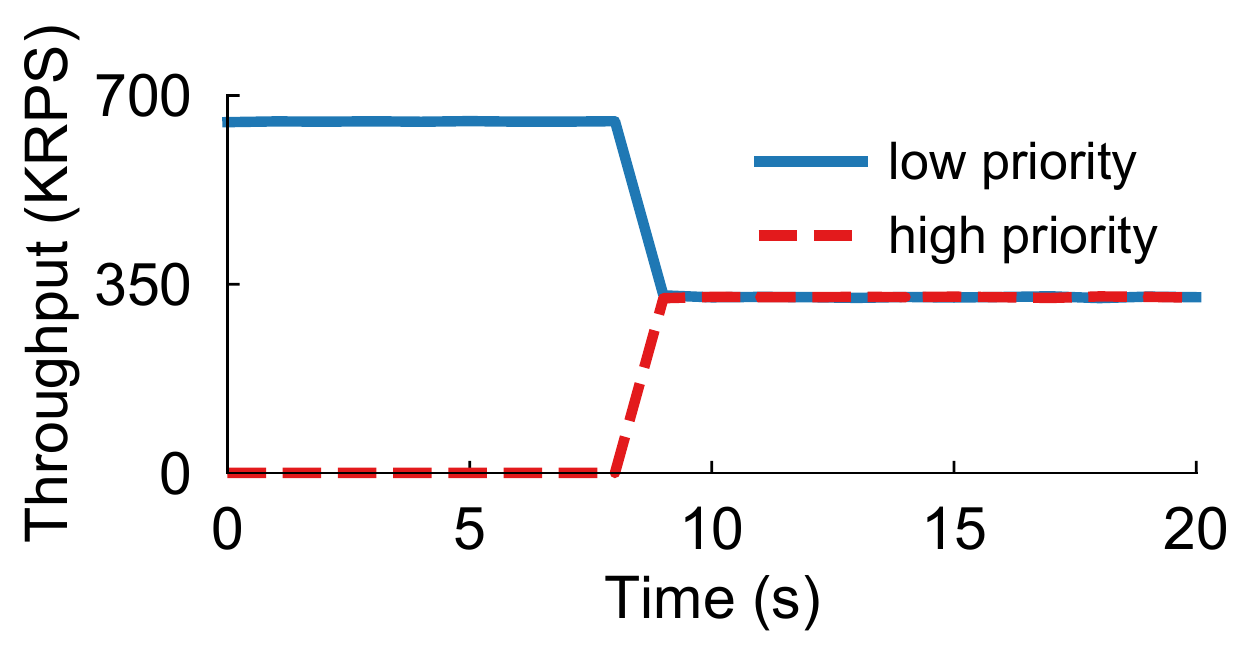}}
    \subfigure[With priority policy.]{
        \label{fig:eval_priority_with}
        \includegraphics[width=0.48\linewidth]{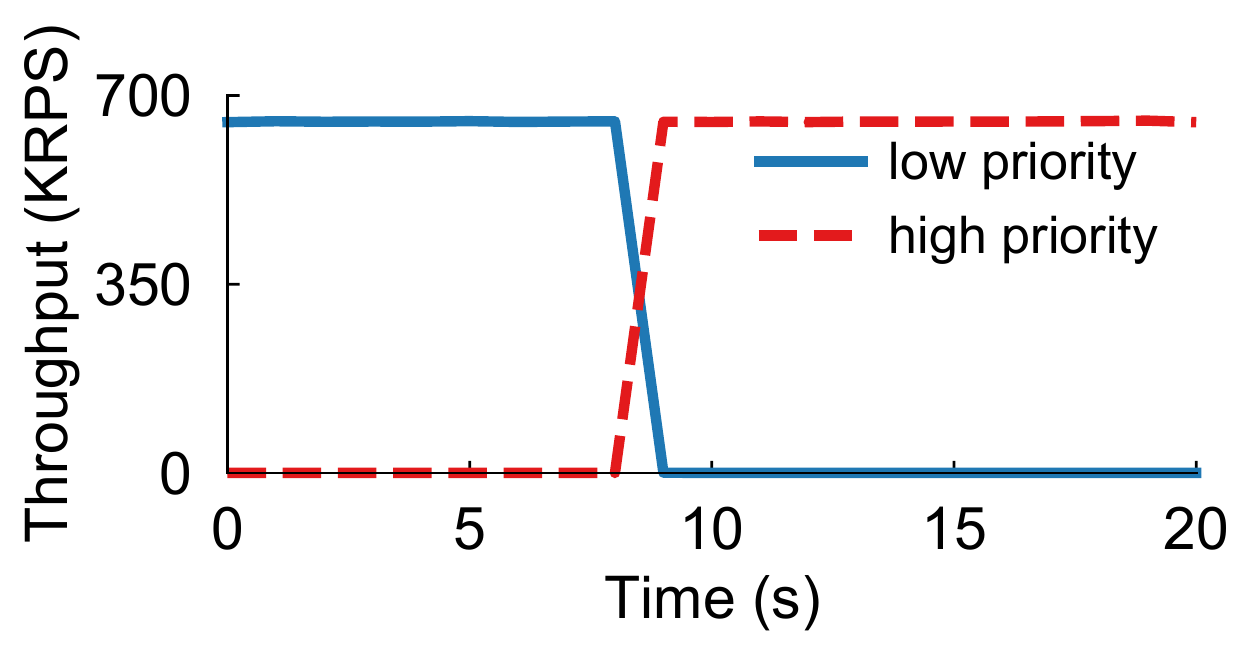}}
    \vspace{-0.1in}
    \caption{\sysname can support strict priority.}
    \vspace{-0.1in}
    \label{fig:eval_priority}
\end{figure}

\para{Step 2. Insensitivity of service distribution.} A combination of special
case analysis and simulation studies~\cite{richa2001power} were used to justify
that $M/G/K/JSQ/PS \approx M/M/K/JSQ/PS$, i.e., the equilibrium of the system is
\emph{insensitive} to the service distribution. First, special case analysis
shows that when $k = 1$, the response time of $M_n/G/1/PS$ is insensitive to the
distribution of the service time. Second, extensive simulations confirm that the
result continues to hold when $K > 1$ and JSQ is the only known policy that is
robust against service time. Thus, C2 is obtained.

\subsection{Analysis for C3}

This section explains C3, i.e., $M/G/K/JSQ/FCFS$ is
optimal when the jobs have low-dispersion (e.g., service time is exponential).
$M/G/K/JSQ/FCFS$ is an extensively examined
policy~\cite{weber1978optimal,winston1977optimality,ephremides1980simple,richa2001power,gupta2007insensitivity}.
The efficacy of this policy for low-dispersion jobs is a known result. When the
service time is an exponential distribution, the policy is \emph{provably
optimal} under multiple metrics, including throughput, average service time, and
longest queue~\cite{winston1977optimality}. The exponential service time
assumption can be relaxed for other low-dispersion settings (e.g., when it is a
random variable with non-decreasing hazard rate)~\cite{weber1978optimal}.

While we do not repeat the analysis in the literature, we explain the key
intuitions here. We need to only understand why PS is unnecessary when the jobs
have low dispersion, and why JSQ is optimal when we use FCFS for each server.

\begin{figure}[t]
    \centering
    \subfigure[Exp(50).]{
        \label{fig:eval_mix_exp}
        \includegraphics[width=0.48\linewidth]{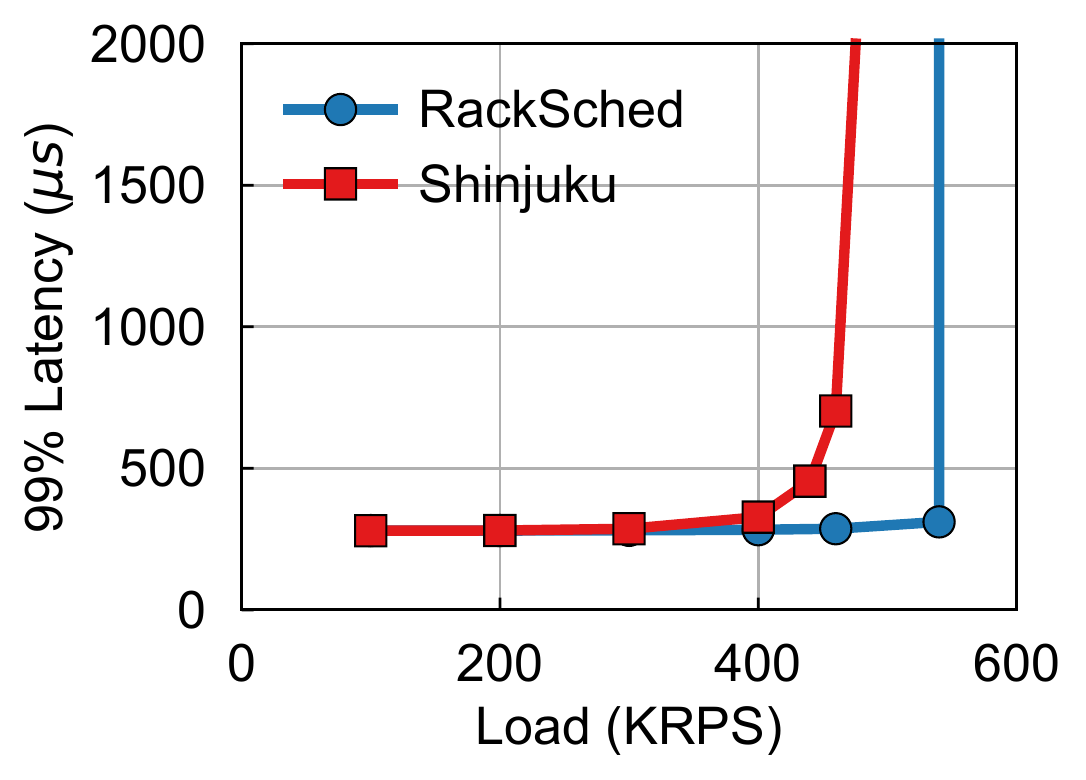}}
    \subfigure[Bimodal(90\%-50,10\%-500).]{
        \label{fig:eval_mix_bimodal}
        \includegraphics[width=0.48\linewidth]{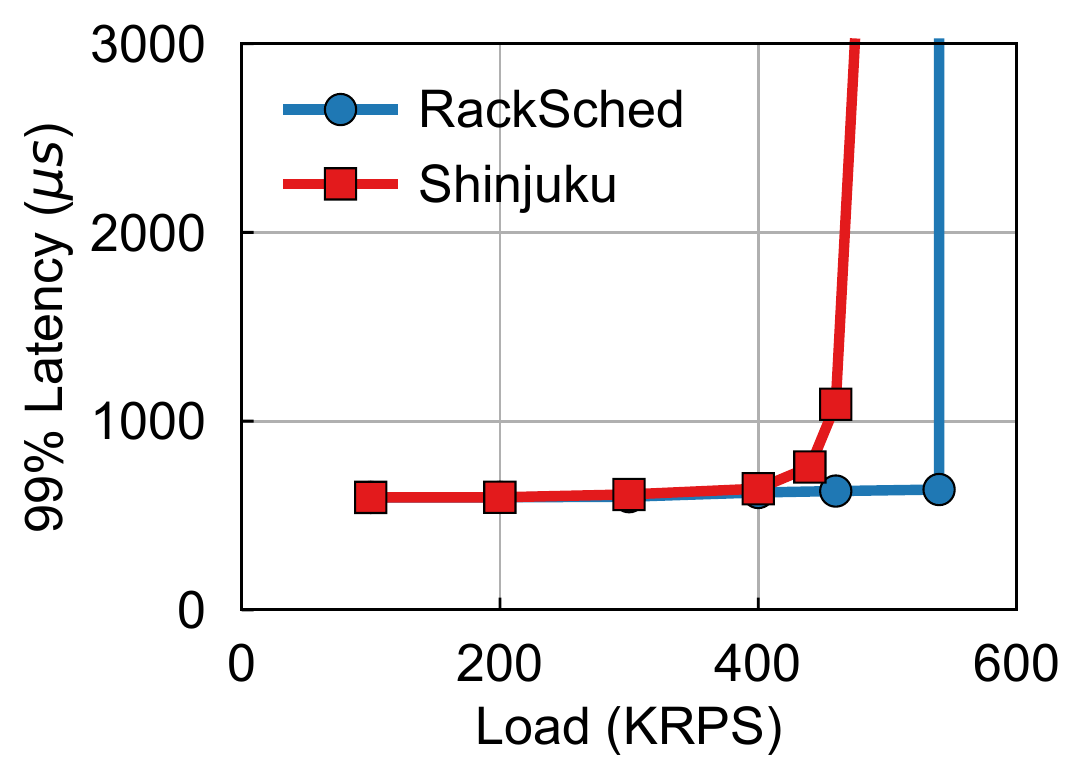}}
    \vspace{-0.1in}
    \caption{\sysname with multiple applications.}
    \vspace{-0.1in}
    \label{fig:eval_mix}
\end{figure}

\noindent{\emph{Optimality for FCFS.}} We first review when PS is useful.
Consider a high-dispersion (heavy-tailed) setting with one server, in which there
are 9 jobs with processing time 1 and 1 job with processing time 10. The total
processing time for all jobs is the same, regardless of using FCFS or PS. The
key edge of PS is its ability to minimize the tail distribution. When we use
FCFS and the jobs arrive in a bad order (e.g., the job with processing time 10
needs to be processed first), all smaller jobs are blocked by the large one, which
causes an increase in median and 90\% quantile processing time. In this case, PS
that continuously shuffles the jobs will reduce median/90\% quantile processing
time.

On the other hand, in a setting where all jobs have uniform processing time, the
blocking phenomenon does not exist so PS will not improve tail distributions. In
practice, the overhead of context switch can become non-negligible so FCFS is
perferred.

\noindent{\emph{Optimality of JSQ.}} Consider a setting, in which JSQ assigns a
job to server 1 with load $x_1 = 3$, whereas a different policy assigns the job
to server 2 with load $x_2 = 20$. Here, $x_1 < x_2$. The loads of using JSQ
becomes $(x_1+1, x_2) = (4, 20)$, whereas the loads of the other policy is
$(x_1, x_2+1) = (3, 21)$. Because jobs have low-dispersion, time to process 3
jobs for server 1 is approximately the same as that for server 2. This means:
\emph{(i)} the new job will always have a shorter waiting time when using JSQ;
therefore, JSQ minimizes average waiting time. \emph{(ii)} it is less likely for
JSQ to have idle servers (e.g., server 1 finishes processing all jobs without
seeing a new one). Therefore, JSQ also maximizes the througput. \emph{(iii)} The
max load of JSQ is smaller (20 vs 21). Therefore JSQ also minimizes the heaviest
load.

Note that the above arguments rely on the assumption that processing time for
any jobs is approximately the same, i.e., jobs have low dispersion. When this
assumption is violated, JSQ with PS is better.

\section{Additional Evaluation Results}
\label{sec:appendix:add_eval}

\subsection{Locality and Placement Constraints}

\sysname can achieve significant improvement over Shinjuku when there are locality constraints. In
this experiment, we use Bimodal(90\%-50, 10\%-500) workload.
We restrict half of the requests (Type-I) to be only processed by four servers,
and other half of the requests (Type-II) can still be processed by all eight servers.
Figure~\ref{fig:eval_locality_total} shows the tail latency across all requests.
We also break down the results for the Type-I and Type-II tail latencies in
Figure~\ref{fig:eval_locality_1} and Figure~\ref{fig:eval_locality_2}
respectively. The x-axis in these figures is the total request load of two types of requests.
Shinjuku does random scheduling for the two types of requests, while
\sysname tends to send the Type-II requests to the four servers without the workload of the Type-I requests,
since the other four servers are already loaded by the Type-I requests.
Thus both the Type-I and Type-II requests can get better performance with \sysname.

\subsection{Priority Policies}

\sysname supports strict priority policies where it can prioritize one type of requests
over others. In this scenario, we only send the low-priority requests with 700KRPS under
Bimodal(90\%-50, 10\%-500) distribution at first. Such workload is enough to saturate our
8-server rack. Then at time 8~s, the high-priority requests with the same service time distribution
enter the system at a load of 700KRPS. Figure~\ref{fig:eval_priority_wo} shows the throughput of high- and
low-priority requests without priority policy enforcement. The requests with
different priorities share the server capacity equally, and they have similar throughputs. With the priority
policy enforced, \sysname prioritizes the high-priority requests.
Figure~\ref{fig:eval_priority_with} shows the throughput of low-priority
requests drops to near zero after the high-priority requests enters the system and
occupies the workers' capability.

\subsection{Multiple Applications}

This experiment shows how \sysname supports multiple applications.
One application processes requests under Exp(50) and runs on all the
eight servers, and the other application processes requests under
Bimodal(90\%-50, 10\%-500) and runs on only four servers in the rack. The clients
send requests to the two applications at the same load.
Figure~\ref{fig:eval_mix} shows that \sysname supports larger request load with lower tail
latency for both applications compared with Shinjuku.
The x-axis is the total request load of two applications.

\end{appendix}

\end{document}